\documentclass[longauth]{aa} 

\newcommand{\pixedfit}{\texttt{piXedfit}}
\newcommand{\bagpipes}{\textsc{BAGPIPES}}
\newcommand{\sextractor}{\textsc{SExtractor}}
\newcommand{\sep}{\textsc{sep}}

\newcommand{\photutils}{\textsc{photutils}}

\newcommand{\galex}{GALEX}
\newcommand{\lsst}{LSST}
\newcommand{\euclid}{\textit{Euclid}}
\newcommand{\jwst}{JWST}
\newcommand{\hst}{HST}
\newcommand{\hsc}{HSC}
\newcommand{\des}{DES}
\newcommand{\ie}{$I_{\rm E}$}
\newcommand{\ye}{$Y_{\rm E}$}
\newcommand{\je}{$J_{\rm E}$}
\newcommand{\he}{$H_{\rm E}$}
\newcommand{\mass}{$M_{*}$}
\newcommand{\msun}{$M_{\odot}$}
\newcommand{\massd}{$\Sigma_{*}$}
\newcommand{\sfrd}{$\Sigma_{\rm SFR}$}
\newcommand{\gledc}{G + L + E}
\newcommand{\ledc}{L + E}
\newcommand{\logten}{\ensuremath{\log_{10}}} 

\usepackage{hyperref}
\hypersetup{
    colorlinks=true,
    linkcolor=blue,       
    citecolor=blue,       
    urlcolor=blue,        
    filecolor=blue,       
    anchorcolor=blue,     
}

\usepackage{amsmath}
\usepackage{float}
\usepackage{graphicx}
\usepackage{euclid}
\usepackage{graphicx}
\usepackage{natbib}
\usepackage{scalerel}
\usepackage{siunitx}
\usepackage{xcolor}
\usepackage{amssymb}
\usepackage{multirow}
\usepackage{tabularx}
\usepackage{txfonts}
\usepackage{orcidlink}

\usepackage{subcaption}         
\usepackage{lscape}             
\usepackage{placeins}           

\usepackage[switch, modulo]{lineno}

\def\msun{\mbox{M$_{\rm \odot}$}}
\def\om{\mbox{$\Omega_{\rm m}$}}
\def\ol{\mbox{$\Omega_{\rm \Lambda}$}}

\begin{document} 

\title{\euclid\ preparation}
\subtitle{LXXIII. Spatially resolved stellar populations of local galaxies with \euclid: A proof of concept using synthetic images with the TNG50 simulation}

\newcommand{\orcid}[1]{}			   
\author{Euclid Collaboration: Abdurro'uf\orcid{0000-0002-5258-8761}\thanks{\email{fabdurr1@jhu.edu}}\inst{\ref{aff1}}
\and C.~Tortora\orcid{0000-0001-7958-6531}\inst{\ref{aff2}}
\and M.~Baes\orcid{0000-0002-3930-2757}\inst{\ref{aff3}}
\and A.~Nersesian\orcid{0000-0001-6843-409X}\inst{\ref{aff4},\ref{aff3}}
\and I.~Kova{\v{c}}i{\'{c}}\orcid{0000-0001-6751-3263}\inst{\ref{aff3}}
\and M.~Bolzonella\orcid{0000-0003-3278-4607}\inst{\ref{aff5}}
\and A.~Lan\c{c}on\orcid{0000-0002-7214-8296}\inst{\ref{aff6}}
\and L.~Bisigello\orcid{0000-0003-0492-4924}\inst{\ref{aff7},\ref{aff8}}
\and F.~Annibali\inst{\ref{aff5}}
\and M.~N.~Bremer\inst{\ref{aff9}}
\and D.~Carollo\orcid{0000-0002-0005-5787}\inst{\ref{aff10}}
\and C.~J.~Conselice\orcid{0000-0003-1949-7638}\inst{\ref{aff11}}
\and A.~Enia\orcid{0000-0002-0200-2857}\inst{\ref{aff12},\ref{aff5}}
\and A.~M.~N.~Ferguson\inst{\ref{aff13}}
\and A.~Ferr\'e-Mateu\orcid{0000-0002-6411-220X}\inst{\ref{aff14},\ref{aff15}}
\and L.~K.~Hunt\orcid{0000-0001-9162-2371}\inst{\ref{aff16}}
\and E.~Iodice\orcid{0000-0003-4291-0005}\inst{\ref{aff2}}
\and J.~H.~Knapen\orcid{0000-0003-1643-0024}\inst{\ref{aff14},\ref{aff15}}
\and A.~Iovino\orcid{0000-0001-6958-0304}\inst{\ref{aff17}}
\and F.~R.~Marleau\orcid{0000-0002-1442-2947}\inst{\ref{aff18}}
\and R.~F.~Peletier\orcid{0000-0001-7621-947X}\inst{\ref{aff19}}
\and R.~Ragusa\inst{\ref{aff2}}
\and M.~Rejkuba\orcid{0000-0002-6577-2787}\inst{\ref{aff20}}
\and A.~S.~G.~Robotham\orcid{0000-0003-0429-3579}\inst{\ref{aff21}}
\and J.~Rom\'an\orcid{0000-0002-3849-3467}\inst{\ref{aff22}}
\and T.~Saifollahi\orcid{0000-0002-9554-7660}\inst{\ref{aff6}}
\and P.~Salucci\orcid{0000-0002-5476-2954}\inst{\ref{aff23},\ref{aff24},\ref{aff25}}
\and M.~Scodeggio\inst{\ref{aff26}}
\and M.~Siudek\orcid{0000-0002-2949-2155}\inst{\ref{aff27},\ref{aff28}}
\and A.~van~der~Wel\orcid{0000-0002-5027-0135}\inst{\ref{aff3}}
\and K.~Voggel\orcid{0000-0001-6215-0950}\inst{\ref{aff6}}
\and B.~Altieri\orcid{0000-0003-3936-0284}\inst{\ref{aff29}}
\and S.~Andreon\orcid{0000-0002-2041-8784}\inst{\ref{aff17}}
\and C.~Baccigalupi\orcid{0000-0002-8211-1630}\inst{\ref{aff23},\ref{aff10},\ref{aff25},\ref{aff24}}
\and M.~Baldi\orcid{0000-0003-4145-1943}\inst{\ref{aff30},\ref{aff5},\ref{aff31}}
\and S.~Bardelli\orcid{0000-0002-8900-0298}\inst{\ref{aff5}}
\and A.~Biviano\orcid{0000-0002-0857-0732}\inst{\ref{aff10},\ref{aff23}}
\and A.~Bonchi\orcid{0000-0002-2667-5482}\inst{\ref{aff32}}
\and D.~Bonino\orcid{0000-0002-3336-9977}\inst{\ref{aff33}}
\and E.~Branchini\orcid{0000-0002-0808-6908}\inst{\ref{aff34},\ref{aff35},\ref{aff17}}
\and M.~Brescia\orcid{0000-0001-9506-5680}\inst{\ref{aff36},\ref{aff2},\ref{aff37}}
\and J.~Brinchmann\orcid{0000-0003-4359-8797}\inst{\ref{aff38},\ref{aff39}}
\and A.~Caillat\inst{\ref{aff40}}
\and S.~Camera\orcid{0000-0003-3399-3574}\inst{\ref{aff41},\ref{aff42},\ref{aff33}}
\and G.~Ca\~nas-Herrera\orcid{0000-0003-2796-2149}\inst{\ref{aff43},\ref{aff44}}
\and V.~Capobianco\orcid{0000-0002-3309-7692}\inst{\ref{aff33}}
\and C.~Carbone\orcid{0000-0003-0125-3563}\inst{\ref{aff26}}
\and J.~Carretero\orcid{0000-0002-3130-0204}\inst{\ref{aff45},\ref{aff46}}
\and S.~Casas\orcid{0000-0002-4751-5138}\inst{\ref{aff47},\ref{aff48}}
\and M.~Castellano\orcid{0000-0001-9875-8263}\inst{\ref{aff49}}
\and G.~Castignani\orcid{0000-0001-6831-0687}\inst{\ref{aff5}}
\and S.~Cavuoti\orcid{0000-0002-3787-4196}\inst{\ref{aff2},\ref{aff37}}
\and K.~C.~Chambers\orcid{0000-0001-6965-7789}\inst{\ref{aff50}}
\and A.~Cimatti\inst{\ref{aff51}}
\and C.~Colodro-Conde\inst{\ref{aff14}}
\and G.~Congedo\orcid{0000-0003-2508-0046}\inst{\ref{aff13}}
\and L.~Conversi\orcid{0000-0002-6710-8476}\inst{\ref{aff52},\ref{aff29}}
\and Y.~Copin\orcid{0000-0002-5317-7518}\inst{\ref{aff53}}
\and F.~Courbin\orcid{0000-0003-0758-6510}\inst{\ref{aff54},\ref{aff55},\ref{aff56}}
\and H.~M.~Courtois\orcid{0000-0003-0509-1776}\inst{\ref{aff57}}
\and M.~Cropper\orcid{0000-0003-4571-9468}\inst{\ref{aff58}}
\and A.~Da~Silva\orcid{0000-0002-6385-1609}\inst{\ref{aff59},\ref{aff60}}
\and H.~Degaudenzi\orcid{0000-0002-5887-6799}\inst{\ref{aff61}}
\and G.~De~Lucia\orcid{0000-0002-6220-9104}\inst{\ref{aff10}}
\and A.~M.~Di~Giorgio\orcid{0000-0002-4767-2360}\inst{\ref{aff62}}
\and J.~Dinis\orcid{0000-0001-5075-1601}\inst{\ref{aff59},\ref{aff60}}
\and H.~Dole\orcid{0000-0002-9767-3839}\inst{\ref{aff63}}
\and F.~Dubath\orcid{0000-0002-6533-2810}\inst{\ref{aff61}}
\and X.~Dupac\inst{\ref{aff29}}
\and S.~Dusini\orcid{0000-0002-1128-0664}\inst{\ref{aff64}}
\and S.~Escoffier\orcid{0000-0002-2847-7498}\inst{\ref{aff65}}
\and M.~Farina\orcid{0000-0002-3089-7846}\inst{\ref{aff62}}
\and R.~Farinelli\inst{\ref{aff5}}
\and S.~Farrens\orcid{0000-0002-9594-9387}\inst{\ref{aff66}}
\and F.~Faustini\orcid{0000-0001-6274-5145}\inst{\ref{aff32},\ref{aff49}}
\and S.~Ferriol\inst{\ref{aff53}}
\and F.~Finelli\orcid{0000-0002-6694-3269}\inst{\ref{aff5},\ref{aff67}}
\and S.~Fotopoulou\orcid{0000-0002-9686-254X}\inst{\ref{aff9}}
\and M.~Frailis\orcid{0000-0002-7400-2135}\inst{\ref{aff10}}
\and E.~Franceschi\orcid{0000-0002-0585-6591}\inst{\ref{aff5}}
\and M.~Fumana\orcid{0000-0001-6787-5950}\inst{\ref{aff26}}
\and S.~Galeotta\orcid{0000-0002-3748-5115}\inst{\ref{aff10}}
\and B.~Gillis\orcid{0000-0002-4478-1270}\inst{\ref{aff13}}
\and C.~Giocoli\orcid{0000-0002-9590-7961}\inst{\ref{aff5},\ref{aff68}}
\and P.~G\'omez-Alvarez\orcid{0000-0002-8594-5358}\inst{\ref{aff69},\ref{aff29}}
\and J.~Gracia-Carpio\inst{\ref{aff70}}
\and A.~Grazian\orcid{0000-0002-5688-0663}\inst{\ref{aff71}}
\and F.~Grupp\inst{\ref{aff70},\ref{aff72}}
\and W.~Holmes\inst{\ref{aff73}}
\and F.~Hormuth\inst{\ref{aff74}}
\and A.~Hornstrup\orcid{0000-0002-3363-0936}\inst{\ref{aff75},\ref{aff76}}
\and P.~Hudelot\inst{\ref{aff77}}
\and K.~Jahnke\orcid{0000-0003-3804-2137}\inst{\ref{aff78}}
\and M.~Jhabvala\inst{\ref{aff79}}
\and E.~Keih\"anen\orcid{0000-0003-1804-7715}\inst{\ref{aff80}}
\and S.~Kermiche\orcid{0000-0002-0302-5735}\inst{\ref{aff65}}
\and A.~Kiessling\orcid{0000-0002-2590-1273}\inst{\ref{aff73}}
\and M.~Kilbinger\orcid{0000-0001-9513-7138}\inst{\ref{aff66}}
\and B.~Kubik\orcid{0009-0006-5823-4880}\inst{\ref{aff53}}
\and M.~K\"ummel\orcid{0000-0003-2791-2117}\inst{\ref{aff72}}
\and M.~Kunz\orcid{0000-0002-3052-7394}\inst{\ref{aff81}}
\and H.~Kurki-Suonio\orcid{0000-0002-4618-3063}\inst{\ref{aff82},\ref{aff83}}
\and A.~M.~C.~Le~Brun\orcid{0000-0002-0936-4594}\inst{\ref{aff84}}
\and S.~Ligori\orcid{0000-0003-4172-4606}\inst{\ref{aff33}}
\and P.~B.~Lilje\orcid{0000-0003-4324-7794}\inst{\ref{aff85}}
\and V.~Lindholm\orcid{0000-0003-2317-5471}\inst{\ref{aff82},\ref{aff83}}
\and I.~Lloro\orcid{0000-0001-5966-1434}\inst{\ref{aff86}}
\and G.~Mainetti\orcid{0000-0003-2384-2377}\inst{\ref{aff87}}
\and D.~Maino\inst{\ref{aff88},\ref{aff26},\ref{aff89}}
\and E.~Maiorano\orcid{0000-0003-2593-4355}\inst{\ref{aff5}}
\and O.~Mansutti\orcid{0000-0001-5758-4658}\inst{\ref{aff10}}
\and O.~Marggraf\orcid{0000-0001-7242-3852}\inst{\ref{aff90}}
\and K.~Markovic\orcid{0000-0001-6764-073X}\inst{\ref{aff73}}
\and M.~Martinelli\orcid{0000-0002-6943-7732}\inst{\ref{aff49},\ref{aff91}}
\and N.~Martinet\orcid{0000-0003-2786-7790}\inst{\ref{aff40}}
\and F.~Marulli\orcid{0000-0002-8850-0303}\inst{\ref{aff12},\ref{aff5},\ref{aff31}}
\and R.~Massey\orcid{0000-0002-6085-3780}\inst{\ref{aff92}}
\and E.~Medinaceli\orcid{0000-0002-4040-7783}\inst{\ref{aff5}}
\and S.~Mei\orcid{0000-0002-2849-559X}\inst{\ref{aff93}}
\and M.~Melchior\inst{\ref{aff94}}
\and Y.~Mellier\inst{\ref{aff95},\ref{aff77}}
\and M.~Meneghetti\orcid{0000-0003-1225-7084}\inst{\ref{aff5},\ref{aff31}}
\and E.~Merlin\orcid{0000-0001-6870-8900}\inst{\ref{aff49}}
\and G.~Meylan\inst{\ref{aff54}}
\and A.~Mora\orcid{0000-0002-1922-8529}\inst{\ref{aff96}}
\and M.~Moresco\orcid{0000-0002-7616-7136}\inst{\ref{aff12},\ref{aff5}}
\and L.~Moscardini\orcid{0000-0002-3473-6716}\inst{\ref{aff12},\ref{aff5},\ref{aff31}}
\and S.-M.~Niemi\inst{\ref{aff43}}
\and J.~W.~Nightingale\orcid{0000-0002-8987-7401}\inst{\ref{aff97}}
\and C.~Padilla\orcid{0000-0001-7951-0166}\inst{\ref{aff98}}
\and S.~Paltani\orcid{0000-0002-8108-9179}\inst{\ref{aff61}}
\and F.~Pasian\orcid{0000-0002-4869-3227}\inst{\ref{aff10}}
\and K.~Pedersen\inst{\ref{aff99}}
\and V.~Pettorino\inst{\ref{aff43}}
\and G.~Polenta\orcid{0000-0003-4067-9196}\inst{\ref{aff32}}
\and M.~Poncet\inst{\ref{aff100}}
\and L.~A.~Popa\inst{\ref{aff101}}
\and L.~Pozzetti\orcid{0000-0001-7085-0412}\inst{\ref{aff5}}
\and F.~Raison\orcid{0000-0002-7819-6918}\inst{\ref{aff70}}
\and A.~Renzi\orcid{0000-0001-9856-1970}\inst{\ref{aff8},\ref{aff64}}
\and J.~Rhodes\orcid{0000-0002-4485-8549}\inst{\ref{aff73}}
\and G.~Riccio\inst{\ref{aff2}}
\and E.~Romelli\orcid{0000-0003-3069-9222}\inst{\ref{aff10}}
\and M.~Roncarelli\orcid{0000-0001-9587-7822}\inst{\ref{aff5}}
\and E.~Rossetti\orcid{0000-0003-0238-4047}\inst{\ref{aff30}}
\and R.~Saglia\orcid{0000-0003-0378-7032}\inst{\ref{aff72},\ref{aff70}}
\and Z.~Sakr\orcid{0000-0002-4823-3757}\inst{\ref{aff102},\ref{aff103},\ref{aff104}}
\and D.~Sapone\orcid{0000-0001-7089-4503}\inst{\ref{aff105}}
\and B.~Sartoris\orcid{0000-0003-1337-5269}\inst{\ref{aff72},\ref{aff10}}
\and M.~Schirmer\orcid{0000-0003-2568-9994}\inst{\ref{aff78}}
\and P.~Schneider\orcid{0000-0001-8561-2679}\inst{\ref{aff90}}
\and T.~Schrabback\orcid{0000-0002-6987-7834}\inst{\ref{aff18}}
\and A.~Secroun\orcid{0000-0003-0505-3710}\inst{\ref{aff65}}
\and E.~Sefusatti\orcid{0000-0003-0473-1567}\inst{\ref{aff10},\ref{aff23},\ref{aff25}}
\and G.~Seidel\orcid{0000-0003-2907-353X}\inst{\ref{aff78}}
\and S.~Serrano\orcid{0000-0002-0211-2861}\inst{\ref{aff106},\ref{aff107},\ref{aff28}}
\and P.~Simon\inst{\ref{aff90}}
\and C.~Sirignano\orcid{0000-0002-0995-7146}\inst{\ref{aff8},\ref{aff64}}
\and G.~Sirri\orcid{0000-0003-2626-2853}\inst{\ref{aff31}}
\and L.~Stanco\orcid{0000-0002-9706-5104}\inst{\ref{aff64}}
\and J.~Steinwagner\orcid{0000-0001-7443-1047}\inst{\ref{aff70}}
\and P.~Tallada-Cresp\'{i}\orcid{0000-0002-1336-8328}\inst{\ref{aff45},\ref{aff46}}
\and A.~N.~Taylor\inst{\ref{aff13}}
\and I.~Tereno\inst{\ref{aff59},\ref{aff108}}
\and S.~Toft\orcid{0000-0003-3631-7176}\inst{\ref{aff109},\ref{aff110}}
\and R.~Toledo-Moreo\orcid{0000-0002-2997-4859}\inst{\ref{aff111}}
\and F.~Torradeflot\orcid{0000-0003-1160-1517}\inst{\ref{aff46},\ref{aff45}}
\and I.~Tutusaus\orcid{0000-0002-3199-0399}\inst{\ref{aff103}}
\and L.~Valenziano\orcid{0000-0002-1170-0104}\inst{\ref{aff5},\ref{aff67}}
\and J.~Valiviita\orcid{0000-0001-6225-3693}\inst{\ref{aff82},\ref{aff83}}
\and T.~Vassallo\orcid{0000-0001-6512-6358}\inst{\ref{aff72},\ref{aff10}}
\and G.~Verdoes~Kleijn\orcid{0000-0001-5803-2580}\inst{\ref{aff19}}
\and A.~Veropalumbo\orcid{0000-0003-2387-1194}\inst{\ref{aff17},\ref{aff35},\ref{aff112}}
\and Y.~Wang\orcid{0000-0002-4749-2984}\inst{\ref{aff113}}
\and J.~Weller\orcid{0000-0002-8282-2010}\inst{\ref{aff72},\ref{aff70}}
\and G.~Zamorani\orcid{0000-0002-2318-301X}\inst{\ref{aff5}}
\and E.~Zucca\orcid{0000-0002-5845-8132}\inst{\ref{aff5}}
\and E.~Bozzo\orcid{0000-0002-8201-1525}\inst{\ref{aff61}}
\and C.~Burigana\orcid{0000-0002-3005-5796}\inst{\ref{aff7},\ref{aff67}}
\and M.~Calabrese\orcid{0000-0002-2637-2422}\inst{\ref{aff114},\ref{aff26}}
\and D.~Di~Ferdinando\inst{\ref{aff31}}
\and J.~A.~Escartin~Vigo\inst{\ref{aff70}}
\and S.~Matthew\orcid{0000-0001-8448-1697}\inst{\ref{aff13}}
\and N.~Mauri\orcid{0000-0001-8196-1548}\inst{\ref{aff51},\ref{aff31}}
\and M.~P\"ontinen\orcid{0000-0001-5442-2530}\inst{\ref{aff82}}
\and C.~Porciani\orcid{0000-0002-7797-2508}\inst{\ref{aff90}}
\and V.~Scottez\inst{\ref{aff95},\ref{aff115}}
\and M.~Tenti\orcid{0000-0002-4254-5901}\inst{\ref{aff31}}
\and M.~Viel\orcid{0000-0002-2642-5707}\inst{\ref{aff23},\ref{aff10},\ref{aff24},\ref{aff25},\ref{aff116}}
\and M.~Wiesmann\orcid{0009-0000-8199-5860}\inst{\ref{aff85}}
\and Y.~Akrami\orcid{0000-0002-2407-7956}\inst{\ref{aff117},\ref{aff118}}
\and V.~Allevato\orcid{0000-0001-7232-5152}\inst{\ref{aff2}}
\and S.~Anselmi\orcid{0000-0002-3579-9583}\inst{\ref{aff64},\ref{aff8},\ref{aff84}}
\and M.~Archidiacono\orcid{0000-0003-4952-9012}\inst{\ref{aff88},\ref{aff89}}
\and F.~Atrio-Barandela\orcid{0000-0002-2130-2513}\inst{\ref{aff119}}
\and M.~Ballardini\orcid{0000-0003-4481-3559}\inst{\ref{aff120},\ref{aff5},\ref{aff121}}
\and D.~Bertacca\orcid{0000-0002-2490-7139}\inst{\ref{aff8},\ref{aff71},\ref{aff64}}
\and A.~Blanchard\orcid{0000-0001-8555-9003}\inst{\ref{aff103}}
\and L.~Blot\orcid{0000-0002-9622-7167}\inst{\ref{aff122},\ref{aff84}}
\and S.~Borgani\orcid{0000-0001-6151-6439}\inst{\ref{aff123},\ref{aff23},\ref{aff10},\ref{aff25},\ref{aff116}}
\and M.~L.~Brown\orcid{0000-0002-0370-8077}\inst{\ref{aff11}}
\and S.~Bruton\orcid{0000-0002-6503-5218}\inst{\ref{aff124}}
\and R.~Cabanac\orcid{0000-0001-6679-2600}\inst{\ref{aff103}}
\and A.~Calabro\orcid{0000-0003-2536-1614}\inst{\ref{aff49}}
\and A.~Cappi\inst{\ref{aff5},\ref{aff125}}
\and F.~Caro\inst{\ref{aff49}}
\and C.~S.~Carvalho\inst{\ref{aff108}}
\and T.~Castro\orcid{0000-0002-6292-3228}\inst{\ref{aff10},\ref{aff25},\ref{aff23},\ref{aff116}}
\and F.~Cogato\orcid{0000-0003-4632-6113}\inst{\ref{aff12},\ref{aff5}}
\and T.~Contini\orcid{0000-0003-0275-938X}\inst{\ref{aff103}}
\and A.~R.~Cooray\orcid{0000-0002-3892-0190}\inst{\ref{aff126}}
\and O.~Cucciati\orcid{0000-0002-9336-7551}\inst{\ref{aff5}}
\and G.~Desprez\orcid{0000-0001-8325-1742}\inst{\ref{aff127}}
\and A.~D\'iaz-S\'anchez\orcid{0000-0003-0748-4768}\inst{\ref{aff128}}
\and S.~Di~Domizio\orcid{0000-0003-2863-5895}\inst{\ref{aff34},\ref{aff35}}
\and A.~G.~Ferrari\orcid{0009-0005-5266-4110}\inst{\ref{aff31}}
\and I.~Ferrero\orcid{0000-0002-1295-1132}\inst{\ref{aff85}}
\and A.~Finoguenov\orcid{0000-0002-4606-5403}\inst{\ref{aff82}}
\and A.~Fontana\orcid{0000-0003-3820-2823}\inst{\ref{aff49}}
\and F.~Fornari\orcid{0000-0003-2979-6738}\inst{\ref{aff67}}
\and K.~Ganga\orcid{0000-0001-8159-8208}\inst{\ref{aff93}}
\and J.~Garc\'ia-Bellido\orcid{0000-0002-9370-8360}\inst{\ref{aff117}}
\and T.~Gasparetto\orcid{0000-0002-7913-4866}\inst{\ref{aff10}}
\and E.~Gaztanaga\orcid{0000-0001-9632-0815}\inst{\ref{aff28},\ref{aff106},\ref{aff48}}
\and F.~Giacomini\orcid{0000-0002-3129-2814}\inst{\ref{aff31}}
\and F.~Gianotti\orcid{0000-0003-4666-119X}\inst{\ref{aff5}}
\and G.~Gozaliasl\orcid{0000-0002-0236-919X}\inst{\ref{aff129}}
\and A.~Gregorio\orcid{0000-0003-4028-8785}\inst{\ref{aff123},\ref{aff10},\ref{aff25}}
\and M.~Guidi\orcid{0000-0001-9408-1101}\inst{\ref{aff30},\ref{aff5}}
\and C.~M.~Gutierrez\orcid{0000-0001-7854-783X}\inst{\ref{aff130}}
\and A.~Hall\orcid{0000-0002-3139-8651}\inst{\ref{aff13}}
\and S.~Hemmati\orcid{0000-0003-2226-5395}\inst{\ref{aff131}}
\and H.~Hildebrandt\orcid{0000-0002-9814-3338}\inst{\ref{aff132}}
\and J.~Hjorth\orcid{0000-0002-4571-2306}\inst{\ref{aff99}}
\and M.~Huertas-Company\orcid{0000-0002-1416-8483}\inst{\ref{aff14},\ref{aff27},\ref{aff133},\ref{aff134}}
\and A.~Jimenez~Mu\~noz\orcid{0009-0004-5252-185X}\inst{\ref{aff135}}
\and J.~J.~E.~Kajava\orcid{0000-0002-3010-8333}\inst{\ref{aff136},\ref{aff137}}
\and Y.~Kang\orcid{0009-0000-8588-7250}\inst{\ref{aff61}}
\and V.~Kansal\orcid{0000-0002-4008-6078}\inst{\ref{aff138},\ref{aff139}}
\and D.~Karagiannis\orcid{0000-0002-4927-0816}\inst{\ref{aff140},\ref{aff141}}
\and C.~C.~Kirkpatrick\inst{\ref{aff80}}
\and S.~Kruk\orcid{0000-0001-8010-8879}\inst{\ref{aff29}}
\and M.~Lattanzi\orcid{0000-0003-1059-2532}\inst{\ref{aff121}}
\and S.~Lee\orcid{0000-0002-8289-740X}\inst{\ref{aff73}}
\and J.~Le~Graet\orcid{0000-0001-6523-7971}\inst{\ref{aff65}}
\and L.~Legrand\orcid{0000-0003-0610-5252}\inst{\ref{aff142},\ref{aff143}}
\and M.~Lembo\orcid{0000-0002-5271-5070}\inst{\ref{aff120},\ref{aff121}}
\and J.~Lesgourgues\orcid{0000-0001-7627-353X}\inst{\ref{aff47}}
\and T.~I.~Liaudat\orcid{0000-0002-9104-314X}\inst{\ref{aff144}}
\and A.~Loureiro\orcid{0000-0002-4371-0876}\inst{\ref{aff145},\ref{aff146}}
\and J.~Macias-Perez\orcid{0000-0002-5385-2763}\inst{\ref{aff135}}
\and M.~Magliocchetti\orcid{0000-0001-9158-4838}\inst{\ref{aff62}}
\and F.~Mannucci\orcid{0000-0002-4803-2381}\inst{\ref{aff16}}
\and R.~Maoli\orcid{0000-0002-6065-3025}\inst{\ref{aff147},\ref{aff49}}
\and J.~Mart\'{i}n-Fleitas\orcid{0000-0002-8594-569X}\inst{\ref{aff96}}
\and C.~J.~A.~P.~Martins\orcid{0000-0002-4886-9261}\inst{\ref{aff148},\ref{aff38}}
\and L.~Maurin\orcid{0000-0002-8406-0857}\inst{\ref{aff63}}
\and R.~B.~Metcalf\orcid{0000-0003-3167-2574}\inst{\ref{aff12},\ref{aff5}}
\and M.~Miluzio\inst{\ref{aff29},\ref{aff149}}
\and P.~Monaco\orcid{0000-0003-2083-7564}\inst{\ref{aff123},\ref{aff10},\ref{aff25},\ref{aff23}}
\and C.~Moretti\orcid{0000-0003-3314-8936}\inst{\ref{aff24},\ref{aff116},\ref{aff10},\ref{aff23},\ref{aff25}}
\and G.~Morgante\inst{\ref{aff5}}
\and K.~Naidoo\orcid{0000-0002-9182-1802}\inst{\ref{aff150}}
\and Nicholas~A.~Walton\orcid{0000-0003-3983-8778}\inst{\ref{aff151}}
\and K.~Paterson\orcid{0000-0001-8340-3486}\inst{\ref{aff78}}
\and L.~Patrizii\inst{\ref{aff31}}
\and A.~Pisani\orcid{0000-0002-6146-4437}\inst{\ref{aff65},\ref{aff152}}
\and V.~Popa\orcid{0000-0002-9118-8330}\inst{\ref{aff101}}
\and D.~Potter\orcid{0000-0002-0757-5195}\inst{\ref{aff153}}
\and I.~Risso\orcid{0000-0003-2525-7761}\inst{\ref{aff154}}
\and P.-F.~Rocci\inst{\ref{aff63}}
\and M.~Sahl\'en\orcid{0000-0003-0973-4804}\inst{\ref{aff155}}
\and E.~Sarpa\orcid{0000-0002-1256-655X}\inst{\ref{aff24},\ref{aff116},\ref{aff25}}
\and A.~Schneider\orcid{0000-0001-7055-8104}\inst{\ref{aff153}}
\and D.~Sciotti\orcid{0009-0008-4519-2620}\inst{\ref{aff49},\ref{aff91}}
\and E.~Sellentin\inst{\ref{aff156},\ref{aff157}}
\and M.~Sereno\orcid{0000-0003-0302-0325}\inst{\ref{aff5},\ref{aff31}}
\and K.~Tanidis\inst{\ref{aff158}}
\and C.~Tao\orcid{0000-0001-7961-8177}\inst{\ref{aff65}}
\and G.~Testera\inst{\ref{aff35}}
\and R.~Teyssier\orcid{0000-0001-7689-0933}\inst{\ref{aff152}}
\and S.~Tosi\orcid{0000-0002-7275-9193}\inst{\ref{aff34},\ref{aff35}}
\and A.~Troja\orcid{0000-0003-0239-4595}\inst{\ref{aff8},\ref{aff64}}
\and M.~Tucci\inst{\ref{aff61}}
\and C.~Valieri\inst{\ref{aff31}}
\and D.~Vergani\orcid{0000-0003-0898-2216}\inst{\ref{aff5}}
\and G.~Verza\orcid{0000-0002-1886-8348}\inst{\ref{aff159},\ref{aff160}}
\and P.~Vielzeuf\orcid{0000-0003-2035-9339}\inst{\ref{aff65}}}
										   
\institute{Johns Hopkins University 3400 North Charles Street Baltimore, MD 21218, USA\label{aff1}
\and
INAF-Osservatorio Astronomico di Capodimonte, Via Moiariello 16, 80131 Napoli, Italy\label{aff2}
\and
Sterrenkundig Observatorium, Universiteit Gent, Krijgslaan 281 S9, 9000 Gent, Belgium\label{aff3}
\and
STAR Institute, University of Li{\`e}ge, Quartier Agora, All\'ee du six Ao\^ut 19c, 4000 Li\`ege, Belgium\label{aff4}
\and
INAF-Osservatorio di Astrofisica e Scienza dello Spazio di Bologna, Via Piero Gobetti 93/3, 40129 Bologna, Italy\label{aff5}
\and
Universit\'e de Strasbourg, CNRS, Observatoire astronomique de Strasbourg, UMR 7550, 67000 Strasbourg, France\label{aff6}
\and
INAF, Istituto di Radioastronomia, Via Piero Gobetti 101, 40129 Bologna, Italy\label{aff7}
\and
Dipartimento di Fisica e Astronomia "G. Galilei", Universit\`a di Padova, Via Marzolo 8, 35131 Padova, Italy\label{aff8}
\and
School of Physics, HH Wills Physics Laboratory, University of Bristol, Tyndall Avenue, Bristol, BS8 1TL, UK\label{aff9}
\and
INAF-Osservatorio Astronomico di Trieste, Via G. B. Tiepolo 11, 34143 Trieste, Italy\label{aff10}
\and
Jodrell Bank Centre for Astrophysics, Department of Physics and Astronomy, University of Manchester, Oxford Road, Manchester M13 9PL, UK\label{aff11}
\and
Dipartimento di Fisica e Astronomia "Augusto Righi" - Alma Mater Studiorum Universit\`a di Bologna, via Piero Gobetti 93/2, 40129 Bologna, Italy\label{aff12}
\and
Institute for Astronomy, University of Edinburgh, Royal Observatory, Blackford Hill, Edinburgh EH9 3HJ, UK\label{aff13}
\and
Instituto de Astrof\'{\i}sica de Canarias, V\'{\i}a L\'actea, 38205 La Laguna, Tenerife, Spain\label{aff14}
\and
Universidad de La Laguna, Departamento de Astrof\'{\i}sica, 38206 La Laguna, Tenerife, Spain\label{aff15}
\and
INAF-Osservatorio Astrofisico di Arcetri, Largo E. Fermi 5, 50125, Firenze, Italy\label{aff16}
\and
INAF-Osservatorio Astronomico di Brera, Via Brera 28, 20122 Milano, Italy\label{aff17}
\and
Universit\"at Innsbruck, Institut f\"ur Astro- und Teilchenphysik, Technikerstr. 25/8, 6020 Innsbruck, Austria\label{aff18}
\and
Kapteyn Astronomical Institute, University of Groningen, PO Box 800, 9700 AV Groningen, The Netherlands\label{aff19}
\and
European Southern Observatory, Karl-Schwarzschild-Str.~2, 85748 Garching, Germany\label{aff20}
\and
ICRAR, M468, University of Western Australia, Crawley, WA 6009, Australia\label{aff21}
\and
Departamento de F{\'\i}sica de la Tierra y Astrof{\'\i}sica, Universidad Complutense de Madrid, Plaza de las Ciencias 2, E-28040 Madrid, Spain\label{aff22}
\and
IFPU, Institute for Fundamental Physics of the Universe, via Beirut 2, 34151 Trieste, Italy\label{aff23}
\and
SISSA, International School for Advanced Studies, Via Bonomea 265, 34136 Trieste TS, Italy\label{aff24}
\and
INFN, Sezione di Trieste, Via Valerio 2, 34127 Trieste TS, Italy\label{aff25}
\and
INAF-IASF Milano, Via Alfonso Corti 12, 20133 Milano, Italy\label{aff26}
\and
Instituto de Astrof\'isica de Canarias (IAC); Departamento de Astrof\'isica, Universidad de La Laguna (ULL), 38200, La Laguna, Tenerife, Spain\label{aff27}
\and
Institute of Space Sciences (ICE, CSIC), Campus UAB, Carrer de Can Magrans, s/n, 08193 Barcelona, Spain\label{aff28}
\and
ESAC/ESA, Camino Bajo del Castillo, s/n., Urb. Villafranca del Castillo, 28692 Villanueva de la Ca\~nada, Madrid, Spain\label{aff29}
\and
Dipartimento di Fisica e Astronomia, Universit\`a di Bologna, Via Gobetti 93/2, 40129 Bologna, Italy\label{aff30}
\and
INFN-Sezione di Bologna, Viale Berti Pichat 6/2, 40127 Bologna, Italy\label{aff31}
\and
Space Science Data Center, Italian Space Agency, via del Politecnico snc, 00133 Roma, Italy\label{aff32}
\and
INAF-Osservatorio Astrofisico di Torino, Via Osservatorio 20, 10025 Pino Torinese (TO), Italy\label{aff33}
\and
Dipartimento di Fisica, Universit\`a di Genova, Via Dodecaneso 33, 16146, Genova, Italy\label{aff34}
\and
INFN-Sezione di Genova, Via Dodecaneso 33, 16146, Genova, Italy\label{aff35}
\and
Department of Physics "E. Pancini", University Federico II, Via Cinthia 6, 80126, Napoli, Italy\label{aff36}
\and
INFN section of Naples, Via Cinthia 6, 80126, Napoli, Italy\label{aff37}
\and
Instituto de Astrof\'isica e Ci\^encias do Espa\c{c}o, Universidade do Porto, CAUP, Rua das Estrelas, PT4150-762 Porto, Portugal\label{aff38}
\and
Faculdade de Ci\^encias da Universidade do Porto, Rua do Campo de Alegre, 4150-007 Porto, Portugal\label{aff39}
\and
Aix-Marseille Universit\'e, CNRS, CNES, LAM, Marseille, France\label{aff40}
\and
Dipartimento di Fisica, Universit\`a degli Studi di Torino, Via P. Giuria 1, 10125 Torino, Italy\label{aff41}
\and
INFN-Sezione di Torino, Via P. Giuria 1, 10125 Torino, Italy\label{aff42}
\and
European Space Agency/ESTEC, Keplerlaan 1, 2201 AZ Noordwijk, The Netherlands\label{aff43}
\and
Institute Lorentz, Leiden University, Niels Bohrweg 2, 2333 CA Leiden, The Netherlands\label{aff44}
\and
Centro de Investigaciones Energ\'eticas, Medioambientales y Tecnol\'ogicas (CIEMAT), Avenida Complutense 40, 28040 Madrid, Spain\label{aff45}
\and
Port d'Informaci\'{o} Cient\'{i}fica, Campus UAB, C. Albareda s/n, 08193 Bellaterra (Barcelona), Spain\label{aff46}
\and
Institute for Theoretical Particle Physics and Cosmology (TTK), RWTH Aachen University, 52056 Aachen, Germany\label{aff47}
\and
Institute of Cosmology and Gravitation, University of Portsmouth, Portsmouth PO1 3FX, UK\label{aff48}
\and
INAF-Osservatorio Astronomico di Roma, Via Frascati 33, 00078 Monteporzio Catone, Italy\label{aff49}
\and
Institute for Astronomy, University of Hawaii, 2680 Woodlawn Drive, Honolulu, HI 96822, USA\label{aff50}
\and
Dipartimento di Fisica e Astronomia "Augusto Righi" - Alma Mater Studiorum Universit\`a di Bologna, Viale Berti Pichat 6/2, 40127 Bologna, Italy\label{aff51}
\and
European Space Agency/ESRIN, Largo Galileo Galilei 1, 00044 Frascati, Roma, Italy\label{aff52}
\and
Universit\'e Claude Bernard Lyon 1, CNRS/IN2P3, IP2I Lyon, UMR 5822, Villeurbanne, F-69100, France\label{aff53}
\and
Institute of Physics, Laboratory of Astrophysics, Ecole Polytechnique F\'ed\'erale de Lausanne (EPFL), Observatoire de Sauverny, 1290 Versoix, Switzerland\label{aff54}
\and
Institut de Ci\`{e}ncies del Cosmos (ICCUB), Universitat de Barcelona (IEEC-UB), Mart\'{i} i Franqu\`{e}s 1, 08028 Barcelona, Spain\label{aff55}
\and
Instituci\'o Catalana de Recerca i Estudis Avan\c{c}ats (ICREA), Passeig de Llu\'{\i}s Companys 23, 08010 Barcelona, Spain\label{aff56}
\and
UCB Lyon 1, CNRS/IN2P3, IUF, IP2I Lyon, 4 rue Enrico Fermi, 69622 Villeurbanne, France\label{aff57}
\and
Mullard Space Science Laboratory, University College London, Holmbury St Mary, Dorking, Surrey RH5 6NT, UK\label{aff58}
\and
Departamento de F\'isica, Faculdade de Ci\^encias, Universidade de Lisboa, Edif\'icio C8, Campo Grande, PT1749-016 Lisboa, Portugal\label{aff59}
\and
Instituto de Astrof\'isica e Ci\^encias do Espa\c{c}o, Faculdade de Ci\^encias, Universidade de Lisboa, Campo Grande, 1749-016 Lisboa, Portugal\label{aff60}
\and
Department of Astronomy, University of Geneva, ch. d'Ecogia 16, 1290 Versoix, Switzerland\label{aff61}
\and
INAF-Istituto di Astrofisica e Planetologia Spaziali, via del Fosso del Cavaliere, 100, 00100 Roma, Italy\label{aff62}
\and
Universit\'e Paris-Saclay, CNRS, Institut d'astrophysique spatiale, 91405, Orsay, France\label{aff63}
\and
INFN-Padova, Via Marzolo 8, 35131 Padova, Italy\label{aff64}
\and
Aix-Marseille Universit\'e, CNRS/IN2P3, CPPM, Marseille, France\label{aff65}
\and
Universit\'e Paris-Saclay, Universit\'e Paris Cit\'e, CEA, CNRS, AIM, 91191, Gif-sur-Yvette, France\label{aff66}
\and
INFN-Bologna, Via Irnerio 46, 40126 Bologna, Italy\label{aff67}
\and
Istituto Nazionale di Fisica Nucleare, Sezione di Bologna, Via Irnerio 46, 40126 Bologna, Italy\label{aff68}
\and
FRACTAL S.L.N.E., calle Tulip\'an 2, Portal 13 1A, 28231, Las Rozas de Madrid, Spain\label{aff69}
\and
Max Planck Institute for Extraterrestrial Physics, Giessenbachstr. 1, 85748 Garching, Germany\label{aff70}
\and
INAF-Osservatorio Astronomico di Padova, Via dell'Osservatorio 5, 35122 Padova, Italy\label{aff71}
\and
Universit\"ats-Sternwarte M\"unchen, Fakult\"at f\"ur Physik, Ludwig-Maximilians-Universit\"at M\"unchen, Scheinerstrasse 1, 81679 M\"unchen, Germany\label{aff72}
\and
Jet Propulsion Laboratory, California Institute of Technology, 4800 Oak Grove Drive, Pasadena, CA, 91109, USA\label{aff73}
\and
Felix Hormuth Engineering, Goethestr. 17, 69181 Leimen, Germany\label{aff74}
\and
Technical University of Denmark, Elektrovej 327, 2800 Kgs. Lyngby, Denmark\label{aff75}
\and
Cosmic Dawn Center (DAWN), Denmark\label{aff76}
\and
Institut d'Astrophysique de Paris, UMR 7095, CNRS, and Sorbonne Universit\'e, 98 bis boulevard Arago, 75014 Paris, France\label{aff77}
\and
Max-Planck-Institut f\"ur Astronomie, K\"onigstuhl 17, 69117 Heidelberg, Germany\label{aff78}
\and
NASA Goddard Space Flight Center, Greenbelt, MD 20771, USA\label{aff79}
\and
Department of Physics and Helsinki Institute of Physics, Gustaf H\"allstr\"omin katu 2, 00014 University of Helsinki, Finland\label{aff80}
\and
Universit\'e de Gen\`eve, D\'epartement de Physique Th\'eorique and Centre for Astroparticle Physics, 24 quai Ernest-Ansermet, CH-1211 Gen\`eve 4, Switzerland\label{aff81}
\and
Department of Physics, P.O. Box 64, 00014 University of Helsinki, Finland\label{aff82}
\and
Helsinki Institute of Physics, Gustaf H{\"a}llstr{\"o}min katu 2, University of Helsinki, Helsinki, Finland\label{aff83}
\and
Laboratoire Univers et Th\'eorie, Observatoire de Paris, Universit\'e PSL, Universit\'e Paris Cit\'e, CNRS, 92190 Meudon, France\label{aff84}
\and
Institute of Theoretical Astrophysics, University of Oslo, P.O. Box 1029 Blindern, 0315 Oslo, Norway\label{aff85}
\and
SKA Observatory, Jodrell Bank, Lower Withington, Macclesfield, Cheshire SK11 9FT, UK\label{aff86}
\and
Centre de Calcul de l'IN2P3/CNRS, 21 avenue Pierre de Coubertin 69627 Villeurbanne Cedex, France\label{aff87}
\and
Dipartimento di Fisica "Aldo Pontremoli", Universit\`a degli Studi di Milano, Via Celoria 16, 20133 Milano, Italy\label{aff88}
\and
INFN-Sezione di Milano, Via Celoria 16, 20133 Milano, Italy\label{aff89}
\and
Universit\"at Bonn, Argelander-Institut f\"ur Astronomie, Auf dem H\"ugel 71, 53121 Bonn, Germany\label{aff90}
\and
INFN-Sezione di Roma, Piazzale Aldo Moro, 2 - c/o Dipartimento di Fisica, Edificio G. Marconi, 00185 Roma, Italy\label{aff91}
\and
Department of Physics, Institute for Computational Cosmology, Durham University, South Road, Durham, DH1 3LE, UK\label{aff92}
\and
Universit\'e Paris Cit\'e, CNRS, Astroparticule et Cosmologie, 75013 Paris, France\label{aff93}
\and
University of Applied Sciences and Arts of Northwestern Switzerland, School of Engineering, 5210 Windisch, Switzerland\label{aff94}
\and
Institut d'Astrophysique de Paris, 98bis Boulevard Arago, 75014, Paris, France\label{aff95}
\and
Aurora Technology for European Space Agency (ESA), Camino bajo del Castillo, s/n, Urbanizacion Villafranca del Castillo, Villanueva de la Ca\~nada, 28692 Madrid, Spain\label{aff96}
\and
School of Mathematics, Statistics and Physics, Newcastle University, Herschel Building, Newcastle-upon-Tyne, NE1 7RU, UK\label{aff97}
\and
Institut de F\'{i}sica d'Altes Energies (IFAE), The Barcelona Institute of Science and Technology, Campus UAB, 08193 Bellaterra (Barcelona), Spain\label{aff98}
\and
DARK, Niels Bohr Institute, University of Copenhagen, Jagtvej 155, 2200 Copenhagen, Denmark\label{aff99}
\and
Centre National d'Etudes Spatiales -- Centre spatial de Toulouse, 18 avenue Edouard Belin, 31401 Toulouse Cedex 9, France\label{aff100}
\and
Institute of Space Science, Str. Atomistilor, nr. 409 M\u{a}gurele, Ilfov, 077125, Romania\label{aff101}
\and
Institut f\"ur Theoretische Physik, University of Heidelberg, Philosophenweg 16, 69120 Heidelberg, Germany\label{aff102}
\and
Institut de Recherche en Astrophysique et Plan\'etologie (IRAP), Universit\'e de Toulouse, CNRS, UPS, CNES, 14 Av. Edouard Belin, 31400 Toulouse, France\label{aff103}
\and
Universit\'e St Joseph; Faculty of Sciences, Beirut, Lebanon\label{aff104}
\and
Departamento de F\'isica, FCFM, Universidad de Chile, Blanco Encalada 2008, Santiago, Chile\label{aff105}
\and
Institut d'Estudis Espacials de Catalunya (IEEC),  Edifici RDIT, Campus UPC, 08860 Castelldefels, Barcelona, Spain\label{aff106}
\and
Satlantis, University Science Park, Sede Bld 48940, Leioa-Bilbao, Spain\label{aff107}
\and
Instituto de Astrof\'isica e Ci\^encias do Espa\c{c}o, Faculdade de Ci\^encias, Universidade de Lisboa, Tapada da Ajuda, 1349-018 Lisboa, Portugal\label{aff108}
\and
Cosmic Dawn Center (DAWN)\label{aff109}
\and
Niels Bohr Institute, University of Copenhagen, Jagtvej 128, 2200 Copenhagen, Denmark\label{aff110}
\and
Universidad Polit\'ecnica de Cartagena, Departamento de Electr\'onica y Tecnolog\'ia de Computadoras,  Plaza del Hospital 1, 30202 Cartagena, Spain\label{aff111}
\and
Dipartimento di Fisica, Universit\`a degli studi di Genova, and INFN-Sezione di Genova, via Dodecaneso 33, 16146, Genova, Italy\label{aff112}
\and
Infrared Processing and Analysis Center, California Institute of Technology, Pasadena, CA 91125, USA\label{aff113}
\and
Astronomical Observatory of the Autonomous Region of the Aosta Valley (OAVdA), Loc. Lignan 39, I-11020, Nus (Aosta Valley), Italy\label{aff114}
\and
ICL, Junia, Universit\'e Catholique de Lille, LITL, 59000 Lille, France\label{aff115}
\and
ICSC - Centro Nazionale di Ricerca in High Performance Computing, Big Data e Quantum Computing, Via Magnanelli 2, Bologna, Italy\label{aff116}
\and
Instituto de F\'isica Te\'orica UAM-CSIC, Campus de Cantoblanco, 28049 Madrid, Spain\label{aff117}
\and
CERCA/ISO, Department of Physics, Case Western Reserve University, 10900 Euclid Avenue, Cleveland, OH 44106, USA\label{aff118}
\and
Departamento de F{\'\i}sica Fundamental. Universidad de Salamanca. Plaza de la Merced s/n. 37008 Salamanca, Spain\label{aff119}
\and
Dipartimento di Fisica e Scienze della Terra, Universit\`a degli Studi di Ferrara, Via Giuseppe Saragat 1, 44122 Ferrara, Italy\label{aff120}
\and
Istituto Nazionale di Fisica Nucleare, Sezione di Ferrara, Via Giuseppe Saragat 1, 44122 Ferrara, Italy\label{aff121}
\and
Center for Data-Driven Discovery, Kavli IPMU (WPI), UTIAS, The University of Tokyo, Kashiwa, Chiba 277-8583, Japan\label{aff122}
\and
Dipartimento di Fisica - Sezione di Astronomia, Universit\`a di Trieste, Via Tiepolo 11, 34131 Trieste, Italy\label{aff123}
\and
California Institute of Technology, 1200 E California Blvd, Pasadena, CA 91125, USA\label{aff124}
\and
Universit\'e C\^{o}te d'Azur, Observatoire de la C\^{o}te d'Azur, CNRS, Laboratoire Lagrange, Bd de l'Observatoire, CS 34229, 06304 Nice cedex 4, France\label{aff125}
\and
Department of Physics \& Astronomy, University of California Irvine, Irvine CA 92697, USA\label{aff126}
\and
Department of Astronomy \& Physics and Institute for Computational Astrophysics, Saint Mary's University, 923 Robie Street, Halifax, Nova Scotia, B3H 3C3, Canada\label{aff127}
\and
Departamento F\'isica Aplicada, Universidad Polit\'ecnica de Cartagena, Campus Muralla del Mar, 30202 Cartagena, Murcia, Spain\label{aff128}
\and
Department of Computer Science, Aalto University, PO Box 15400, Espoo, FI-00 076, Finland\label{aff129}
\and
Instituto de Astrof\'\i sica de Canarias, c/ Via Lactea s/n, La Laguna 38200, Spain. Departamento de Astrof\'\i sica de la Universidad de La Laguna, Avda. Francisco Sanchez, La Laguna, 38200, Spain\label{aff130}
\and
Caltech/IPAC, 1200 E. California Blvd., Pasadena, CA 91125, USA\label{aff131}
\and
Ruhr University Bochum, Faculty of Physics and Astronomy, Astronomical Institute (AIRUB), German Centre for Cosmological Lensing (GCCL), 44780 Bochum, Germany\label{aff132}
\and
Universit\'e PSL, Observatoire de Paris, Sorbonne Universit\'e, CNRS, LERMA, 75014, Paris, France\label{aff133}
\and
Universit\'e Paris-Cit\'e, 5 Rue Thomas Mann, 75013, Paris, France\label{aff134}
\and
Univ. Grenoble Alpes, CNRS, Grenoble INP, LPSC-IN2P3, 53, Avenue des Martyrs, 38000, Grenoble, France\label{aff135}
\and
Department of Physics and Astronomy, Vesilinnantie 5, 20014 University of Turku, Finland\label{aff136}
\and
Serco for European Space Agency (ESA), Camino bajo del Castillo, s/n, Urbanizacion Villafranca del Castillo, Villanueva de la Ca\~nada, 28692 Madrid, Spain\label{aff137}
\and
ARC Centre of Excellence for Dark Matter Particle Physics, Melbourne, Australia\label{aff138}
\and
Centre for Astrophysics \& Supercomputing, Swinburne University of Technology,  Hawthorn, Victoria 3122, Australia\label{aff139}
\and
School of Physics and Astronomy, Queen Mary University of London, Mile End Road, London E1 4NS, UK\label{aff140}
\and
Department of Physics and Astronomy, University of the Western Cape, Bellville, Cape Town, 7535, South Africa\label{aff141}
\and
DAMTP, Centre for Mathematical Sciences, Wilberforce Road, Cambridge CB3 0WA, UK\label{aff142}
\and
Kavli Institute for Cosmology Cambridge, Madingley Road, Cambridge, CB3 0HA, UK\label{aff143}
\and
IRFU, CEA, Universit\'e Paris-Saclay 91191 Gif-sur-Yvette Cedex, France\label{aff144}
\and
Oskar Klein Centre for Cosmoparticle Physics, Department of Physics, Stockholm University, Stockholm, SE-106 91, Sweden\label{aff145}
\and
Astrophysics Group, Blackett Laboratory, Imperial College London, London SW7 2AZ, UK\label{aff146}
\and
Dipartimento di Fisica, Sapienza Universit\`a di Roma, Piazzale Aldo Moro 2, 00185 Roma, Italy\label{aff147}
\and
Centro de Astrof\'{\i}sica da Universidade do Porto, Rua das Estrelas, 4150-762 Porto, Portugal\label{aff148}
\and
HE Space for European Space Agency (ESA), Camino bajo del Castillo, s/n, Urbanizacion Villafranca del Castillo, Villanueva de la Ca\~nada, 28692 Madrid, Spain\label{aff149}
\and
Department of Physics and Astronomy, University College London, Gower Street, London WC1E 6BT, UK\label{aff150}
\and
Institute of Astronomy, University of Cambridge, Madingley Road, Cambridge CB3 0HA, UK\label{aff151}
\and
Department of Astrophysical Sciences, Peyton Hall, Princeton University, Princeton, NJ 08544, USA\label{aff152}
\and
Department of Astrophysics, University of Zurich, Winterthurerstrasse 190, 8057 Zurich, Switzerland\label{aff153}
\and
INAF-Osservatorio Astronomico di Brera, Via Brera 28, 20122 Milano, Italy, and INFN-Sezione di Genova, Via Dodecaneso 33, 16146, Genova, Italy\label{aff154}
\and
Theoretical astrophysics, Department of Physics and Astronomy, Uppsala University, Box 515, 751 20 Uppsala, Sweden\label{aff155}
\and
Mathematical Institute, University of Leiden, Einsteinweg 55, 2333 CA Leiden, The Netherlands\label{aff156}
\and
Leiden Observatory, Leiden University, Einsteinweg 55, 2333 CC Leiden, The Netherlands\label{aff157}
\and
Department of Physics, Oxford University, Keble Road, Oxford OX1 3RH, UK\label{aff158}
\and
Center for Cosmology and Particle Physics, Department of Physics, New York University, New York, NY 10003, USA\label{aff159}
\and
Center for Computational Astrophysics, Flatiron Institute, 162 5th Avenue, 10010, New York, NY, USA\label{aff160}}

\date{Received ; accepted }

\abstract{

The European Space Agency's \euclid\ mission will observe approximately $14\,000 \ \rm{deg}^{2}$ of the extragalactic sky and deliver high-quality imaging of a large number of galaxies. The depth and high spatial resolution of the data will enable a detailed analysis of the stellar population properties of local galaxies through spatially resolved spectral energy distribution (SED) fitting.

In this study, we test our pipeline for spatially resolved SED fitting using synthetic images of \euclid, \lsst, and \galex\ generated from the TNG50 simulation using the SKIRT 3D radiative transfer code. Our pipeline uses functionalities in \pixedfit\ for processing the simulated data cubes and carrying out SED fitting. We apply our pipeline to 25 simulated galaxies at $z\sim 0$ to recover their resolved stellar population properties. For each galaxy, we produce three types of data cubes: \galex\ + \lsst\ + \euclid, \lsst\ + \euclid, and \euclid-only.

We performed the SED fitting tests with two stellar population synthesis (SPS) models in a Bayesian framework. Because the age, metallicity ($Z$), and dust attenuation estimates are biased when applying only classical formulations of flat priors (even with the combined \galex\ + \lsst\ + \euclid\ data), we examined the effects of additional physically motivated priors in the forms of mass-age and mass-metallicity relations, constructed using a combination of empirical and simulated data.
Stellar-mass surface densities can be recovered well using any of the three data cubes, regardless of the SPS model and prior variations. The new priors then significantly improve the measurements of mass-weighted age and $Z$ compared to results obtained without priors, but they may play an excessive role compared to the data in determining the outcome when no ultraviolet (UV) data is available.  
Compared to varying the spectral extent of the data cube or including and discarding the additional priors, replacing one SPS model family with the other has little effect on the results.

The spatially resolved SED fitting method is powerful for mapping the stellar population properties of many galaxies with the current abundance of high-quality imaging data. Our study re-emphasizes the gain added by including multi-wavelength data from ancillary surveys and the roles of priors in Bayesian SED fitting.  
With the \euclid\ data alone, we will be able to generate complete and deep stellar mass maps of galaxies in the local Universe ($z\lesssim 0.1$), exploiting the telescope's wide field, near-infrared sensitivity, and high spatial resolution. 
}
   

    \titlerunning{Spatially resolved stellar populations with \euclid}
    \authorrunning{Euclid Collaboration: Abdurro'uf et al.}
   
\maketitle

\section{Introduction}

Multi-wavelength photometric and spectroscopic surveys over the past few decades have greatly contributed to our understanding of galaxy formation and evolution. Some major galaxy surveys (including the Sloan Digital Sky Survey, SDSS; \citealt{2000York}; The 2dF Galaxy Redshift Survey; \citealt{2001Colless}; Galaxy and Mass Assembly, GAMA; \citealt{Baldry2010},\citealt{2022Driver}) have observed many galaxies and provided important insights into the study of galaxy evolution. The depth and high spatial resolution of the photometric instruments of galaxy surveys, such as the \textit{Hubble} Space Telescope (\hst), the Dark Energy Survey (\des; \citealt{DES2016}), the Hyper Suprime-Cam SSP Survey (\hsc; \citealt{Aihara2018}), and the \textit{James Webb} Space Telescope (\jwst; \citealt{Gardner2023}), have enabled detailed observations of galaxies up to high redshifts, probing further back in the history of galaxy evolution. 

The current and upcoming galaxy surveys (e.g.~the Vera Rubin Observatory Legacy Survey of Space and Time, LSST; \citealt{Ivezic2019}, the Dark Energy Spectroscopic Instrument, DESI; \citealt{2016DESI}, and the Nancy Grace Roman Space Telescope; \citealt{2015Spergel_roman}) provide even more data, opening a new era of big data in astronomy. The abundance of astronomical data requires sophisticated techniques for analysing and interpreting them. One of the main ways of measuring the physical properties of galaxies from their light is through so-called spectral energy distribution (SED) fitting \citep[e.g.][]{Sawicki1998, Bolzonella2000, Sawicki2012, 2011Walcher, Conroy2013, Pacifici2023}. Major developments have been made in this technique over the last two decades, thanks to advancements in both the modelling of SEDs and the statistical methods of fitting data with the models. In the former aspect, we have seen the inclusion of more complex and physically realistic components in the models. This has enabled the modelling of a wide range of physical processes that occur in galaxies, making it possible to simulate galaxy spectra across a wide range of wavelengths, from X-ray to radio \citep[e.g.][]{Bruzual2003, Conroy2009, daCunha2008, Boquien2019, Yang2022}. As for statistical techniques, we have seen the emergence of the Bayesian fitting technique with sophisticated posterior sampling methods, such as the Markov chain Monte Carlo (MCMC) and nested sampling \citep[e.g.][]{Acquaviva2011, Han2014, Leja2017, Carnall2018, Abdurrouf2021, Pacifici2023} and even the applications of machine learning in SED fitting \citep[e.g.][]{Lovell2019, Dobbels2020, Gilda2021, Hahn2022, Bisigello2023, 2025Kovacic_euclid}.    

Despite the abundance of data, most studies on galaxies have focused on the global integrated properties. In the case of spectroscopic observations, a galaxy-integrated spectrum is obtained with spectroscopy (e.g.~SDSS, GAMA). In the case of photometric observations, the integrated SED of a galaxy is usually measured within an aperture of a certain radius centred on the galaxy. These analyses have revealed many important properties of galaxy populations across a wide range of redshifts and provided significant insights into our understanding of galaxy structure, formation, and evolution \citep[e.g.][]{Tremonti2004, Brinchmann2004, Salucci+08, Blanton2009}. However, these studies overlook the crucial insights that spatially resolved data can offer into the physical processes within galaxies.  

The integral field spectroscopy (IFS) surveys in the past few decades have revolutionized the study of galaxy evolution. Some major IFS surveys are SAURON \citep{deZeeuw2002}, ATLAS$^{\rm{3D}}$ \citep{Cappellari2011}, the Calar Alto Legacy Integral Field Area survey \citep[CALIFA;][]{Sanchez2012}, Mapping nearby Galaxies at Apache Point
Observatory \citep[MaNGA;][]{Bundy2015,Abdurrouf2022manga}, Sydney-AAO Multi-object Integral field spectrograph \citep[SAMI;][]{Croom2012}, and KMOS$^{3\rm{D}}$ \citep{Wisnioski2015}. These surveys have enabled the study of galaxies in great detail, thanks to the ability to observe the spectra of spatially resolved regions inside them. One of the important findings from these spatially resolved studies is that key scaling relations among galaxies on a global scale (e.g.~the star-forming main sequence, \citealt{Brinchmann2004,Noeske2007,Speagle2014,2023Popesso,Q1-SP031-Enia}; the mass-metallicity relation, \citealt{Tremonti2004}) likely originate from similar scaling relations on kiloparsec and sub-kiloparsec scales \citep[see e.g.][]{Sanchez2020}.       

While the IFS surveys arguably provide the most detailed insights into the properties of each spatially resolved element, to date only a limited number of them have observed a large number of galaxies over a wide area of the sky. Those surveys mostly focus on the local Universe (e.g.~CALIFA, MaNGA, and SAMI). The largest IFS survey to date is the MaNGA survey, which observed around 10\,000 galaxies at $0.01<z<0.15$. Unfortunately, IFS surveys tend to have a lower spatial resolution than imaging data, and unlike imaging data, the integral field units usually do not cover the entire optical region of galaxies. 
        
Recent improvements in the depth, spatial resolution, and wavelength coverage of imaging surveys provide opportunities for conducting SED fitting on spatially resolved scales in 2D images of galaxies, namely the spatially resolved SED fitting method. Previous studies have shown the great potential of this method in spatially resolving stellar population properties in galaxies across a wide range of redshifts \citep[e.g.][]{Conti2003, 2009Zibetti, Welikala2011, 2012Wuyts, Wuyts2013, Viaene2014, 2018Sorba, Abdurrouf2017, Abdurrouf2018, 2022Robotham, Abdurrouf2022a, Abdurrouf2022c, Abdurrouf2023, 2024Bellstedt}. \citet{2012Wuyts, Wuyts2013} applied this method to $0.5<z<2.5$ star-forming galaxies in the GOODS-South field using imaging data from \hst. They studied the radial gradients of rest-frame colours, surface density of stellar mass (\massd), age of the stellar population, and dust attenuation. 
\citet{Abdurrouf2022a, Abdurrouf2022c} utilized multi-wavelength imaging data across more than 20 bands, ranging from the far-ultraviolet (FUV) to the far-infrared (FIR), obtained from various telescopes, to create maps of stellar populations and dust properties of galaxies on kiloparsec scales. Recently, some works have employed \jwst\ data to study the stellar populations of galaxies at very high redshifts (up to $z = 9$, e.g. \citealt{Abdurrouf2023, Gimenez-Arteaga+23}).
The abundance of high-quality (i.e.~deep and high spatial resolution) multi-wavelength imaging data from the current and upcoming surveys opens up avenues for the spatially resolved SED analysis of a large number of galaxies. The wide wavelength range of modern data can provide coverage from the rest-frame UV to the near-infrared (NIR), which is essential for SED fitting across a wide redshift range. Although it does not replace spectroscopy, it offers precious complementary constraints on stellar population properties. 

A major step forwards in large galaxy surveys is the European Space Agency's \euclid\ mission \citep{Laureijs2011, Mellier+24}, launched on July 1st, 2023, which will observe approximately 14\,000 $\rm{deg}^{2}$ of extragalactic sky within the six years of its operation.
It will deliver high-quality imaging data for about $1.5$ billion galaxies at an optical wavelength with the visible imaging instrument (VIS; \citealt{2016ropper, 2024Cropper_euclid}; passband \ie) and in the NIR with the NISP instrument (the Near-Infrared Spectrometer and Photometer, \citealt{2022Maciaszek,Schirmer2022,2024Jahnke_euclid,2024Hormuth_euclid}; passbands \ye, \je, and \he). \euclid\ will provide an unprecedented homogeneous view of the local Universe, re-observing thousands of known galaxies at a high spatial resolution ($\sim 1$ kpc), as well as newly discovered low-surface brightness galaxies. The recent early-release observations have highlighted the exceptional potential of \euclid\ data (e.g. \citealt{Cuillandre+24_Perseus-LF, Hunt+24_showcase, Marleau+24, Saifollahi+24}). 

For a wider wavelength coverage, the \Euclid\ data can be combined with other wide-area and deep imaging surveys in the optical, such as the future Rubin/\lsst\ \citep{Ivezic2019} that will observe the southern hemisphere, and the UNIONS survey \citep{Ibata2017}, a combined effort by the CFHT, Pan-STARRS, and Subaru telescopes, which observe the northern hemisphere. In combination, these datasets extend from the near-UV to the NIR. For galaxies that are covered by the GALEX Medium Imaging Survey (MIS) or Deep Imaging Survey (DIS), we will incorporate GALEX images to get even wider wavelength coverage extending to FUV. The wide-area coverage will make it possible to analyse many galaxies residing in a wide range of local density environments (from a low-density field to a dense galaxy cluster) that enable a more comprehensive analysis of the effects of internal and external factors in the evolution of galaxies.     

In this work, we test a pipeline designed for spatially resolved SED fitting of local galaxies using imaging data primarily from \euclid, complemented by optical data from surveys such as \lsst\ and UNIONS, as well as UV data from GALEX's medium- and deep-imaging surveys. The pipeline mainly utilizes various functionalities from \pixedfit\ package \citep{Abdurrouf2021, Abdurrouf2022b_pixedfitcode}. For this testing, we use synthetic images generated from the TNG50 cosmological magneto-hydrodynamical simulations \citep{Marinacci2018, Naiman2018, Nelson2018, Pillepich2019, Springel2018}. We aim to test to what extent \euclid\ imaging data (with 4 bands: \ie, \ye, \je, \he) alone can provide constraints on some basic but important stellar population properties (stellar mass, age, and metallicity) and how additional imaging data in the optical and UV can improve the fitting results.

This paper is structured as follows. Section~\ref{sec:generate_synthetic_images} describes the procedures for generating synthetic images of \euclid, \lsst, and \galex. The analysis pipeline and the SED fitting procedures are described in Sect.~\ref{sec:simulate_test_res_SEDfits}. In Sect.~\ref{sec:results} we present our results and further discuss them in Sect.~\ref{sec:discussion}, focusing on some factors that drive the SED fitting output. Finally, we provide our conclusions and outlooks in Sect.~\ref{sec:conclusion}. Throughout this paper, we assume the \citet{Chabrier2003} initial mass function (IMF) with a mass range of $0.1-100\,\msun$ and cosmological parameters of $\om=0.3$, $\ol=0.7$, and $H_0=70\,\kmsMpc$.

\section{Generating synthetic images of \euclid, \lsst, and \galex}
\label{sec:generate_synthetic_images}

In this section, we first describe the synthetic images from the TNG50-SKIRT Atlas generated using the radiative transfer method (Sect.~\ref{sec:gen_noisefree_images}). We then describe the procedures for adding some observational factors to the synthetic images in Sect.~\ref{sec:adding_obs_factors}. 

\subsection{Synthetic images from the TNG50-SKIRT Atlas}
\label{sec:gen_noisefree_images}

We use the synthetic images from the TNG50-SKIRT Atlas data products (\citealt{Baes2024_a,Baes2024_b}; \citealt{2025Kovacic_euclid}) that were generated from the IllustrisTNG cosmological magneto-hydrodynamical simulations \citep{Marinacci2018, Naiman2018, Nelson2018, Pillepich2019, Springel2018} using the SKIRT radiative transfer code \citep{Camps2015, Camps2020}. The synthetic images were created for galaxies in the TNG50 simulation \citep{Nelson2019}, which has the highest resolution among the three IllustrisTNG runs with a baryonic mass resolution of $8.5\times 10^{4}\,\msun$, though covering the smallest cubic volume of 51.7 comoving Mpc on the side. The TNG50 simulation uses the moving-mesh code AREPO \citep{Springel2010}. It includes a wide range of physical processes in the galaxies, including gas heating and cooling, stochastic star formation, stellar evolution, chemical enrichment in the interstellar medium (ISM), seeding and growth of the supermassive black holes (SMBHs), feedback from the supernova and the active galactic nucleus (AGN), and the magnetic fields.

\citet{Baes2024_a} generated high-resolution multi-wavelength images of 1\,160 galaxies from the $z=0$ snapshot of the TNG50 simulation and released the dataset publicly as the TNG50-SKIRT Atlas database. Initially, the synthetic images were created in 18 bands (\galex\ FUV and NUV, Johnson \textit{UBVRI}, \lsst\ \textit{ugrizy}, 2MASS \textit{JHK$_{\rm s}$}, WISE \textit{W1} and \textit{W2}), then \citet{2025Kovacic_euclid} extended the database with additional synthetic images in the four \euclid\ broad-band filters (\ie, \ye, \je, \he). These images come with stellar mass surface density maps, stellar-mass-weighted metallicity and age maps, and
dust mass surface density maps.

The simulated galaxies have a wide range of stellar mass (\mass) from $10^{9.8}$ to $10^{12}\,\msun$. To generate synthetic images, the 3D radiative transfer SKIRT code simulates the propagation of stellar light through the dust distribution in galaxies, which directly accounts for the absorption and scattering of light by dust. A detailed description of the radiative transfer post-processing is given in \citet{Baes2024_a}. In the following, we only briefly overview the synthetic imaging data.

In the radiative transfer analysis, the primary sources of radiation are the stellar particles in the simulated galaxy, which are obtained from the TNG50 database. The stellar particles are assigned the SED of a simple stellar population (SSP) based on their masses, ages, and metallicities. Stellar particles older than 10 Myr are assigned an SED generated from the \citet[][hereafter BC03]{Bruzual2003} stellar population synthesis (SPS) model.  
The stellar particles younger than 10 Myr are assigned an SED from the MAPPINGS III templates for star-forming regions \citep{2008Groves}. The dust density is determined by assuming a constant dust-to-metal fraction, $f_{\rm dust}=0.2$, adopted from \citet{Trcka2022}, which conducted a thorough calibration of the free parameters in radiative transfer simulation using a sample of nearby galaxies from the DustPedia project \citep{Davies2017}. This value is consistent with observational estimates based on nearby galaxy samples \citep[e.g.][]{DeVis2019,Galliano2021,Zabel2021}. The diffuse ISM THEMIS model \citep{Jones2017} is used as the dust grain model at every location in the galaxy. The radiative transfer analysis does not include dust emission modelling because it only focuses on producing synthetic UV to NIR images. Each galaxy is projected at five different viewing angles and 22-band images are generated for each projection with a field of view of 160 kpc $\times$ 160 kpc, centred on the galaxy. The synthetic images in all filters have a physical pixel size of 100 pc\footnote{This physical scale of 100 pc corresponds to the VIS pixel scale at a distance of approximately 206 Mpc ($z\sim 0.046$).} and pixel value in units of megajanskys per steradian. In this paper, we only use synthetic images in the following filters: FUV, NUV (\galex), $u$, $g$, $r$, $i$, $z$ (\lsst), \ie, \ye, \je, and \he\ (\euclid).

\subsection{Adding observational effects to the synthetic images}
\label{sec:adding_obs_factors}

The synthetic images from the TNG50-SKIRT Atlas do not include observational effects, only small random fluctuations generated in the Monte Carlo process by the SKIRT code. To create realistic images that mimic the actual imaging data from the \euclid, \lsst, and \galex\ surveys, we added some observational effects to the synthetic images. First, we assigned a redshift of $z=0.03$ and resampled (i.e. regridded) the images to have the same angular pixel size as that of the cameras: \ang{;;1.5} for \galex, \ang{;;0.1} for the \euclid\ VIS band \ie, \ang{;;0.3} for \euclid\ NISP bands, and \ang{;;0.2} for \lsst. This requires one to calculate the cosmological angular-diameter distance to convert the angular scale on the sky to an actual physical scale and adjust the resampling factor for the images. We used the 2D cubic spline interpolation method to resample the images. We assigned the same redshift to all simulated galaxies in the sample that we use for our testing (to be described in Sect.~\ref{sec:sample_galaxies}).

We adopted a uniform redshift of $z = 0.03$ in this test for several reasons. First, the TNG50-SKIRT ATLAS images are currently only available for galaxies in snapshot 99 ($z = 0$) of the TNG50 simulation. As a result, we cannot reliably redshift these images to significantly higher redshifts, where cosmological effects become non-negligible. Moreover, the primary goal of our future work with Euclid data is to spatially resolve stellar populations on sub-kiloparsec scales in local galaxies, limiting our target sample to $z \lesssim 0.1$. Given this low-redshift focus, we consider our current testing set-up appropriate. Redshifting between $z = 0$ and $z = 0.1$ does not substantially affect spectral coverage, so we expect the trends in fitting results and the recovery of stellar population properties to remain valid.

The next step is simulating the observational noise and injecting it into the images. The noise includes both the photon shot noise and the sky background noise. For the \euclid\ and \lsst\ images, we follow the same procedure as \citet{Merlin2023} and \citet{Martinet2019}.  
Overall, the noise simulation is designed to achieve a given signal-to-noise ratio ($\rm{S}/\rm{N}$) at a given limiting magnitude by adjusting the instrumental zero points (ZPs). The ZP at a $1$ second exposure was estimated using the following equation adapted from \citet{Merlin2023}:
\begin{equation}
    {\text{ZP}} = 2.5\,\logten \left[ \frac{({\rm S}/{\rm N})^{2}\pi r^{2} g}{t_{\rm exp}} \right] + 2\,m_{\rm lim} - \rm{SB}_{\rm bkg}.
\end{equation}
We adopt the observational parameters listed by \citet[][see Table 1 therein]{Merlin2023}, which were made to mimic the imaging quality of the Euclid Wide Survey \citep{Scaramella2022} and the \lsst\ survey in its early data release.
The parameters include the $10\sigma$ (i.e.~$\rm{S}/\rm{N}=10$) limiting magnitudes ($m_{\rm lim}$) that are expected within an aperture of radius $r=\ang{;;1}$, the background surface brightness ($\rm{SB}_{\rm bkg}$), and the exposure time ($t_{\rm exp}$). These parameters are summarized in Table~\ref{tab:mock_images}. The detector gains (in units of electrons$/$ADU) are $g=3.1$ for \ie\ \citep{Martinet2019}, $g=2.0$ for \euclid\ NISP bands \citep{Secroun2018}, and $g=1.06$ for \lsst\ \citep{OConnor2019}. The estimated ZP is then used for converting pixel values in the original units of megajanskys per steradian into the observational units; in other words, analog-to-digital units (ADUs). This is done before simulating the photon shot noise, where the pixel values in the noiseless images are perturbed by random noise generated from a Poisson distribution. After perturbing the pixel values, we then convert the images back to megajanskys per steradian.   

The sky background noise was simulated as a Gaussian noise with zero mean and a standard deviation that is defined by the depth of the final simulated images. The zero mean reflects that the simulated images are background-subtracted. We adopted the standard deviation values for the background noise from \cite{Merlin2023} and private communication, and created blank sky maps containing only Gaussian noise. We added these maps to the synthetic images to get the final science images. 

In addition to the science images, we needed to generate variance images. For this, we first calculated RMS maps. The noise per pixel is a contribution from the background noise and the photon shot noise from the source. The background noise is reflected in the standard deviation of the Gaussian noise in the empty background maps (RMS$_{\rm sky}$), while the photon shot noise is Poissonian in nature. We used the following equation to calculate flux uncertainty of a 1-second exposure of individual pixels,
\begin{equation}
    N = \sqrt{ {\rm{RMS}}^{2}_{\rm{sky}} + \frac{C}{t_{\rm{exp}}} },
\end{equation}
where $C$ and $t_{\rm exp}$ are the source counts and the exposure time, respectively. The variance images were then created by taking a square of the RMS maps. We have confirmed that our \euclid\ and \lsst\ images reach similar depths at $10\sigma$ as those of the images produced by \citet{Merlin2023}, which are summarized in Table~\ref{tab:mock_images}. We checked this by looking at the correlations between $\rm{S}/\rm{N}$ and magnitudes of $2''$-diameter apertures that are randomly drawn across an image stamp, similar to Fig.~5 in \citet{Merlin2023}.

We followed a similar procedure to the one described above to simulate the observational noise of the \galex\ images with an exception in the calculation of the background RMS, which we derive from the real observational images. We downloaded four tiles of background-subtracted images for each band from the \galex\ website\footnote{\url{https://galex.stsci.edu/gr6/}}. We specifically choose images from the medium imaging survey (MIS) of \galex, which has an exposure time of $\sim 1500$ seconds \citep{Bianchi2014} because we intend to combine \euclid\ and \lsst\ images with the \galex\ images from this type of survey whenever available in our future analyses. To obtain the background RMS of an image tile, first, we remove contamination from the sources within the field by cropping a circular region within the Kron radius around each detected source. The co-ordinates and Kron radii of the sources are taken from the source catalogue associated with the field that we also downloaded from the \galex\ website and was generated using \sextractor\ \citep{Bertin1996}.    
We then take the standard deviation of pixel values in each blank sky image. To get the final background RMS, we take an average of the standard deviations from the four fields. 

After adding the simulated noise to the science images and producing the variance images, we convolve all the images with the point spread functions (PSFs) of the corresponding cameras. For \euclid, we use model PSF images from the \euclid\ Mission Database models, while for \lsst\ we use custom simulated PSFs from \citet{Merlin2023} that were created using \texttt{PhoSim} \citep{Peterson2015}. For \galex, we use empirical PSF images of \galex\ generated by \citet{Abdurrouf2021} that are publicly available.\footnote{\url{https://github.com/aabdurrouf/empPSFs_GALEXSDSS2MASS}} The PSF full width at half maximum (FWHM) and pixel size of the images are summarized in Table~\ref{tab:mock_images}. 

In this analysis, the convolution with the PSF is performed after adding simulated noise. We have checked that reversing the order of operations--applying PSF convolution before noise injection--yields images with similar characteristics and a minimal overall flux excess (approximately 1\%), which does not significantly alter the SED at the pixel scale. We also emphasize that the synthetic images produced here have been cropped around each simulated galaxy, not in the form of wide-field imaging as obtained from observations.

\begin{table*}[h!]
\caption{Basic information about the synthetic imaging datasets and some assumptions for simulating the different sources of noise. \label{tab:mock_images}}
\centering
\begin{tabular}{c c c c c c c c c} 
\hline \hline \\ [-5pt]
Survey & Instrument & Filter & Wavelength & PSF FWHM & Pixel size & $m_{\rm lim}$ & $\rm{SB}_{\rm bkg}$ & $t_{\rm exp}$ \\
 & & & (\AA) & (arcsec) & (arcsec) & (mag) & (mag) & (s) \\ [+3pt]
\hline \\ [-5pt]
\galex\ & ... & FUV & 1340--1810 & $4.48$ & 1.5 & ... & ... & 1500 \\
\galex\ & ... & NUV & 1693--3008 & $5.05$ & 1.5 & ... & ... & 1500 \\ 
\lsst\ & ... & $u$ & 3206--4082 & $1.17$ & $0.2$ & $23.6$ & $22.70$ & 150 \\
\lsst\ & ... & $g$ & 3876--5665 & $1.13$ & $0.2$ & $24.5$ & $22.00$ & 150 \\
\lsst\ & ... & $r$ & 5377--7055 & $1.00$ & $0.2$ & $23.9$ & $20.80$ & 150 \\
\lsst\ & ... & $i$ & 6766--8325 & $1.00$ & $0.2$ & $23.6$ & $20.30$ & 150 \\
\lsst\ & ... & $z$ & 8035--9375 & $1.00$ & $0.2$ & $23.4$ & $19.40$ & 150 \\
Euclid wide & VIS &\ie\ & 4971--9318 & $0.17$ & $0.1$ & $24.6$ & $22.33$ & 4$\times$590 \\
Euclid  wide & NISP &\ye\ & 9382--12435 & $0.40$ & $0.3$ & $23.0$ & $22.10$ & 4$\times$88 \\
Euclid  wide & NISP &\je\ & 11523--15952 & $0.45$ & $0.3$ & $23.0$ & $22.11$ & 4$\times$90 \\
Euclid  wide & NISP &\he\ & 14972--20568 & $0.50$ & $0.3$ & $23.0$ & $22.28$ & 4$\times$54 \\
\hline
\end{tabular}
\tablefoot{$m_{\rm lim}$, $\rm{SB}_{\rm bkg}$, and $t_{\rm exp}$ are $10\sigma$ ($\rm{S}/\rm{N}=10$) limiting magnitude within $2''$ aperture, the background surface brightness, and total exposure time of the final mosaic, respectively. 
The values for \euclid\ are not updated to the latest estimates of the actual in-flight values because that information was not available at the time of our analysis.
For the \lsst\ images, we adopt the depths from a single visit (i.e.~exposure), which will be available for a wide area soon after the survey is started, not the final depths after 10 years of the survey.}
\end{table*}

\section{Simulating and testing the spatially resolved SED fitting pipeline}
\label{sec:simulate_test_res_SEDfits}

Within the Euclid Collaboration, we plan to map the spatially resolved stellar population properties of a large number (100\,000s) of local galaxies at $z$ below 0.1 using upcoming imaging data from \euclid\ combined with imaging data from other large surveys. 
For nearby galaxies (within 100\,Mpc distance), we plan to include deep imaging data from \galex\ surveys to get coverage in the UV. For achieving this goal, we will use \pixedfit\ package \citep{Abdurrouf2021, Abdurrouf2022b_pixedfitcode}. In other efforts, we will apply SED fitting with stepwise SFH models (Euclid Collaboration: Nersesian et al., in prep.) and a machine learning method \citep{2025Kovacic_euclid}. One of the main objectives of this paper is to test to what extent \euclid-only images can be used to robustly map main stellar population properties using the spatially resolved SED fitting, and what improvements can be achieved by adding imaging data in the optical and UV. Synthetic images introduced in Sect.~\ref{sec:generate_synthetic_images} are used for this analysis. In this section, we will first describe the criteria for selecting our sample of simulated galaxies in Sect.~\ref{sec:sample_galaxies}, and then describe our analysis pipeline in Sect.~\ref{sec:analysis_pipeline}. 

\subsection{Sample of simulated galaxies}
\label{sec:sample_galaxies}

\begin{figure}[ht!]
\centering
\includegraphics[width=0.5\textwidth]{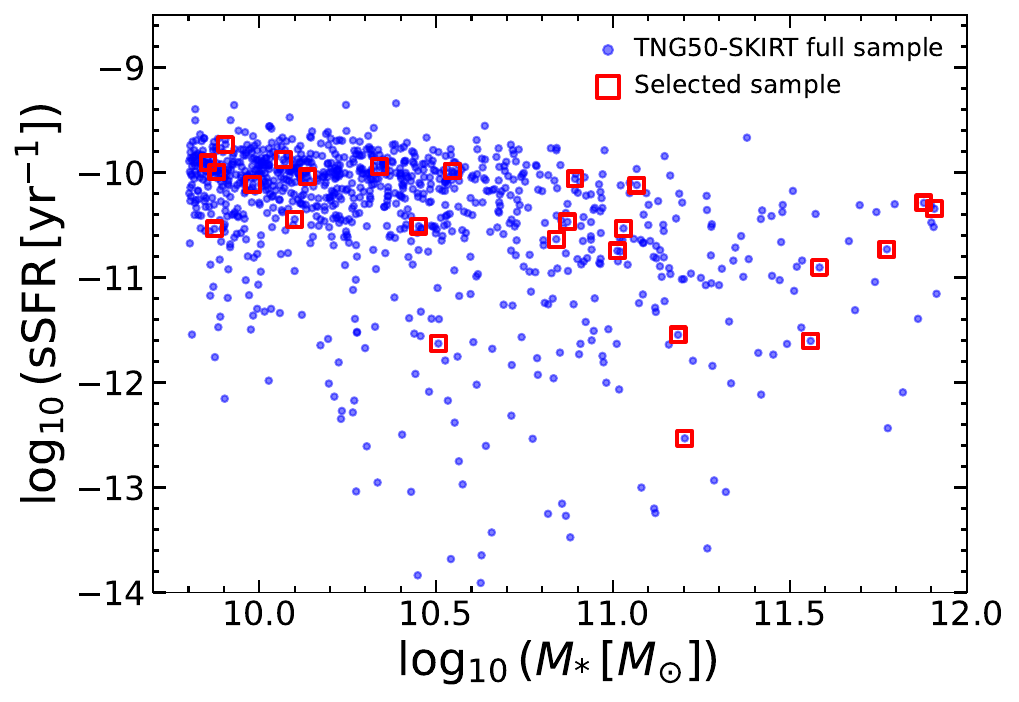}
\caption{The Sample of 25 simulated galaxies used for the analyses in this paper. The selected sample and the parent sample from the TNG50-SKIRT Atlas are shown with red squares and blue dots, respectively. The sample is selected to cover the entire mass range in the TNG50-SKIRT catalogue and a wide range of sSFR.}
\label{fig:plot_sample}
\end{figure}

We select a subset of galaxies from the parent sample of the TNG50-SKIRT Atlas. The sample is selected to cover a wide range of stellar mass ($10^{9.8}$--$10^{12}$\,\msun) and specific star-formation rate (sSFR; $10^{-13}$--$10^{-9}\,{\rm yr}^{-1}$). To construct the sample, first, we divide the entire mass range into five bins: $10^{9.8}$--$10^{10.0}$, $10^{10.0}$--$10^{10.5}$, $10^{10.5}$--$10^{11.0}$, and $10^{11.0}$--$10^{11.5}$, $10^{11.5}$--$10^{12.0}$\,\msun. After that, we further divide the galaxies in each mass bin into five quantile groups with increasing sSFR and then randomly choose one galaxy in each quantile. This results in a sample of 25 galaxies for our analysis in this paper. The distribution of our sample galaxies and the parent sample on the \mass-sSFR plane are shown in Fig.~\ref{fig:plot_sample}.  In addition to this sample selection, for each galaxy, we choose the most face-on configuration among the five viewing angles to minimize the systematic effects of the increasing dust attenuation in an edge-on view. This type of selection has been adopted in multiple spatially resolved studies to minimize projection effects that can otherwise bias the reconstruction of radial profiles from 2D maps of physical properties \citep[e.g.][]{Abdurrouf2017,Abdurrouf2018, Abdurrouf2022a}. Given the large number of galaxies that \euclid\ will observe, limiting the sample to relatively face-on galaxies will not significantly impact the conclusions of structural studies of galaxies. This is because, statistically, there is no expected intrinsic difference in the physical properties of galaxies viewed face-on versus edge-on.

The selected sample galaxies encompass various galaxy types, ranging from star-forming to passive galaxies. However, starburst galaxies are not included, and the imposition of a strict mass cutoff has resulted in the absence of dwarf galaxies. 

\begin{figure*}[h!]
\centering
\includegraphics[width=1.0\textwidth]{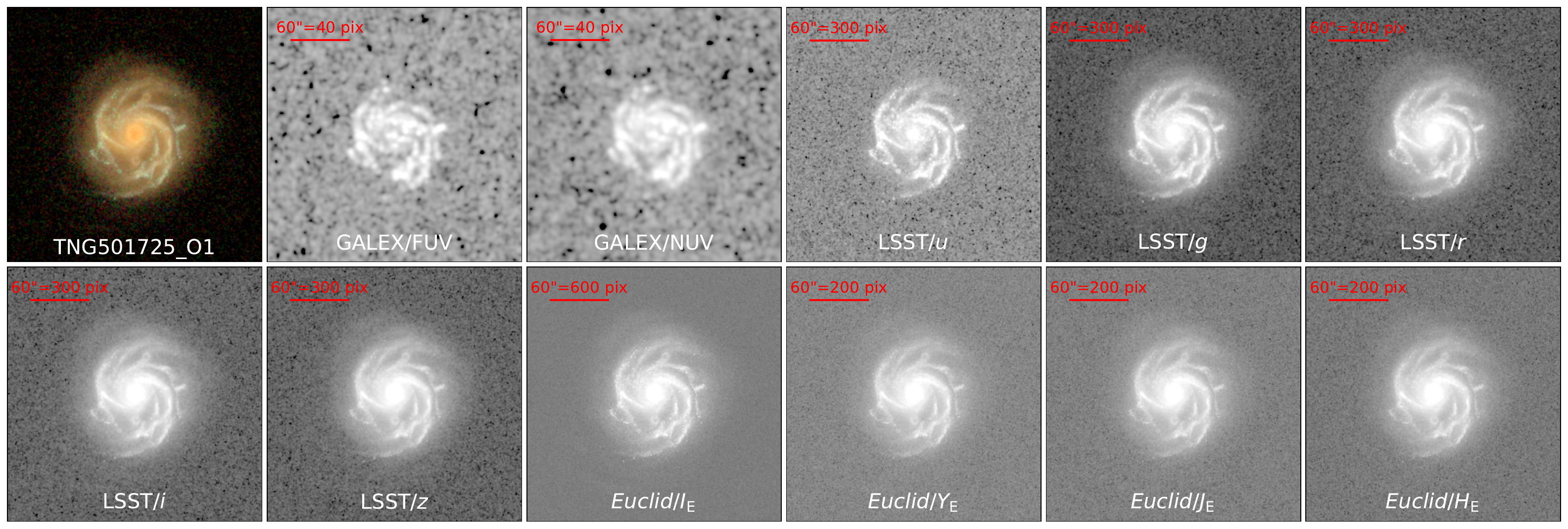}
\includegraphics[width=1.0\textwidth]{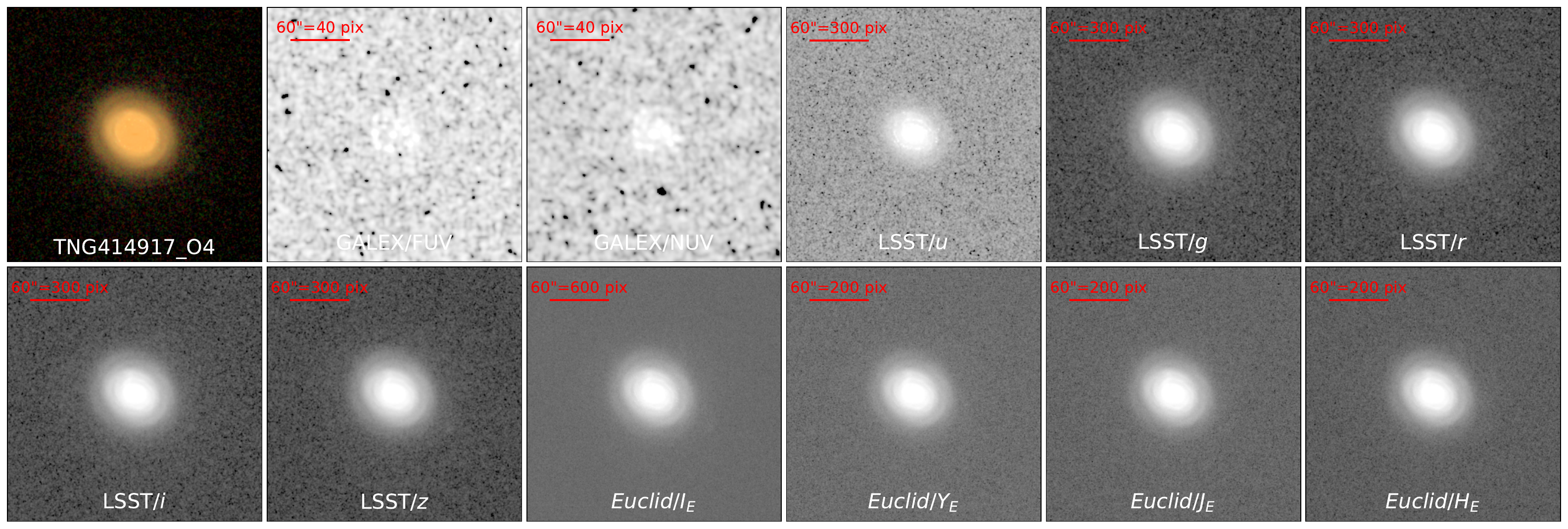}
\caption{Examples of simulated images of \galex, \lsst, and \euclid\ to which observational effects have been added, including spatial resampling (i.e.~regridding), simulated noise, and convolution with the PSF of the corresponding cameras. The galaxies here are TNG501725 with O1 orientation index and TNG414917 with O4 orientation index. The original synthetic images are taken from the TNG50-SKIRT Atlas database and are generated from the TNG50 simulation using the SKIRT radiative transfer code. The variation in spatial resolution and depth can be seen in the images with \galex\ images being the most blurred and shallowest, and \ie\ images being the sharpest and deepest. The pixel size difference is demonstrated with the red horizontal line (\ang{;;60} long). The PSF FWHM and pixel size of the images are summarized in Table~\ref{tab:mock_images}.  The colour images are made from the composite of \lsst\ $g$, $r$, and $i$ images.}
\label{fig:stamp_images}
\end{figure*}

The synthetic images of two example galaxies shown in Fig.~\ref{fig:stamp_images} clearly demonstrate the variation in spatial resolution and depth.
The \galex\ images are the shallowest and have the lowest spatial resolution, while the \ie\ image is the deepest and sharpest. TNG501725 is a spiral galaxy with prominent spiral arm structures, while TNG414917 is an elliptical galaxy. TNG501725 appears to be clumpy in the \galex\ images as these images are dominated by the light from young stars. In the longer-wavelength images, the galaxy appears to be less clumpy and the bulge becomes more prominent. TNG414917 is barely detected in \galex\ images but appears to be bright in \lsst\ and \euclid\ images, as expected for an elliptical galaxy, which has a lack of young stars.    

\subsection{Analysis pipeline with \pixedfit}
\label{sec:analysis_pipeline}

We used \pixedfit\ \citep{Abdurrouf2021,Abdurrouf2022b_pixedfitcode}, a state-of-the-art code that provides a comprehensive set of functionalities for analysing spatially resolved SEDs of galaxies. 
Given a set of multi-wavelength imaging data, \pixedfit\ performs the three following steps:
\begin{itemize}
\item image processing, consisting of PSF matching, spatial resampling, reprojection to match the spatial resolution (i.e.~PSF size), sampling (i.e.~pixel size), and projection of the images based on the world co-ordinate system (WCS) information in their headers
\item pixel binning to increase the $\rm{S}/\rm{N}$ of the spatially resolved SEDs
\item fit those SEDs with models to get the estimates of physical properties on spatially resolved scales.
\end{itemize}
Detailed descriptions of this code and robustness tests are given in \citet{Abdurrouf2021}. In the following, we test \pixedfit\ using realistic synthetic images of \euclid, \lsst, and \galex, which is the first such test using zoom-in hydrodynamical simulation performed for \pixedfit. This work intends to test how good the code is in recovering the maps of stellar population properties of simulated galaxies, given that we know the ground truth.

To investigate the effects of wavelength coverage and variation in the imaging datasets, we performed analyses on three different sets of imaging data cubes: (1) a combination of \galex, \lsst, and \euclid\ (11 bands) that cover UV to NIR (hereafter called \gledc\ data cube); (2) \lsst\ and \euclid\ (9 bands) that cover optical to NIR (hereafter called \ledc); and (3) \euclid\-only images that mostly cover NIR (hereafter called E). In the following, we describe briefly the three main steps of our analysis introduced before. We refer the reader to \citet{Abdurrouf2021} for a more detailed description of each step.

\subsubsection{Image processing and generation of photometric data cubes}
\label{sec:image_processing}

In the image processing with \pixedfit, input multi-wavelength images are matched in spatial resolution (i.e.~PSF size), spatial sampling (i.e.~pixel size), and projection (i.e.~a common WCS). PSF matching is performed to homogenize the spatial resolution of the multi-wavelength imaging data, which is performed by convolving higher-resolution images with pre-calculated kernels that can degrade their spatial resolutions to match those of the lowest-resolution image. We generate convolution kernels based on the PSF images in each filter using the \photutils\ package \citep{Bradley2022_photutils}. We use the same PSF images as the ones we used for convolving the synthetic images (see Sect.~\ref{sec:adding_obs_factors}). After PSF matching, the images are resampled (i.e.~regridded) into the same spatial sampling, which is chosen as the largest pixel size among the imaging data being involved. The final PSF size and the pixel size of the post-processed images produced from the image processing depend on the set of imaging data used. The final PSF FWHMs and pixel sizes of the three types of data cubes are summarized in Table~\ref{tab:image_processing}, showing that \euclid\ images alone result, obviously, in the highest spatial resolution.

\begin{table}[h!]
\caption{The Final PSF FWHM and pixel size of three types of data cubes produced in this work. \label{tab:image_processing}}
\centering
\begin{tabular}{l c c} 
\hline \hline \\ [-5pt]
 & Final & Final \\
Data cube & PSF FWHM & pixel size \\
 & (arcsec) & (arcsec) \\ [+3pt]
\hline \\ [-5pt]
\galex\ + \lsst\ + \euclid\ & $5.05$ & $1.5$ \\
\lsst\ + \euclid\ & $1.17$ & $0.3$ \\
\euclid\ & $0.50$ & $0.3$ \\
\hline
\end{tabular}
\end{table}

\begin{figure*}[h!]
\includegraphics[width=1.0\textwidth]{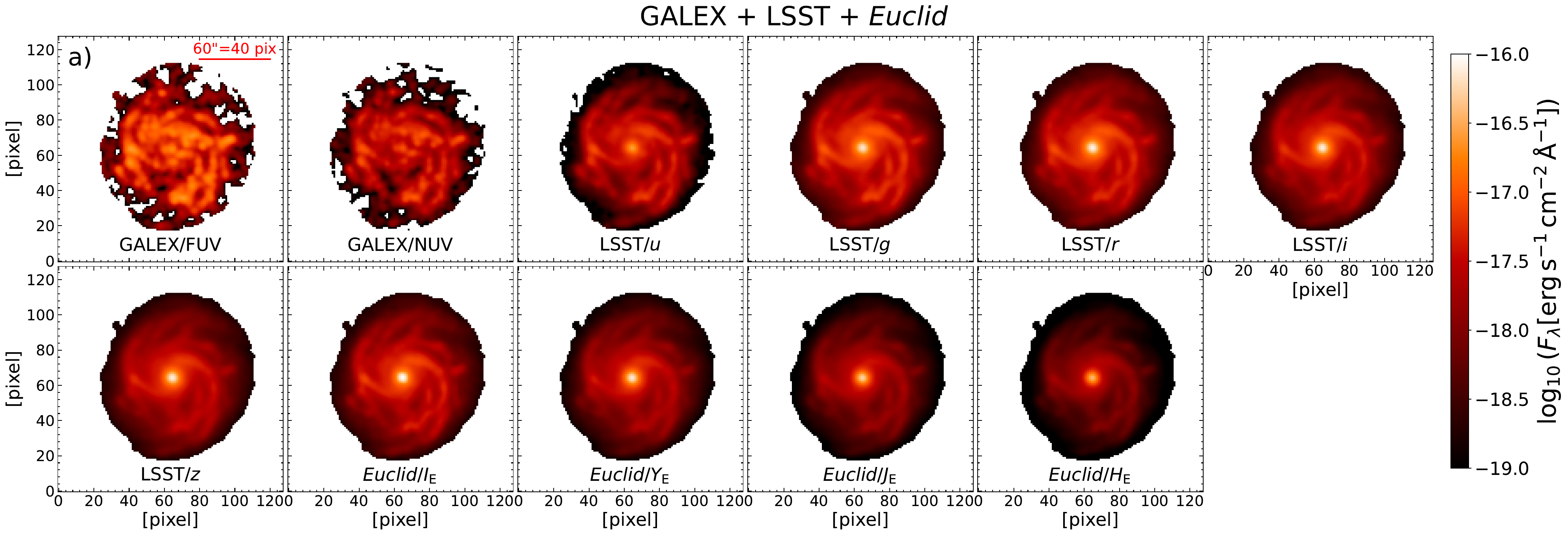}
\includegraphics[width=0.85\textwidth]{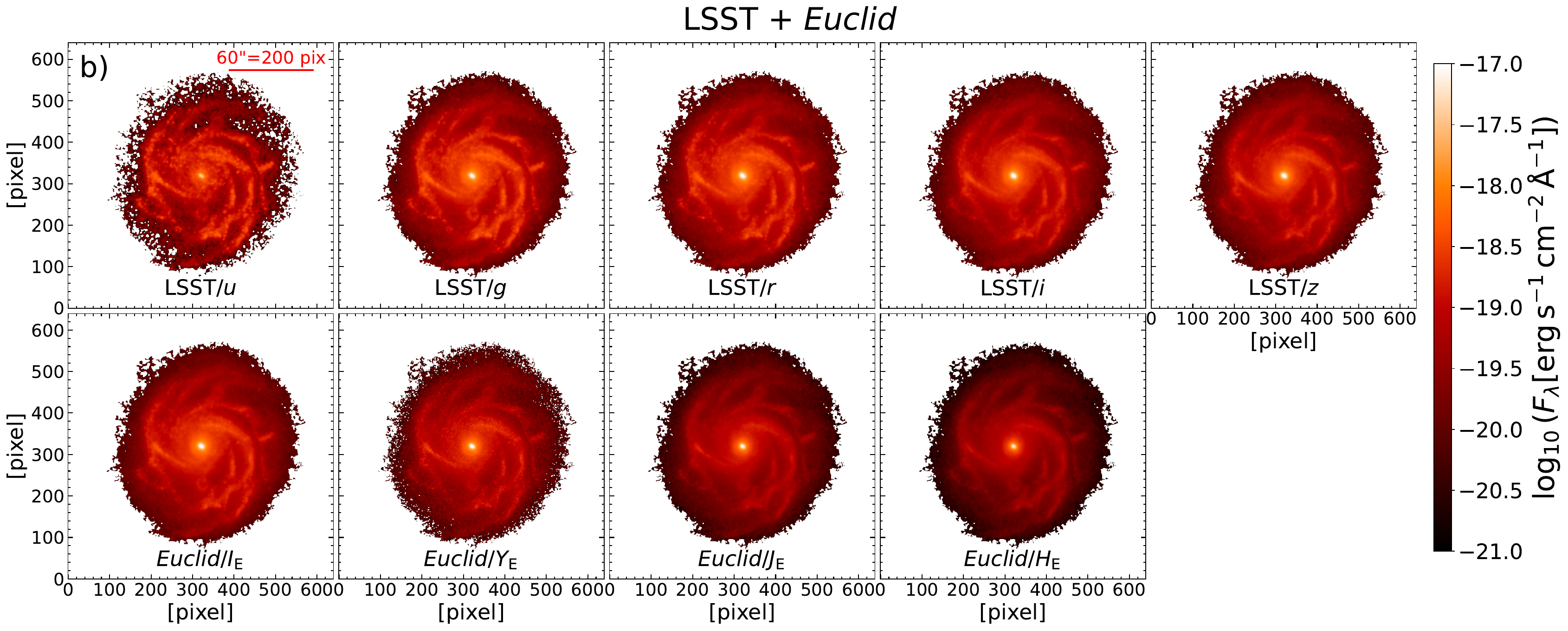}
\includegraphics[width=0.69\textwidth]{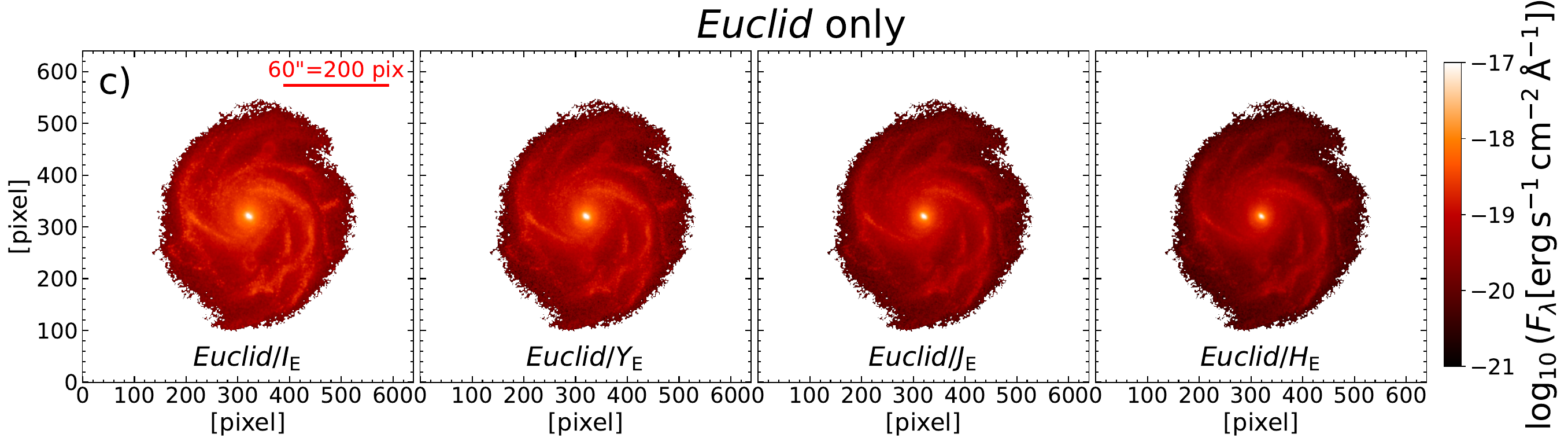}
\caption{Examples of the maps of flux densities produced from the image processing on the sets of three data cubes: \gledc\ (a), \ledc\ (b), and E (c). This example of the galaxy is TNG501725 with the O1 orientation index. The final PSF size and pixel size of the 3 data cubes are summarized in Table~\ref{tab:image_processing}. The \gledc\ data cube has the lowest spatial resolution, trading the high spatial resolution of \euclid\ and \lsst\ with the wider wavelength coverage from UV to NIR.}
\label{fig:img_process_comb}
\end{figure*}

Since all the synthetic images from the TNG50-SKIRT Atlas have the same projection, and we do not assign co-ordinates and WCS information to them, we only resample the images. In an analysis with real imaging data, \pixedfit\ can perform the reprojection and resampling processes simultaneously, which retain the WCS information of the images. The images are then cropped around the target galaxy to produce final stamp images.   
After that, the region of interest (RoI) of the galaxy, within which the spatially resolved SED fitting will be performed, is defined. First, segmentation maps are obtained using \sep\ \citep{Barbary2016_sep}, which is called within \pixedfit, with the detection threshold (\texttt{thresh}) of $2.0$ relative to the background RMS. The segmentation maps in all filters are then merged (i.e.~combined) together to get the final RoI of the galaxy. After that, the flux densities of pixels within the RoI are calculated in units of ${\rm erg}\,{\rm s}^{-1}\,{\rm cm}^{-2}\,\AA^{-1}$. This process produces 3D data cubes that are stored in FITS files. We perform the image processing to all the sample galaxies with the three sets of imaging data (i.e.~\gledc, \ledc, and \euclid\ only), resulting in 75 data cubes, 3 data cubes for each sample galaxy. Examples of the maps of flux densities of the galaxy TNG501725 from the 3 types of data cubes are shown in Fig.~\ref{fig:img_process_comb}.

\begin{figure}[h!]
\centering
\includegraphics[width=0.5\textwidth]{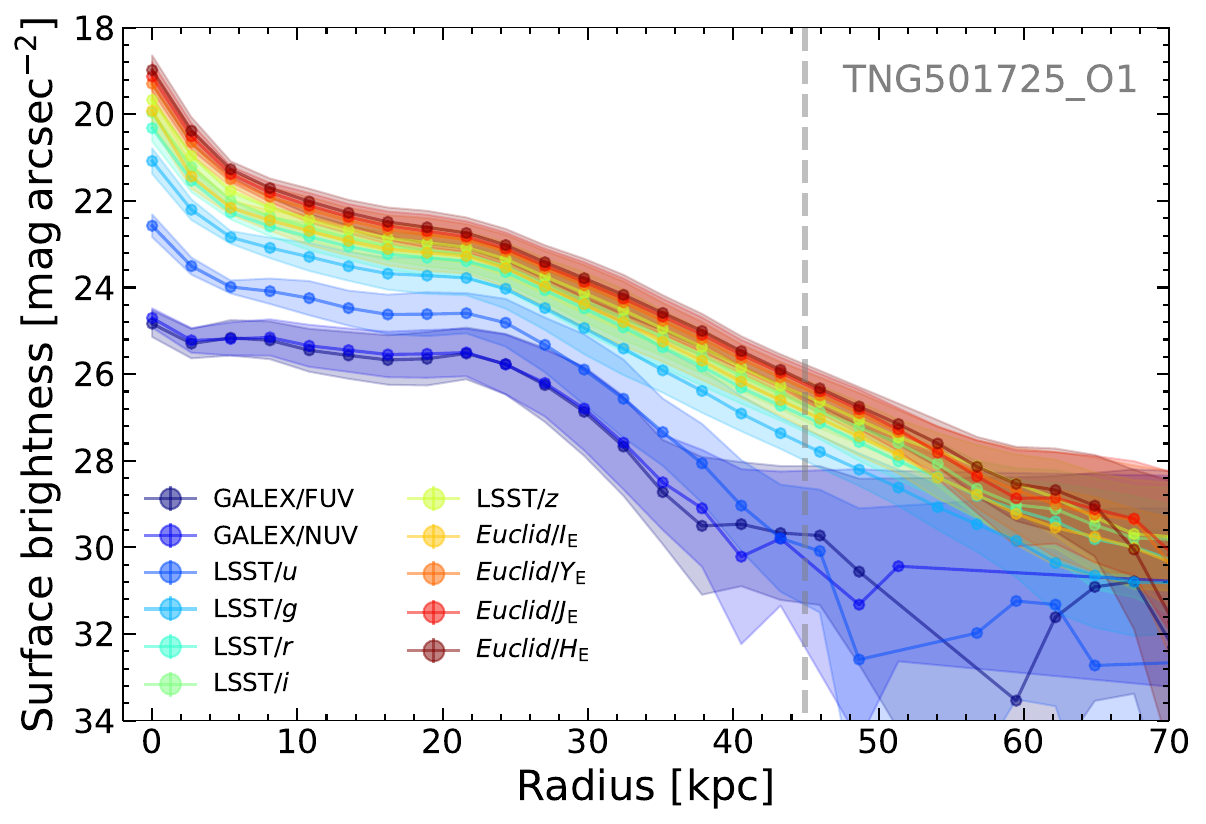}
\caption{Surface brightness radial profiles of TNG501725 derived from the stamp images of \gledc\ data. The vertical dashed grey line indicates the radial extent of the data cube after the galaxy's RoI is defined, roughly corresponding to a detection threshold of $2$ above the background RMS of Euclid images.}
\label{fig:plot_radprof_SB}
\end{figure}

Since the galaxy's RoI is defined by combining segmentation maps in multiple bands, the deepest images control the final region because they usually have a larger segmentation area. Figure~\ref{fig:plot_radprof_SB} shows surface brightness radial profiles of TNG501725 derived from the stamp images of \gledc\ data. The deep \lsst\ and \euclid\ images extend the galaxy's region up to a radius of around $60$ kpc and reach a surface brightness of up to $29$ $\rm{mag}$ $\rm{arcsec}^{-2}$ before sky background noise dominate, while shallower \galex\ images only reach up to around $40$ kpc with a surface brightness of up to $28$ $\rm{mag}$ $\rm{arcsec}^{-2}$. The galaxy's RoI extends up to a radius of $\sim 45$ kpc for this particular galaxy, as indicated by the vertical dashed grey line. As we can see from this figure, the galaxy's RoI is mainly defined by the \lsst\ and \euclid\ images, and the flux maps of \galex\ are extended such that their outskirt regions are noisy, which can also be seen in Fig.~\ref{fig:img_process_comb}. In the \galex\ map, the missing pixels at the galaxy's outer regions are caused by negative flux values. Due to the brighter overall surface brightness in the longer wavelengths in this local galaxy, the lowest surface brightness reached in the outskirts of the galaxy's RoI is deeper in the shorter wavelengths. For TNG501725, the minimum surface brightness of \galex\ (FUV and NUV), \lsst\ ($u$, $g$, $r$, $i$, $z$), and \euclid\ (\ie, \ye, \je, \he) are $28.7$, $28.9$, $29.1$, $27.6$, $26.9$, $26.7$, $26.4$, $26.9$, $26.3$, $26.2$, and $26.1$ mag $\rm{arcsec}^{-2}$, respectively. For all galaxies in the sample, the median values are $28.7$, $28.7$, $29.1$, $27.8$, $27.2$, $27.0$, $26.8$, $27.3$, $26.7$, $26.6$, and $26.6$ mag $\rm{arcsec}^{-2}$. We note that these values depend on the assumed exposure times in our noise simulations.  
Moreover, the RoI can also be enlarged by decreasing \texttt{thresh} to include the fainter outskirt region of the galaxies.

Given the nature of spatially resolved analyses that involve multi-band imaging data from various telescopes with differing spatial resolutions, there is a concern that PSF matching may alter a galaxy's surface brightness. To assess this, we examined the effect of PSF convolution on surface brightness radial profiles and found that the impact is minimal, even when convolving to the broadest resolution of the NUV band.

We also investigated the effect on peak surface brightness (i.e.~the central value) and found no significant overall change due to PSF convolution. However, a slight systematic effect emerges when convolving to the NUV resolution, where the peak surface brightness in the convolved images is decreased by $\sim 0.2$ dex compared to the original images. This effect is not present in the FUV and NUV images themselves but is observed in the other bands.

\subsubsection{Pixel binning}\label{sec:pixel_binning}

The fluxes of pixels usually have low $\rm{S}/\rm{N}$ and cannot provide sufficient constraints on the model SEDs. \pixedfit\ provides a pixel binning feature that can maximize the $\rm{S}/\rm{N}$ of SEDs on spatially resolved scales by binning neighbouring pixels that have similar SED shapes. This scheme of \pixedfit\ can retain important information on the pixel level while binning pixels together, which degrades the spatial resolution. This is because it takes into account the similarity of SED shape among the pixels such that only pixels that have similar SED shape are binned together.      
The pixel binning scheme takes four input parameters: (1) the $\rm{S}/\rm{N}$ thresholds that can be set for multiple filters; (2) reduced $\chi^{2}$ value for evaluating the similarity of SED shape; (3) the minimum diameter of spatial bins; and (4) reference filter for sorting pixels based on their brightness. We refer the reader to \citet{Abdurrouf2021} for a more detailed description.  

We experiment to get optimum binning parameters that can give a sufficient $\rm{S}/\rm{N}$ without losing spatial resolution and detailed spatial structures of our sample galaxies. We converge to the following binning parameters:
\vspace*{-2pt}
\begin{itemize}
    \item[-] The minimum $\rm{S}/\rm{N}$ of 5 in all filters other than FUV, NUV, and $u$;
    \item[-] Maximum reduced $\chi^{2}=5$ for the evaluation of the similarity in SED shape among pixels;
    \item[-] The smallest bin's diameter of 5 pixels; 
    \item[-] \ie\ is chosen as the reference filter for sorting the pixel brightness. 
\end{itemize}
We do not set the $\rm{S}/\rm{N}$ threshold for FUV, NUV, and $u$ filters because pixels in those images have low $\rm{S}/\rm{N}$ such that setting constraints on them can result in bigger overall bin sizes as compared to the binning map obtained with the current setting. This can cause the loss of detailed spatial information in the galaxies. We apply these binning parameters to all 75 photometric data cubes from our three analyses (see Sect.~\ref{sec:image_processing}). The total number of spatial bins from these data cubes is 219\,158.

\begin{figure*}[h!]
\centering
\includegraphics[width=1.0\textwidth]{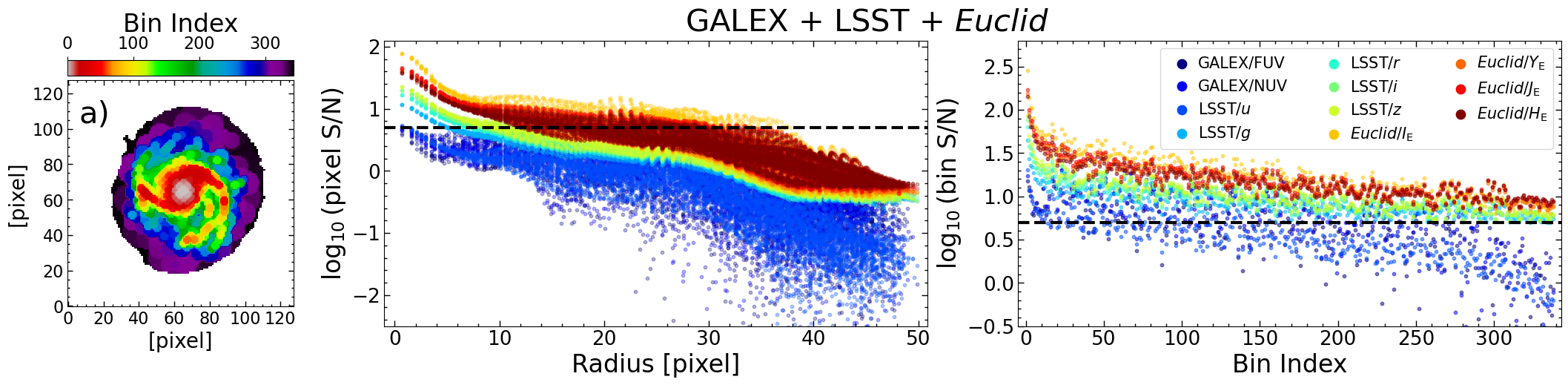}
\includegraphics[width=1.0\textwidth]{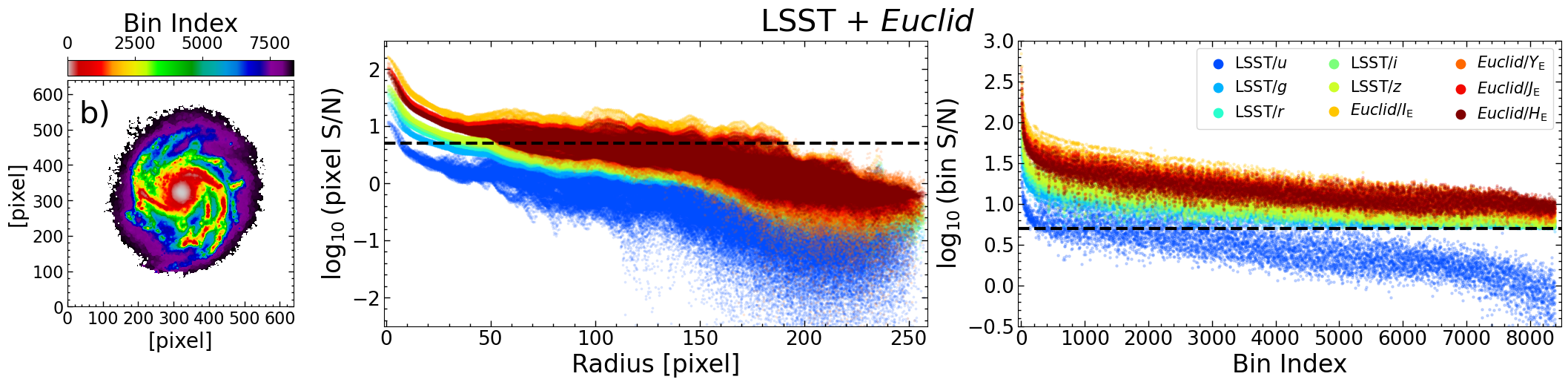}
\includegraphics[width=1.0\textwidth]{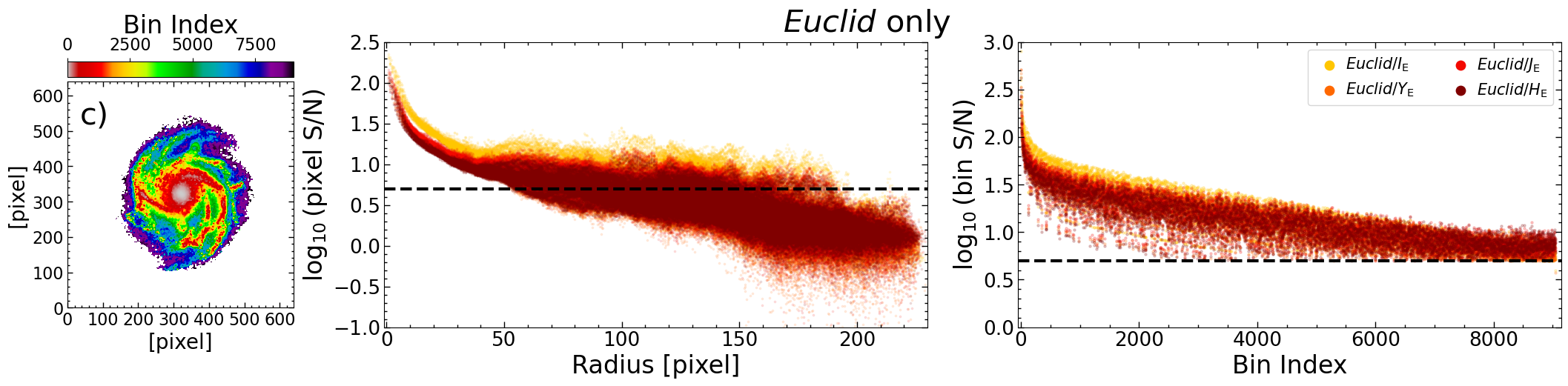}
\caption{Examples of the pixel binning maps of TNG501725 obtained with the three data cubes: \gledc\ (a), \ledc\ (b), and \euclid\ only (c). In each row, the three panels from left to right show the binning map, $\rm{S}/\rm{N}$ radial profile of pixels, and $\rm{S}/\rm{N}$ radial profile of spatial bins. The pixel binning scheme of \pixedfit\ can achieve the $\rm{S}/\rm{N}$ threshold of $5$ (indicated by the dashed black lines) for the $g$, $r$, $i$, $z$, \ie, \ye, \je, and \he\ filters.}
\label{fig:pixbin}
\end{figure*}

Binning maps and $\rm{S}/\rm{N}$ plotted as a function of pixels and spatial bins for TNG501725, obtained with the three data cubes, are shown in Fig.~\ref{fig:pixbin}. As we can see from the plots, the pixel binning by \pixedfit\ can achieve the $\rm{S}/\rm{N}$ thresholds. The \gledc\ data cube has the least number of spatial bins (343) mainly due to the larger pixel size of this as compared to the other two data cubes. The \ledc\ and E data cubes have a total number of spatial bins of 8391, and 9044, respectively. The smaller number of spatial bins in the former, although they both have the same pixel size, is caused by the slightly lower average $\rm{S}/\rm{N}$ per pixel of the \lsst\ images.

\subsubsection{SED fitting set-up}
\label{sec:sed_fitting}

After performing pixel binning to the data cubes, we proceed with SED fitting. For each galaxy, \pixedfit\ fits the SED of spatial bins in the galaxy and obtains the estimates of the physical properties of the underlying stellar populations. Since our synthetic imaging data only covers UV to NIR, we only include the stellar emission modelling in our SED fitting. We turn off the modelling of dust emission because we do not have data in MIR and FIR that can constrain the models. We also turn off the nebular emission modelling in this test analysis because the broad-band photometry is expected to be weakly affected by the nebular emission, especially for our sample of local galaxies that are forming stars only moderately.
We will implement this modelling in future analyses with real data. A detailed description of the SED modelling and fitting method is given in \citet{Abdurrouf2021}. In the following, we briefly outline the priors that we assume in our SED modelling and fitting.  

We assume the \citet{Chabrier2003} IMF, the \citet{Calzetti2000} dust attenuation law, and a parametric star-formation history (SFH) model in the form of a double power law. It has been shown in \citet{Abdurrouf2021} that this SFH model has sufficient flexibility to be able to reconstruct the shape of the true SFH when tested with mock SEDs of simulated galaxies from the TNG100 simulation. The SED modelling has six free parameters and three dependent parameters. We summarize these parameters along with the assumed priors in Table~\ref{tab:sedfits_params}. The SFR is defined as the recent instantaneous rate determined from the SFH at the time of observation. We have verified that this SFR is consistent with the value obtained by averaging over a specific timescale (e.g.~100 Myr), which is a common approach in the literature.

For the fitting method, we use the Bayesian inference method with the random dense sampling of parameter space (RDSPS) as the posterior sampling method. This method samples the parameter space by generating random values for each parameter that are uniformly distributed within the assumed range. We choose this method because it is much faster than the MCMC method.\footnote{RDSPS takes around 20 s to fit a single SED with 100\,000 random model SEDs, running in parallel on a 20-core computer, while the MCMC takes 15--20 minutes.} We refer the reader to \citet[][Sec. 4.2.2 therein]{Abdurrouf2021} for more detailed descriptions of this method. In this fitting method, \mass\ is not a free parameter. For each model, \mass\ is determined by the best model normalization derived by minimizing the $\chi^2$ \citep[see e.g.~][their Eq.~6]{Abdurrouf2017}. The mass-weighted age $t_{\rm M}$, which is a representative age of a stellar population, is directly calculated from the model's SFH using the following equation:
\begin{equation}
t_{\rm M} = \frac{\int_{t_{\rm form}}^{t_z} t\,\mathrm{SFR}(t)\,\diff t}{\int_{t_{\rm form}}^{t_z} \mathrm{SFR}(t)\,\diff t},
\label{eq:eq_mw_age} 
\end{equation}
where $t_z$ is the age of the Universe at the galaxy's redshift and $t_{\rm form}$ is the cosmic time when star formation is started in the galaxy. We refer to this quantity interchangeably as `age' or `mass-weighted age' in the paper. It is worth mentioning that the assumed flat priors induce a `hidden' non-flat prior to the mass-weighted age, peaking around 4--5 Gyr. This age distribution closely resembles the observed distribution of spatially resolved stellar population age distribution in local galaxies by MaNGA \citep{Abdurrouf2022manga,Sanchez2022,2022Lacerda}. 
It is also a distribution found in the priors of a variety of previous studies, as shown in the comparative study of \citet{Leja2019}.

\begin{table*}[h!]
\caption{Parameters and assumed priors in the SED modelling and fitting processes. \label{tab:sedfits_params}}
\centering
\begin{tabular}{lp{6cm}p{5cm}l} 
\hline \hline \\ [-5pt]
Free parameter & Description & Prior & Sampling/Scale \\
\hline \\ [-5pt]
$Z_{*}$ & Stellar metallicity & Uniform: $\log_{10}(Z_{*,\rm{min}}/Z_{\odot})=-1.0$, $\log_{10}(Z_{*,\rm{max}}/Z_{\odot})=0.2$ & Logarithmic \\
$t$ & Time since the onset of star formation\textsuperscript{a} & Uniform: $\log_{10}(t_{\rm{min}})=\log_{10}(t_{z})-0.3$, $\log_{10}(t_{\rm{max}})=\log_{10}(t_{z})$\textsuperscript{b} & Logarithmic \\
$\tau$ & Parameter that controls the peak time in the double power-law SFH model\textsuperscript{a} & Uniform: $\log_{10}(\tau_{\rm min})=0.3$, $\log_{10}(\tau_{\rm max})=0.9$ & Logarithmic \\
$\alpha$ & Falling slope of the SFH model\textsuperscript{a} & Uniform: $\log_{10}(\alpha_{\rm min})=-0.8$, $\log_{10}(\alpha_{\rm max})=0.8$ & Logarithmic \\
$\beta$ & Rising slope of the SFH model\textsuperscript{a} & uniform: $\log_{10}(\beta_{\rm min})=-0.8$, $\log_{10}(\beta_{\rm max})=0.8$ & Logarithmic \\
$\hat{\tau}_{2}$ & Dust optical depth in the \citet{Calzetti2000} dust attenuation law & Uniform: $\rm{min}=0.0$, $\rm{max}=2.5$ & Linear \\

\hline \\ [-5pt]
Dependent parameter & Description & Method & Sampling/Scale \\
\hline \\ [-5pt]
\mass\ & Stellar mass & Best-fit model normalization obtained from minimizing $\chi^{2}$ & Logarithmic \\
SFR & Star formation rate & Recent instantaneous SFR determined from SFH at the present time &  Logarithmic \\
Age & Representative age of a stellar population & Mass-averaging the look-back time in the SFH (Eq.~\ref{eq:eq_mw_age}) & Logarithmic \\
\hline
\end{tabular}
\tablefoot{\textsuperscript{a} The mathematical formula of the double power-law SFH can be seen in \citet[][Equation 7 therein]{Abdurrouf2021}. \textsuperscript{b} $t_{z}$ is the age of the Universe at the galaxy's redshift. All the simulated galaxies in our sample are put at $z=0.03$. $t$, $t_{z}$ are in units of Gyr. $\hat{\tau}_{2}$ is in units of magnitude.} 
\end{table*}

For each data cube, we performed multiple SED fitting runs by varying two factors: the adopted SPS models and the priors. We consider two SPS models: the flexible stellar population synthesis \citep[FSPS;][]{Conroy2009, Conroy2010} and the 2016 version of \citet{Bruzual2003}, hereafter BC16. The former is the native SPS model used in \pixedfit. With this, we can learn possible systematic effects caused by the SPS model on the fitting results. In the FSPS model, we use the Padova isochrones \citep{Girardi2000, Marigo2007, Marigo2008} and stellar spectral library from the Medium-resolution Isaac Newton Telescope library of empirical spectra \citep[MILES;][]{Sanchez-Blazquez2006, Falcon-Barroso2011,Rock+16}.\footnote{We have verified that the BaSeL spectral library produces overall consistent UV-to-NIR colours compared to those from the MILES library. Since our SED fitting results are primarily driven by these colours, we expect that the choice of spectral library--at least between MILES and BaSeL--does not significantly affect the SED fitting outcomes.} 

For generating model SEDs based on the BC16 model that assumes the double power-law SFH, we used \bagpipes\ \citep{Carnall2018}. This code applies the same isochrones and stellar spectral library as those used in FSPS, but a different IMF from \citet{Kroupa2002}. However, the expected difference in the estimated \mass\ between \citet{Kroupa2002} and \citet{Chabrier2003} IMFs is very small (within $\sim 0.03$ dex, see e.g.~\citealt{Speagle2014}). In both the SPS models, each SED is characterized by a constant metallicity.

For the variation in priors, we defined two settings: the first corresponds to the classical flat priors described in Table~\ref{tab:sedfits_params}, and the second adds a special prior in the form of mass-$Z$ and mass-age relations (hereafter called mass-$Z$-age prior). We have constructed the latter prior from the combination of empirical relations taken from the MaNGA survey \citep{Bundy2015} and theoretical predictions from the TNG50 simulation. We describe this in more detail in Appendix~\ref{sec:appendix_priors}. The extra prior is particularly relevant when the wavelength coverage is limited and physical information available in the data is scarce, as will be further discussed later.

\section{Results}
\label{sec:results}

\subsection{Examples of the maps of stellar population properties}
\label{sec:example_maps_properties}

\begin{figure*}[h!]
\centering
\includegraphics[width=0.85\textwidth]{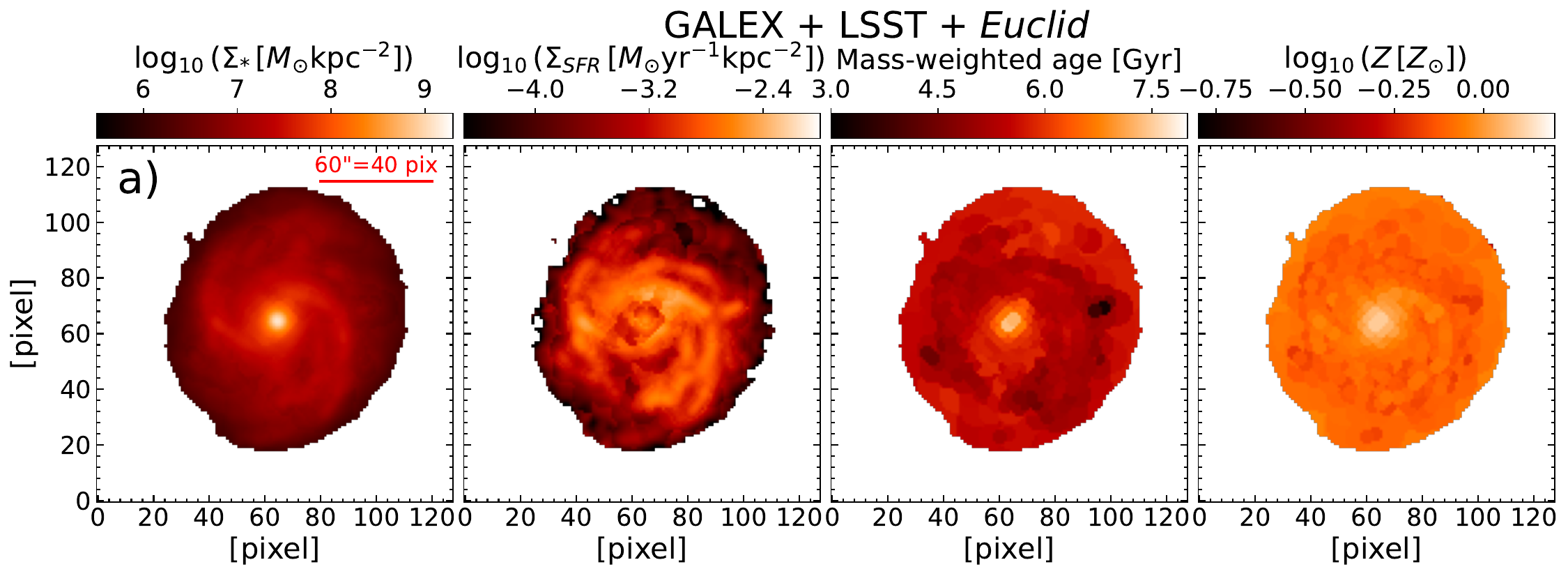}
\includegraphics[width=0.85\textwidth]{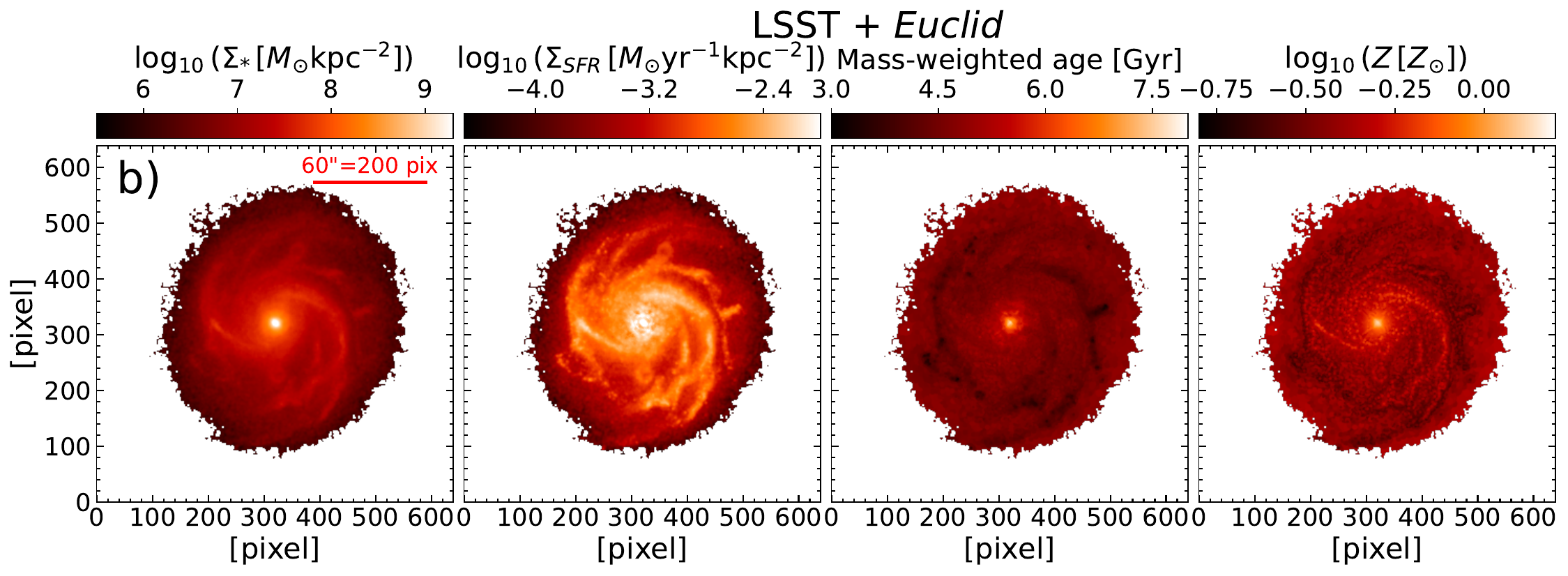}
\includegraphics[width=0.85\textwidth]{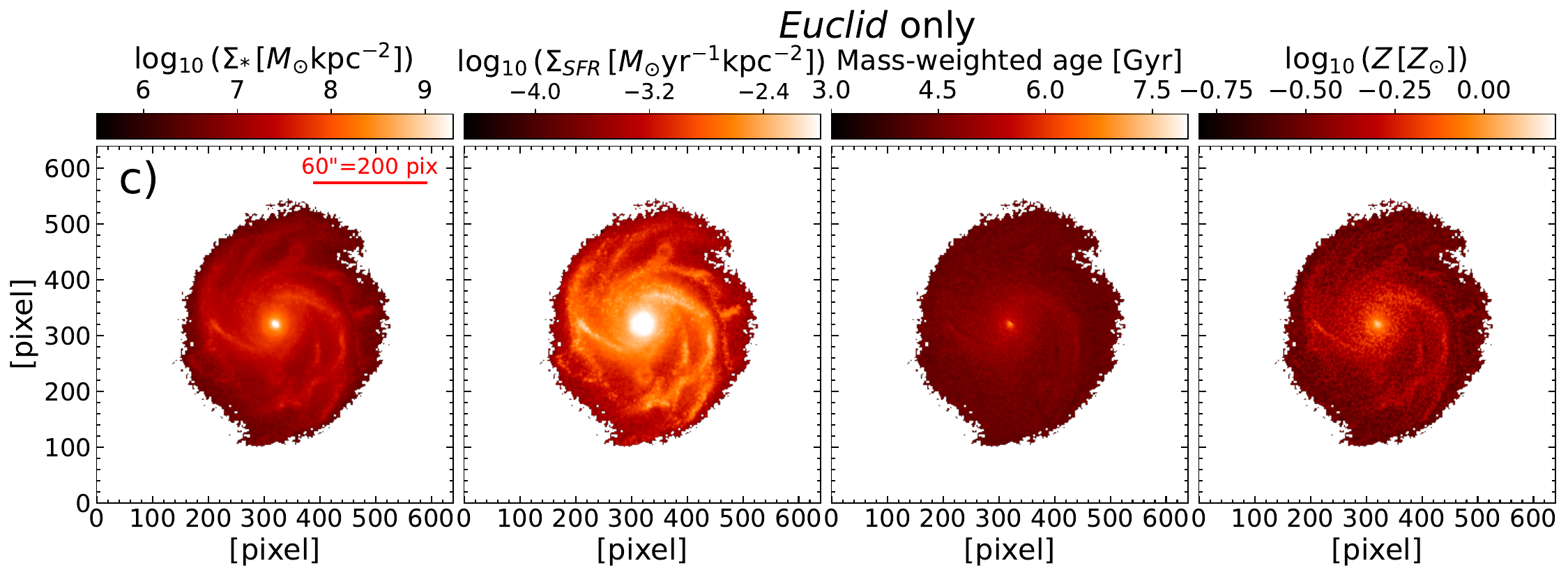}
\caption{Examples of the maps of the stellar population properties of TNG501725 obtained from the spatially resolved SED fitting on the three data cubes: \gledc\ (a), \ledc\ (b), and E (c). The maps derived from the analysis with the \ledc\ and E data cubes alone have a high spatial resolution and reveal detailed spatial variation in the physical properties compared to those obtained from the analysis with the \gledc\ data cubes.}
\label{fig:maps_props_example}
\end{figure*}

After SED fitting is performed on all the spatial bins in a galaxy, 2D maps of stellar population properties can be constructed. The properties include stellar mass surface density (\massd), SFR surface density (\sfrd), mass-weighted age, and metallicity ($Z$).  
Examples of the maps of stellar population properties of TNG501725 obtained from the analyses with the three data cubes are shown in Fig.~\ref{fig:maps_props_example}. The analyses with \ledc\ and E data cubes yield maps that can resolve the spiral arms and bulge structures in great detail, in contrast to the \gledc\ due to its worse resolution. These structures appear in all the maps of properties, not only in the \massd\ and \sfrd. From the maps, especially the ones obtained with \ledc\ data cube, we can see that overall, bulge and spiral arms have higher \massd, \sfrd, metallicity, and younger stellar populations than the interarm regions. The maps of the rest of our sample galaxies are presented in Appendix~\ref{sec:maps_properties_all}.       

\subsection{Recovering spatially resolved stellar population properties}
\label{sec:recov_spatial_resolved_SP}

\begin{table*}[h!]
\caption{The reliability of stellar population properties derived from SED fitting that applies the mass-$Z$-age prior. \label{tab:robustness_fits_with_priors}}
\centering
\begin{tabular}{lcccccc}
\hline \hline \\ [-5pt]
 & \multicolumn{6}{c}{Data cube} \\
\cline{2-7} \\ [-5pt]
 & \multicolumn{2}{c}{\galex\ + \lsst\ + \euclid} & \multicolumn{2}{c}{\lsst\ + \euclid} & \multicolumn{2}{c}{\euclid-only} \\ 
\cline{2-7} \\ [-5pt]
 & BC16 & FSPS & BC16 & FSPS & BC16 & FSPS \\
\hline \\ [-5pt]
$\Sigma_{*}$ & & & & & & \\
\hline \\ [-5pt]
$\mu$\textsuperscript{a} & 0.05 & 0.02 & 0.05 & 0.01 & 0.05 & 0.01 \\ 
$\sigma$\textsuperscript{b} & 0.10 & 0.10 & 0.12 & 0.08 & 0.11 & 0.11 \\
$\rho$\textsuperscript{c} & 0.98 & 0.98 & 0.97 & 0.99 & 0.97 & 0.97 \\
\hline \\ [-5pt]
$\Sigma_{\rm SFR}$ & & & & & & \\
\hline \\ [-5pt]
$\mu$\textsuperscript{a} & 0.06/$-$0.39 & 0.12/$-$0.37 & 1.87/$-$0.15 & 1.85/$-$0.16 & 2.03/$-$0.17 & 2.00/-0.19 \\
$\sigma$\textsuperscript{b} & 0.77/0.24 & 0.82/0.22 & 3.61/0.34 & 3.60/0.31 & 3.40/0.41 & 3.39/0.41 \\
$\rho$\textsuperscript{c} & 0.92/0.68 & 0.91/0.72 & 0.59/0.54 & 0.62/0.55 & 0.48/0.39 & 0.48/0.39 \\
\hline \\ [-5pt]
Age & & & & & & \\
\hline \\ [-5pt]
$\mu$\textsuperscript{a} & $-$0.06 & $-$0.07 & $-$0.10 & $-$0.10 & $-$0.11 & $-$0.12 \\
$\sigma$\textsuperscript{b} & 0.09 & 0.09 & 0.09 & 0.09 & 0.10 & 0.10 \\
$\rho$\textsuperscript{c} & 0.42 & 0.43 & 0.62 & 0.63 & 0.46 & 0.47 \\
\hline \\ [-5pt]
$Z$ & & & & & & \\
\hline \\ [-5pt]
$\mu$\textsuperscript{a} & $-$0.04 & $-$0.01 & $-$0.11 & $-$0.12 & $-$0.14 & $-$0.16 \\
$\sigma$\textsuperscript{b} & 0.08 & 0.08 & 0.11 & 0.11 & 0.13 & 0.13 \\
$\rho$\textsuperscript{c} & 0.69 & 0.72 & 0.11 & 0.11 & 0.13 & 0.13 \\
\hline
\end{tabular}
\tablefoot{\textsuperscript{a} Systematic offset defined as the mean of the logarithmic ratios between the parameter derived from SED fitting and the true value: $\log_{10}({\rm fit}/{\rm true})$. \textsuperscript{b} The standard deviation of $\log_{10}({\rm fit}/{\rm true})$. \textsuperscript{c} The Spearman rank-order correlation coefficient. For $\Sigma_{\rm SFR}$, the values to the right of the slash symbol are calculated after applying a threshold of $\Sigma_{\rm SFR}>10^{-3.0}\,M_{\odot}\,\rm{yr}^{-1}\,\rm{kpc}^{-2}$.}
\end{table*}

In this section, we analyse the robustness of our analysis pipeline in recovering the true stellar population properties on spatially resolved scales. We investigate the effects of three important factors on the fitting results: wavelength coverage of the data cube, SPS model, and assumed priors. For the prior variations, we compare between flat (i.e.~uniform) and the mass-$Z$-age priors. First, in this section, we directly compare the inferred parameters of all spatial bins in the sample galaxies derived from SED fitting and the true values from the TNG50 simulations. We focus on four main properties in this paper: \massd, \sfrd, age, and $Z$. 

\begin{figure*}[h!]
\centering
\includegraphics[width=1.0\textwidth]{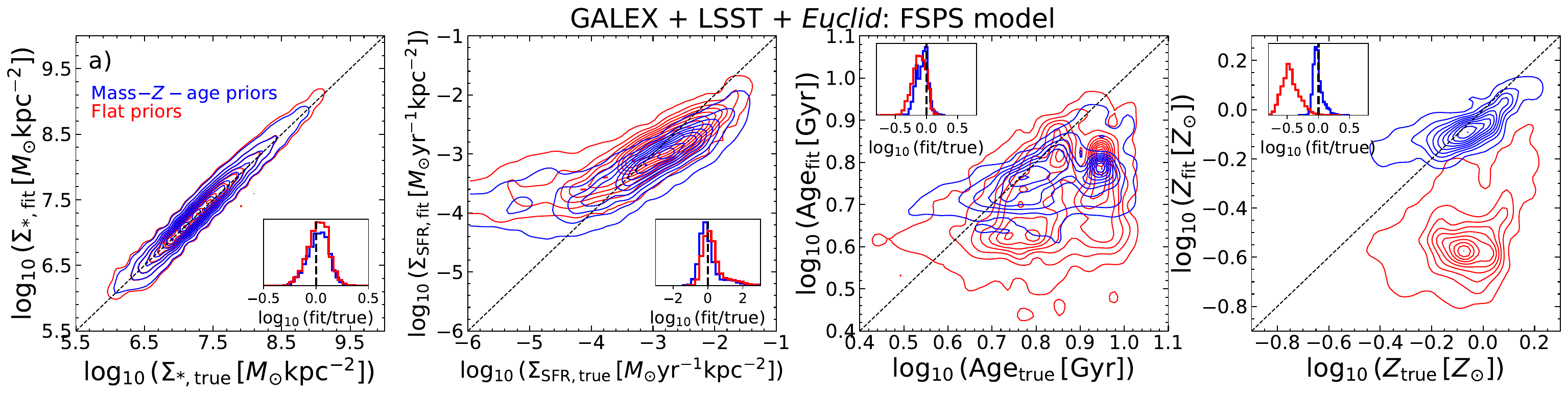}
\includegraphics[width=1.0\textwidth]{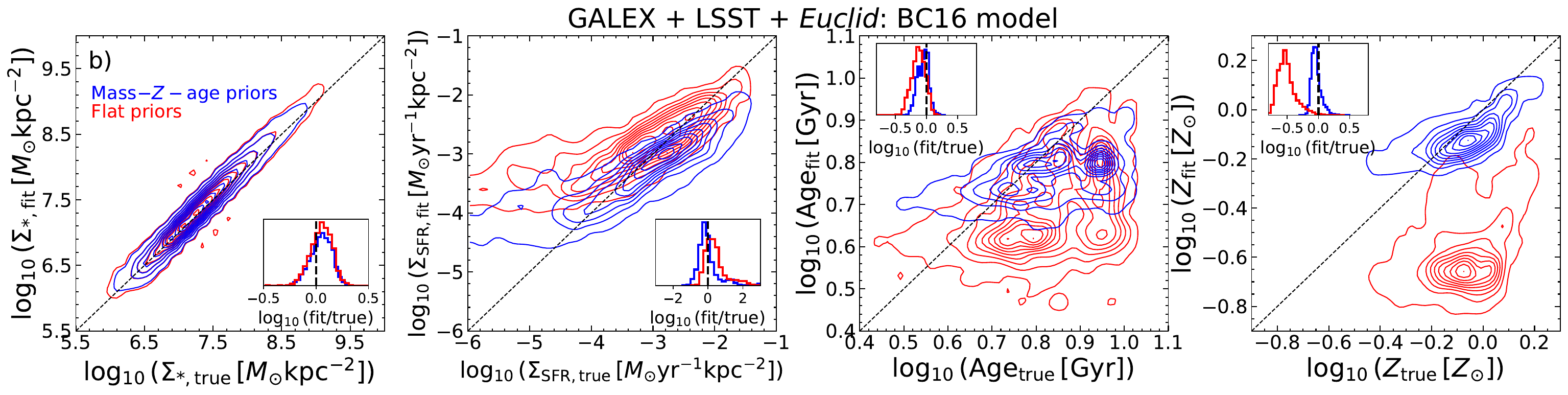}
\caption{Comparisons of the spatially resolved stellar population properties derived from SED fitting on the \gledc\ data cubes (ordinate) and the true properties (abscissa). These include all spatial bins from the whole sample of galaxies. We focus on four main properties in this test analysis: \massd, $Z$, age, and \sfrd. The top row (a) shows results obtained with the FSPS model, while the bottom row (b) shows results using the BC16 model. The blue and red colours represent results obtained with and without applying the mass-$Z$-age prior.}
\label{fig:comp_simfits_GLE}
\end{figure*}

\begin{figure*}[h!]
\centering
\includegraphics[width=1.0\textwidth]{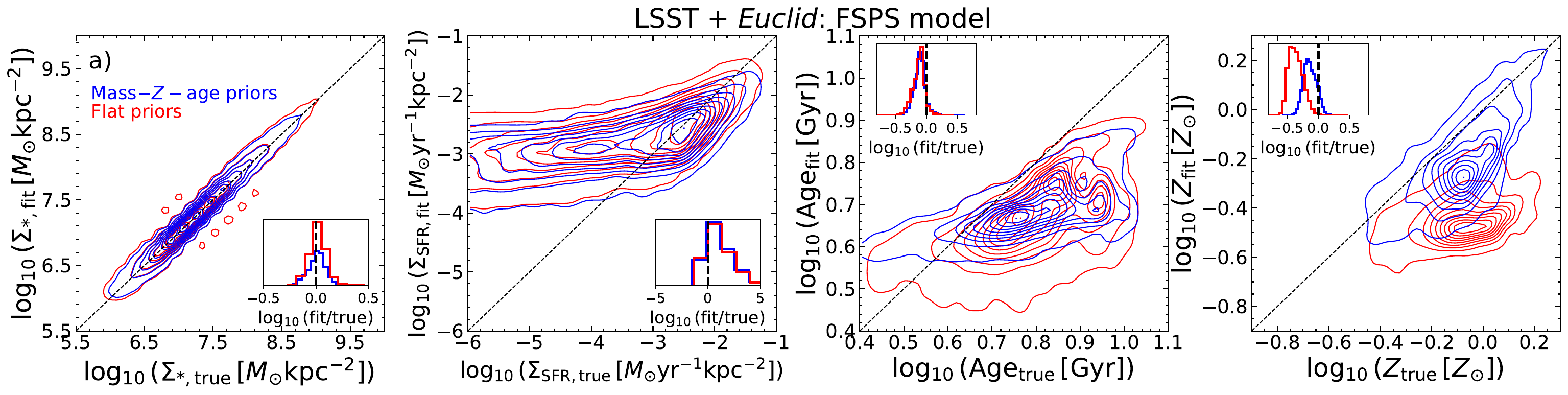}
\includegraphics[width=1.0\textwidth]{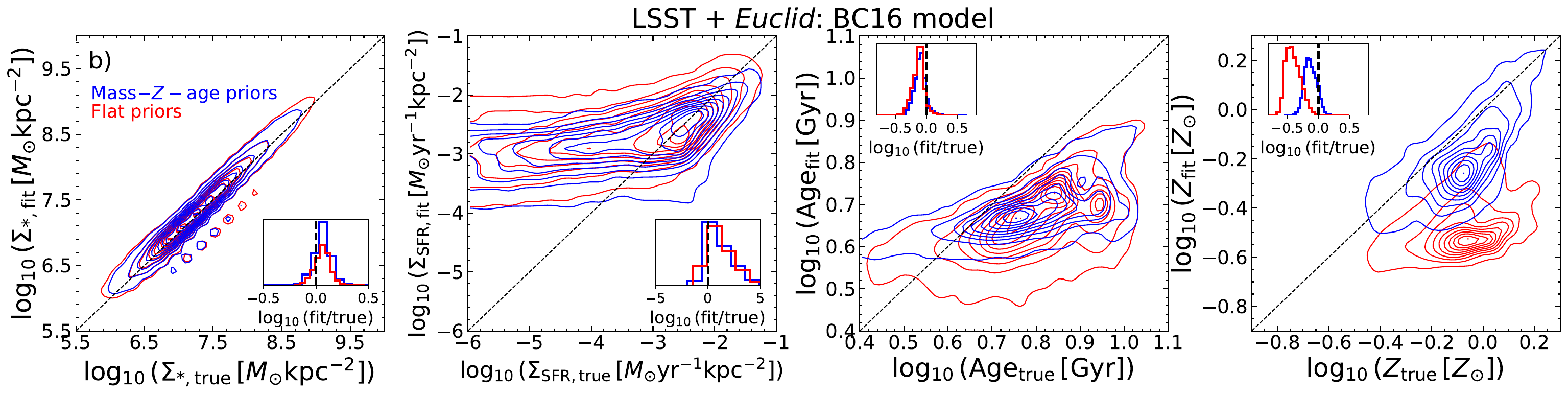}
\caption{Similar to Fig.~\ref{fig:comp_simfits_GLE} but for results of SED fitting on the \ledc\ data cubes.}
\label{fig:comp_simfits_LE}
\end{figure*}

\begin{figure*}[h!]
\centering
\includegraphics[width=1.0\textwidth]{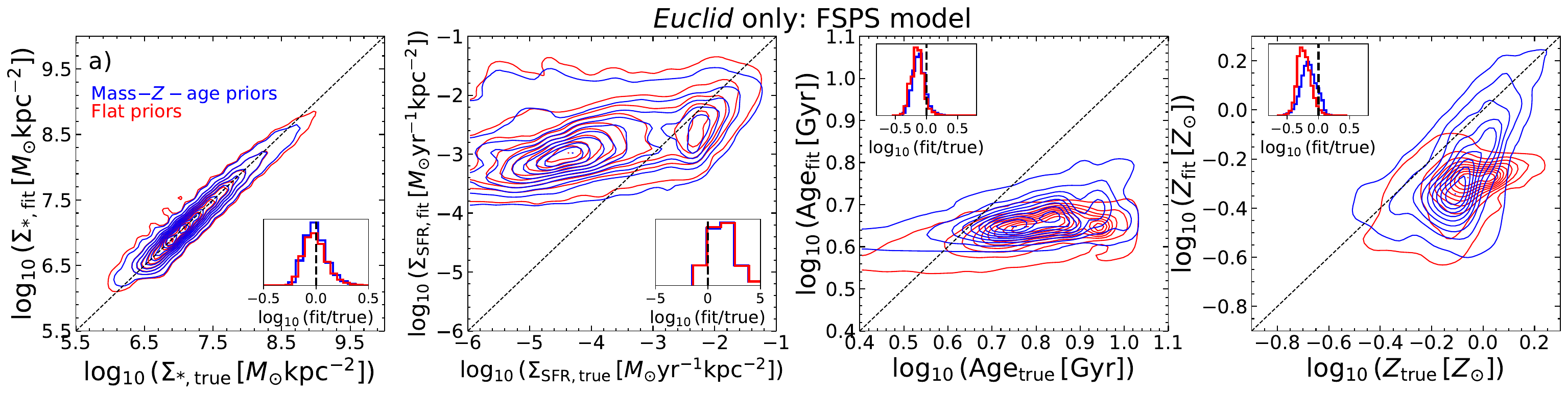}
\includegraphics[width=1.0\textwidth]{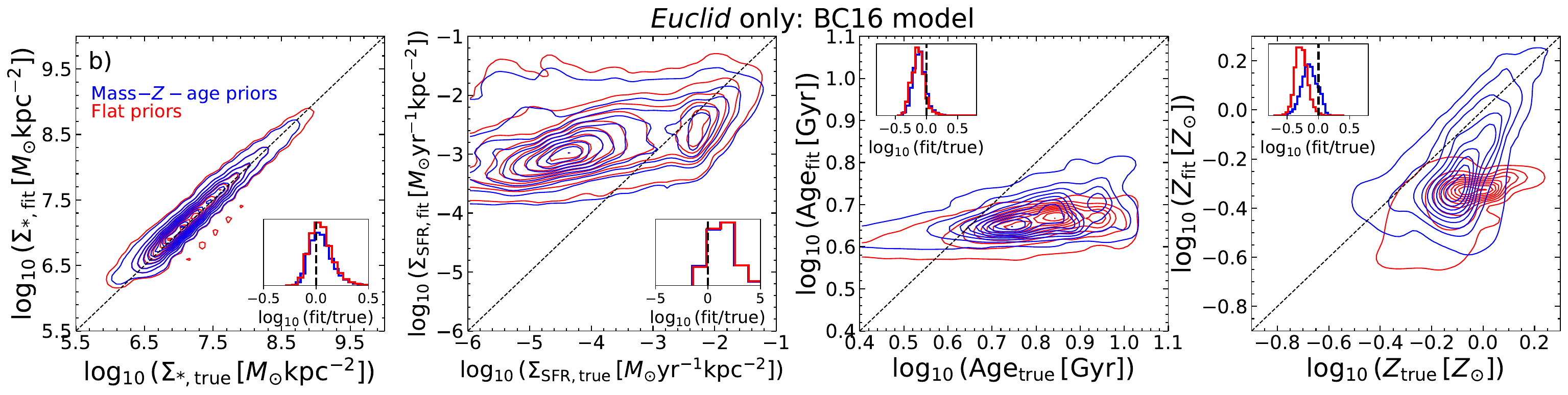}
\caption{Similar to Fig.~\ref{fig:comp_simfits_GLE} but for results of SED fitting on the E data cube.}
\label{fig:comp_simfits_E}
\end{figure*}

In Figs.~\ref{fig:comp_simfits_GLE},~\ref{fig:comp_simfits_LE}, and~\ref{fig:comp_simfits_E}, we show the comparisons between our fitting results and the ground truth from the analyses with the three imaging data cubes. The contours show the distributions of all spatial bins in the sample galaxies. We show the results obtained with the FSPS and BC16 models, with and without the mass-$Z$-age prior. To make quantitative comparisons, we show the histograms of the logarithmic ratio between the best-fit parameters and the ground truth in the figures and calculate the mean offset ($\mu$), scatter (i.e.~standard deviation; $\sigma$), and the Spearman rank-order correlation coefficient ($\rho$). $\rho$ is a nonparametric measure of the monotonicity of a relationship between two datasets and can indicate the significance of a correlation. The derived parameters are summarized in Table~\ref{tab:robustness_fits_with_priors} for the fitting that applies the mass-$Z$-age prior and Table~\ref{tab:robustness_fits_no_priors} for the one that does not apply the special prior.  

\begin{table*}[h!]
\caption{The reliability of stellar population properties derived from SED fitting without applying the mass-$Z$-age prior \label{tab:robustness_fits_no_priors}}
\centering
\begin{tabular}{lcccccc}  
\hline \hline \\ [-5pt]
 & \multicolumn{6}{c}{Data cube} \\ 
\cline{2-7} \\ [-5pt]
 & \multicolumn{2}{c}{\galex\ + \lsst\ + \euclid} & \multicolumn{2}{c}{\lsst\ + \euclid} & \multicolumn{2}{c}{\euclid-only} \\ 
\cline{2-7} \\ [-5pt]
 & BC16 & FSPS & BC16 & FSPS & BC16 & FSPS \\
\hline \\ [-5pt]
$\Sigma_{*}$ & & & & & & \\
\hline \\ [-5pt]
$\mu$ & 0.05 & 0.02 & 0.06 & 0.02 & 0.05 & 0.01 \\ 
$\sigma$ & 0.10 & 0.10 & 0.08 & 0.08 & 0.11 & 0.11 \\
$\rho$ & 0.98 & 0.98 & 0.99 & 0.99 & 0.97 & 0.97 \\
\hline \\ [-5pt]
$\Sigma_{\rm SFR}$ & & & & & & \\
\hline \\ [-5pt]
$\mu$\textsuperscript{a} & 0.52/$-$0.01 & 0.35/$-$0.16 & 2.02/0.02 & 1.95/$-$0.03 & 2.10/$-0.08$ & 2.08/$-$0.09 \\
$\sigma$\textsuperscript{b} & 0.85/0.24 & 0.84/0.23 & 3.60/0.34 & 3.64/0.31 & 3.39/0.43 & 3.39/0.42 \\
$\rho$\textsuperscript{c} & 0.87/0.69 & 0.90/0.71 & 0.57/0.53 & 0.64/0.57 & 0.48/0.39 & 0.49/0.40 \\
\hline \\ [-5pt]
Age & & & & & & \\
\hline \\ [-5pt]
$\mu$ & $-$0.13 & $-$0.12 & $-$0.13 & $-$0.13 & $-$0.13 & $-$0.15 \\
$\sigma$ & 0.11 & 0.11 & 0.10 & 0.10 & 0.10 & 0.10 \\
$\rho$ & 0.36 & 0.35 & 0.58 & 0.60 & 0.49 & 0.50 \\
\hline \\ [-5pt]
$Z$ & & & & & & \\
\hline \\ [-5pt]
$\mu$ & $-$0.51 & $-$0.45 & $-$0.40 & $-$0.36 & $-$0.27 & $-$0.23 \\
$\sigma$ & 0.15 & 0.14 & 0.12 & 0.11 & 0.10 & 0.11 \\
$\rho$ & 0.17 & 0.14 & 0.34 & 0.33 & 0.51 & 0.45 \\
\hline
\end{tabular}
\tablefoot{\textsuperscript{a} Systematic offset defined as the mean of the logarithmic ratios between the parameter derived from SED fitting and the true value: $\log_{10}({\rm fit}/{\rm true})$. \textsuperscript{b} The standard deviation of $\log_{10}({\rm fit}/{\rm true})$. \textsuperscript{c} The Spearman rank-order correlation coefficient. For $\Sigma_{\rm SFR}$, the values to the right of the slash symbol are calculated after applying a threshold of $\Sigma_{\rm SFR}>10^{-3.0}\,M_{\odot}\,\rm{yr}^{-1}\,\rm{kpc}^{-2}$.}
\end{table*}

Not too surprisingly, from Figs.~\ref{fig:comp_simfits_GLE}, ~\ref{fig:comp_simfits_LE}, and ~\ref{fig:comp_simfits_E}, we can see that \massd\ is recovered very well in all the analyses with the three types of data cubes, corroborated by small offset ($\mu \lesssim 0.06$ dex), small scatter ($\sigma \lesssim 0.12$ dex), and high Spearman rank-order correlation coefficient ($\rho \gtrsim 0.97$). This is the case for both SPS models and whether or not the mass-$Z$-age prior is applied. This result suggests that stellar mass can be well recovered on spatially resolved scales using photometry data that cover rest-frame NIR. This is encouraging and shows the great potential of \euclid\ for mapping \massd\ of local galaxies.    

Star formation rate density can be recovered reasonably well with the \gledc\ data cube but poorly with the \ledc\ and E data cubes. Notably, most data points cluster along the one-to-one line for \ledc\ data cube. A critical threshold emerges at an \sfrd\ of approximately $10^{-3}\,M_{\odot}\,\rm{yr}^{-1}\,\rm{kpc}^{-2}$, where SFR recovery becomes markedly differentiated--performing well above this density and poorly below it, particularly for \ledc\ and E data cubes. This differentiation is evidenced by the mean ($\mu$) and standard deviation ($\sigma$) of the logarithmic ratio between true and inferred $\Sigma_{\rm SFR}$ as shown in Tables~\ref{tab:robustness_fits_with_priors} and~\ref{tab:robustness_fits_no_priors}, with values for spatial bins above $10^{-3}\,M_{\odot}\,\rm{yr}^{-1}\,\rm{kpc}^{-2}$ noted after slash symbol.

These results show that our pipeline can recover spatially resolved SFR well with \gledc\ data irrespective of how active star formation in the spatial region is, while with \ledc\ and E data cube, it can recover SFR well only in star-forming regions and fails in passive regions. The critical SFR density of $10^{-3}\,M_{\odot}\,\rm{yr}^{-1}\,\rm{kpc}^{-2}$ is relatively low, even by local galaxy standards, a threshold typically exceeded in spiral arm regions. The less accurate \sfrd\ estimates derived from \ledc\ and E data cubes are expected, primarily due to their lack of UV coverage. UV data are crucial as indicators of SFR and dust attenuation. Although \ledc\ data offers a marginal improvement by capturing the $4000\,\AA$ break, a valuable age indicator, \euclid-only data provides the most limited constraints on SFR, dust properties, and stellar age.

Without the mass-$Z$-age prior, the inferred metallicities and ages are systematically underestimated even in the \gledc\ data cube. 
The biases may have their origin in the traditional flat priors (combined with parameter-estimates based on marginalized posterior distributions rather than best-fits), since a flat distribution in $\rm log_{10}(Z)$ contains more low-Z than high-Z models, and a flat distribution of $\hat{\tau}_{2}$ contains more low values of $\rm exp(\rm -\hat{\tau}_{2})$ than high values. It is also worth recalling the impact of the non-flat distribution of the mass-weighted ages. 

Once the prior is applied, age is recovered fairly well with the \gledc\ and \ledc\ data cubes, but barely recovered with the E data cube. The trends obtained with the two SPS models are similar regardless of the data cubes being used, which suggests a weak effect of the SPS model variation. In the analysis with the \gledc\ data, the results obtained by applying the special prior have a smaller systematic offset ($\mu \sim -0.07$) as compared to those obtained without applying the priors ($\mu \sim -0.13$). Interestingly, the fitting results without applying the priors do not deviate significantly from the one-to-one line, as is the case for $Z$. The systematic offset becomes slightly larger in the analysis with the \ledc\ data. The deviation from the true value is more significant for the older stellar populations than the younger ones. We also note the similarity between the results obtained for age with and without priors in the analysis with the \ledc\ data. This might indicate that the special priors cannot improve the estimate of age further for this dataset that only covers the rest-frame optical-to-NIR, and missing the UV data, which is an important constraint for age. As we can see from Fig.~\ref{fig:comp_simfits_E}, the E data cube struggles to recover age, as shown by the broadly flat distribution of the fitting results across a wide range of the true values, flatter than the results with \ledc\ data (Fig.~\ref{fig:comp_simfits_LE}). This is understandable because of the absence of UV and optical data. Although it has \ie\ filter, its wavelength range does not cover the Balmer break, which is an important age indicator.  

For metallicity, the best ground truth recovery is achieved using the \gledc\ data cube and applying the mass-$Z$-age prior to the fitting process. The variation in the SPS model does not seem to affect the results, unlike the priors that play a big role. With the help of the mass-$Z$-age prior, the result is encouraging, with a small absolute offset of $\lesssim 0.05$ dex, a scatter of around $0.08$ dex, and $\rho$ greater than $0.6$. Removing \galex\ and \lsst\ data, but still applying the prior, results in a slightly larger offset than that achieved with the \gledc\ data cube. The \ledc\ gives $\mu \sim -0.12$, $\sigma \sim 0.11$, and $\rho \sim 0.6$ with the two SPS models, while the E data cube gives slightly larger offset and scatter but similar $\rho$. These results show the significant influence of priors in the SED fitting with the Bayesian method. 

In summary, it is important to emphasize the joint role of priors and data. For example, the comparison of offsets in Tables~\ref{tab:robustness_fits_with_priors} and \ref{tab:robustness_fits_no_priors} demonstrates that the priors are essential for accurately recovering age and metallicity. However, incorporating LSST data and, subsequently, GALEX data also plays a significant role. Specifically, for the BC16 case, applying the special priors improves the age offsets from $-0.13$ to $-0.11$, $-0.13$ to $-0.11$, and $-0.13$ to $-0.06$ for the E, \ledc, and \gledc\ data cubes, respectively. The most significant improvement is naturally achieved when all datasets are used. When only E data are used, the inferred age and metallicity are mainly driven by the priors.

To get a better look at the influences of priors on the fitting results, we plot the distributions of derived age and metallicities on the mass-$Z$-age prior in Fig.~\ref{fig:priors_vs_data}. Overall, we can conclude that the influence of priors becomes more prominent when fewer data points (covering a narrower wavelength range) are used, as expected.   

When using only the E data with flat priors, the estimated ages exhibit a very narrow distribution centred around $\log_{10}(\rm Age_{\rm true}\rm [Gyr]) \sim 0.6-0.7$. This corresponds to the peak of the mass-weighted age distribution of the SPS model libraries used in the SED fitting procedure. This result confirms that with limited data and without imposing the mass-Z-age prior, the `hidden' mass-weighted age prior can bias the fitting results.

\subsection{Offset maps: Locating the good and poor fitting results}
\label{sec:offset_maps}

Next, we analyse the maps of spatially resolved offsets between the fitting results and the true values to investigate the spatial regions where good and poor fitting results are located and try to get more insights into the contributing factors to the discrepancies. We show offset maps of a spiral galaxy TNG501725, obtained from the analyses with the \gledc, \ledc, and E data cubes. We only show results from SED fitting with the FSPS model in this example because the BC16 model produces offset maps that are similar overall. The mass-$Z$-age prior is applied.

The offset maps show that our SED fitting slightly overestimates \massd\ in the spiral arm regions of TNG501725, though the median offset is small. This is also the case for other spiral galaxies in our sample.  
In contrast to \massd, the offset map of age is roughly flat across the galaxy's region in all data types. Both good estimates of age (i.e.~nearly zero offset) and overestimation can be seen in the spiral arms. For metallicity, the \gledc\ data give good estimates in almost the whole region, except the outer shell where $Z$ is overestimated. However, the offset maps are roughly flat for the results with \ledc\ and E data cubes, showing an overall underestimation of $Z$. In contrast to \massd, SFR is underestimated in the spiral arm regions and overestimated in the interarm regions. The colour map shows that the absolute offset is larger for SFR than other properties, exceeding $0.4$ dex (the range set for consistency with other maps) for results using \ledc\ and E data cubes.

\begin{figure*}[h!]
\centering
\includegraphics[width=0.8\textwidth]{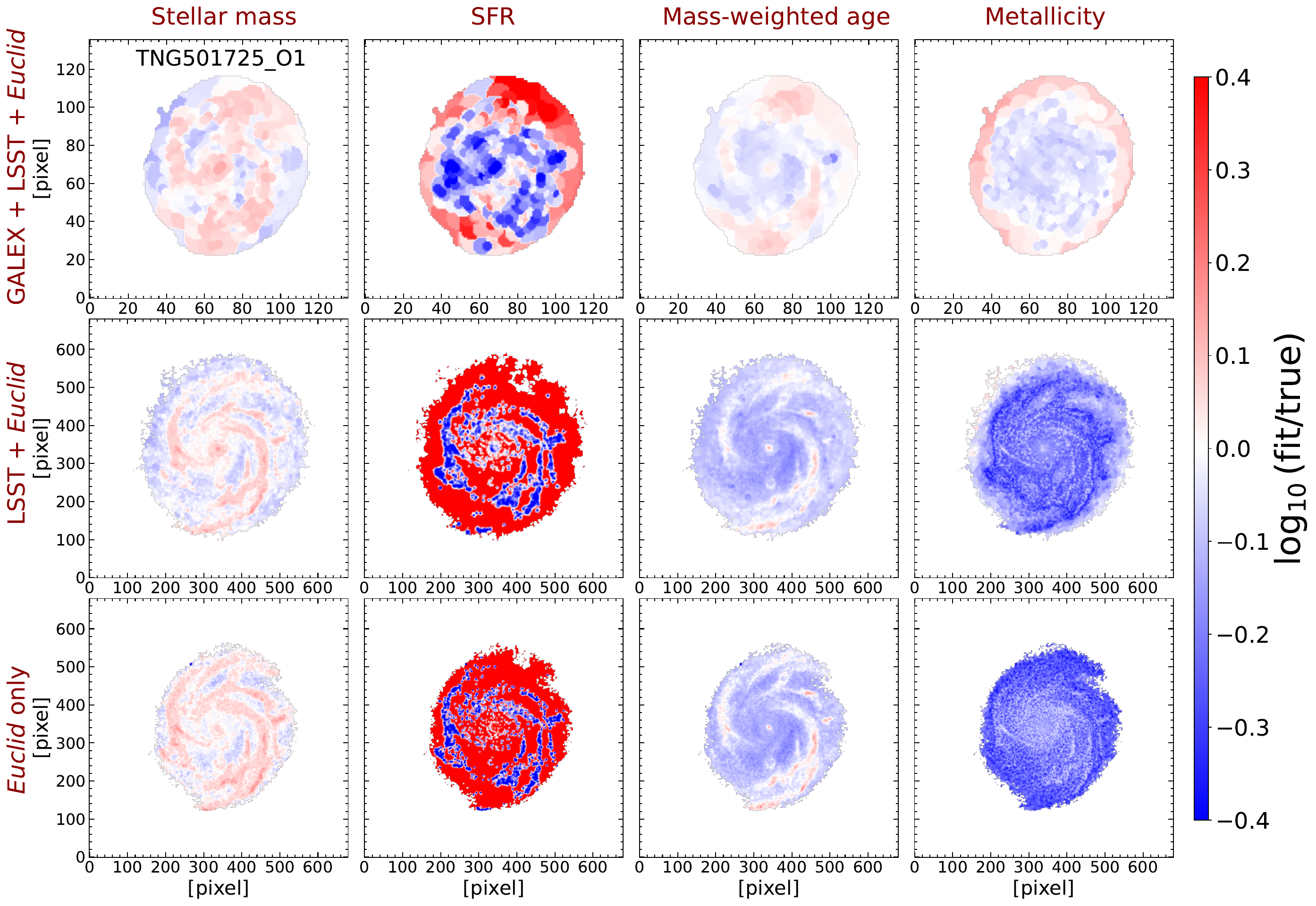}
\caption{Examples of spatially resolved distributions of the systematic offsets (in logarithmic scale) between the parameters derived from SED fitting and the true values. Offset maps of a galaxy TNG501725, are shown here for example. 
The first, second, and third rows are for analyses with \gledc, \ledc, and E data cubes, respectively. Only results obtained with the FSPS model and applying the mass-$Z$-age prior are shown.}
\label{fig:offset_map1}
\end{figure*}

\subsection{Recovering radial profiles of stellar population properties}
\label{sec:recov_radial_profile}

In galaxies, dynamical timescales, gas properties relevant to star formation, the history of stellar feedback, and the influence of neighbouring galaxies, all depend on radius. The radial profiles of stellar population properties are often used for describing the local distribution of galaxy properties because their variations with distance from the galactic centre can inform us about the growth history of the galaxies. For instance, the negative gradient of the age radial profile and the positive gradient of specific SFR is thought to be the consequence of the growth history of the galaxy structure that happens in an inside-out manner \citep[e.g.][]{Abdurrouf2022a,Abdurrouf2023}. Here we check the robustness of the radial profiles of $\Sigma_{*}$, age, and $Z$. For simplicity, we only analyse the radial profiles from the fitting results with the FSPS model and apply the mass-$Z$-age prior because this fitting set-up gives the best recovery of the ground truth as shown in Sect.~\ref{sec:recov_spatial_resolved_SP}. 

\begin{figure*}[h!]
\centering
\includegraphics[width=0.9\textwidth]{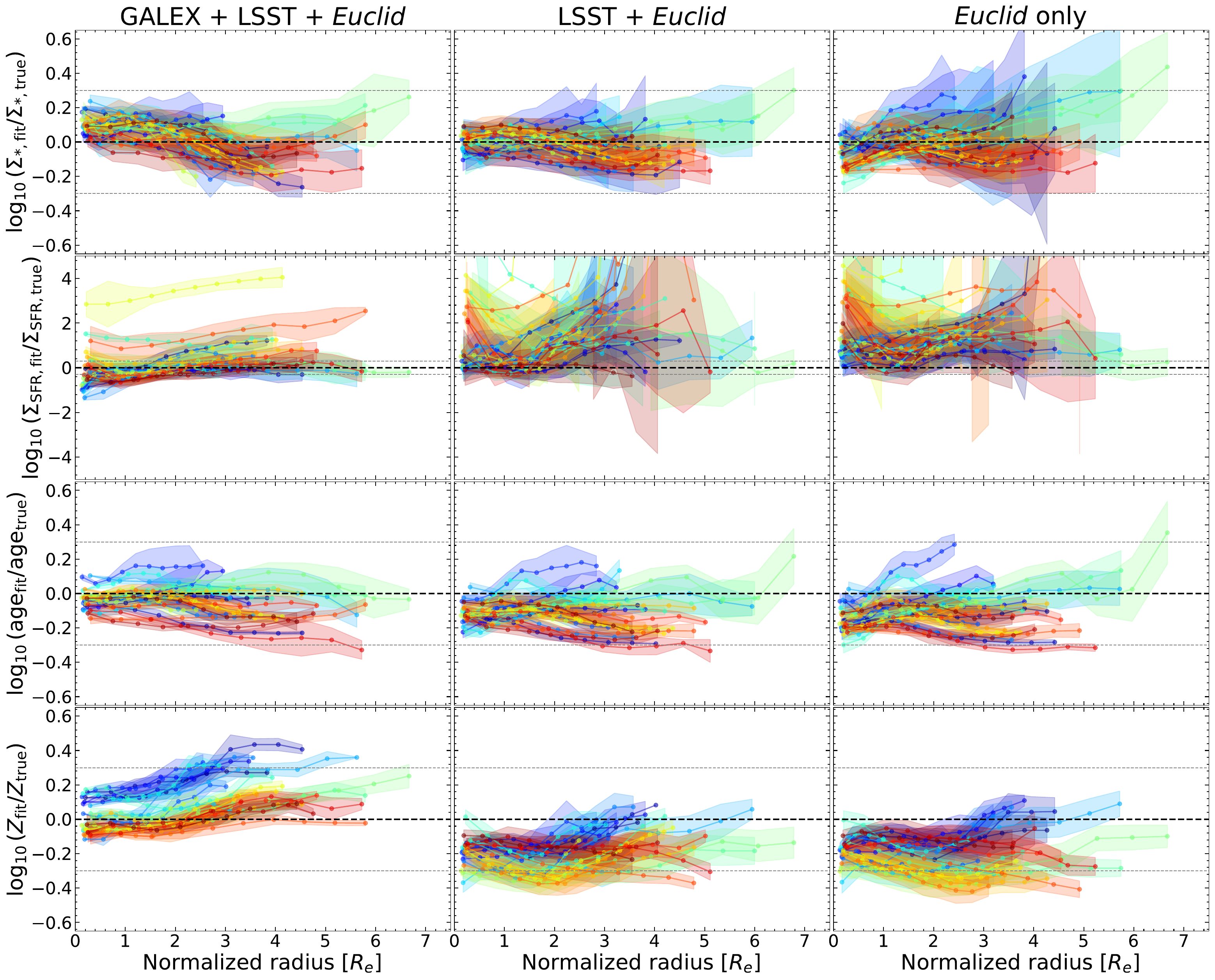}
\caption{The Systematic offsets between estimated and input properties ($\mu$) as a function of the radial distance from the galactic centre. The radial profiles of the offsets, as obtained from fits with the FSPS models and with the mass-$Z$-age prior, are shown for all sample galaxies with three types of data cubes. Columns from left to right show results from the analysis with the \gledc, \ledc, and E data cubes. Each galaxy's radial extent is divided into ten radial bins equally spaced and radial profiles are then derived by taking the median of $\mu$ in circular annuli. The radius is normalized to the half-mass radius that is taken from the catalogue of the TNG50-SKIRT Atlas. The two horizontal dashed grey lines mark $\pm 0.3$ dex offset.}
\label{fig:rp_residuals_combine}
\end{figure*}

The radial profiles of the systematic offset of all galaxies in our sample are shown in Fig.~\ref{fig:rp_residuals_combine}.  
As we can see from the figure, the absolute offset varies with radius in the majority of the galaxies in our sample. The offset is generally smaller in the central region than in the outskirts, which is reasonable given the overall higher S/N in the central region. The $\Sigma_{*}$ offset is centred around zero and has a scatter that increases with radius. However, the overall absolute offset is smaller than $0.3$ dex across the radial extent. No galaxy exhibits $\Sigma_{*}$ values that systematically deviate from the true values over the entire galaxy's radial extent (see, for example, Fig.~\ref{fig:offset_map1}). In contrast, some galaxies show consistent deviations from the ground truth in age and $Z$ throughout their entire regions.
In those galaxies, the offset radial profile tends to be flat (i.e.~constant across the galaxy's region). Despite the full deviation, the overall offset is mostly within $\pm 0.3$ dex.  

\subsection{Recovering the integrated stellar mass}
\label{sec:recover_integrated_properties}

Previous studies have found some indications that global \mass\ derived from fitting to the integrated SED of galaxies could be underestimated as compared to the integrated \mass\ derived from spatially resolved SED fitting \citep[e.g.][]{2009Zibetti,2015Sorba,2018Sorba}. The underestimation of \mass\ is thought to be due to the domination of light from young stellar populations that make up the integrated SED of a galaxy. This outshining effect can bias the fitting of integrated SED towards inferring younger and less massive stellar population properties, especially if the assumed parametric SFH model is too simple and not flexible enough to capture the star-formation activity that occurred farther back in time. This effect becomes more significant at high redshifts,  as young stars mostly dominate the stellar populations in galaxies. Spatially resolved SED fitting can alleviate this effect because it can directly disentangle young and old stellar populations in the process. Here, we compare integrated \mass\ derived from the two methods (i.e.~spatially resolved and integrated SED fitting) and the true values from TNG50. The integrated SEDs are obtained by summing up fluxes of all pixels within the galaxy's RoI, then SED fitting is performed on them. The set-up for SED modelling and all the assumed priors in the fitting process are the same as those used in the spatially resolved SED fitting (see Sect.~\ref{sec:sed_fitting} and Table~\ref{tab:sedfits_params}). For the true \mass, we adopt the values from the TNG50-SKIRT catalogue, which are the total mass of stellar particles in the subhalo of the galaxies.      

\begin{figure*}[h!]
\centering
\includegraphics[width=1.0\textwidth]{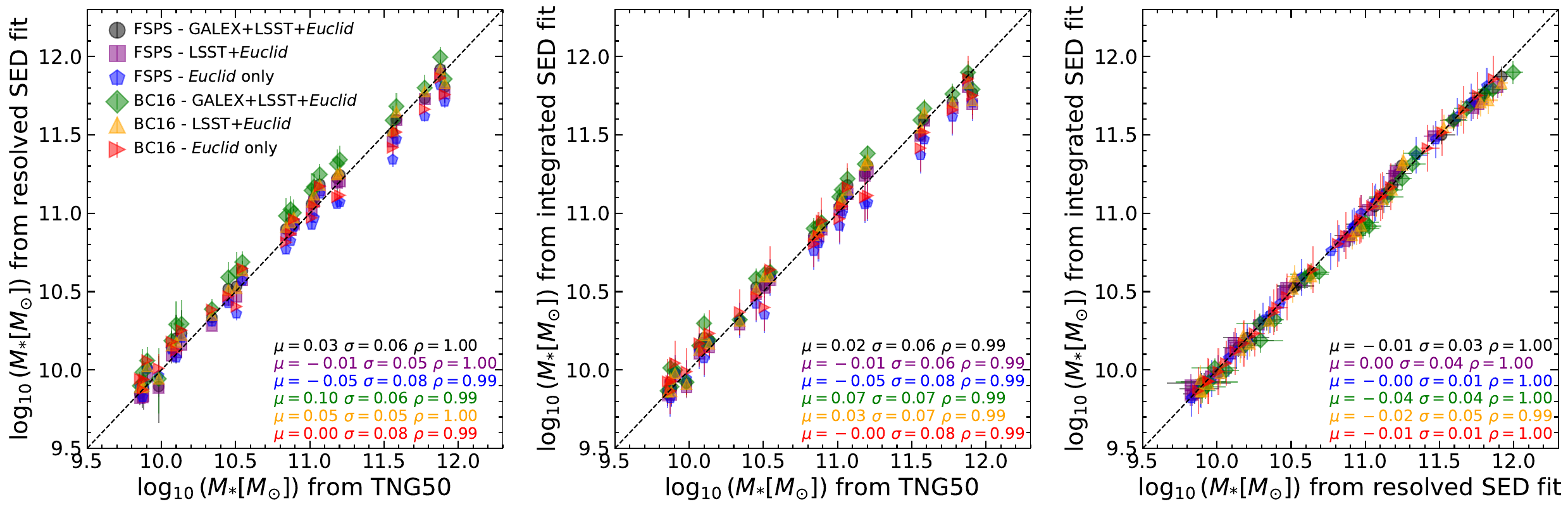}
\caption{The Robustness of integrated \mass\ derived from the spatially resolved SED fitting and the integrated SED fitting compared to the true \mass\ from the TNG50 simulation. \emph{Left panel}: Comparison between \mass\ derived from spatially resolved SED fitting (i.e.~by summing up \mass\ of pixels within the ROI of the galaxies) and the true values. \emph{Middle panel}: Comparison between \mass\ derived from integrated SED fitting and the true values. \emph{Right panel}: Comparison between integrated \mass\ from spatially resolved and integrated SED fitting.}
\label{fig:comp_integrated_mass_fitting}
\end{figure*}

We show the comparisons between the two estimates of the global stellar masses and the ground truth in Fig.~\ref{fig:comp_integrated_mass_fitting}. For \mass\ derived from SED fitting, we show the six variations in the fitting, consisting of two SPS models and three sets of imaging data cubes. These variations are performed to both the integrated and spatially resolved SED fitting. From this figure, we can see that overall, integrated \mass\ is well recovered by all the methods, which are shown by the very small systematic offset and scatter in the one-to-one relations. We also compare the integrated \mass\ derived from the integrated and spatially resolved SED fitting. Considering the $\mu$ values, stellar masses from integrated SED fitting are only imperceptibly lower than those from resolved SED fitting. Within the uncertainties, there are no noticeable differences between the two estimates, which suggests that our integrated SED fitting is not affected by the outshining effects of young stellar populations in the galaxies, in contrast to the results of previous studies \citep[e.g.][]{2009Zibetti,2015Sorba,2018Sorba}. Other previous studies have also found consistent stellar mass derived from spatially resolved SED fitting and integrated one \citep[e.g.][]{2012Wuyts,2024Bellstedt}, consistent with our finding. This can be caused in part by two factors. First, our sample galaxies are forming stars only moderately, and their stellar populations are not dominated by young stars. Therefore, the outshining effect is expected to be minor. The outshining effect is expected to be more significant in galaxies that are actively forming stars, which are more prevalent in high-redshift samples \citep[e.g.][]{2018Sorba}. Second, our SFH model in the form of the double power-law function and the assumed priors for the associated parameters are flexible enough to capture the SFH of the galaxies, especially those that occurred farther back in time, which formed the old stars. \citet{Abdurrouf2021} has shown that this SFH model is flexible enough to reconstruct the true SFH of TNG100 galaxies in a fitting test with mock SEDs.       

\section{Discussion}
\label{sec:discussion}

In our tests, we have made variations in some of the main and basic factors involved in SED fitting, including the SPS model, the wavelength coverage of the photometric data, and priors. In the following, we further discuss the influences of these three factors on the fitting results. We also put our results in the context of findings from previous studies in the literature.   

\subsection{The influence of SPS model}

The variation in the SPS model is made to accommodate the SPS model used in generating the synthetic images by the TNG50-SKIRT Atlas, which is BC16 (see Sect.~\ref{sec:gen_noisefree_images}), while \pixedfit\ uses FSPS by default. We use the two models in our experiment for two reasons: (a) by applying the same BC16 model, the systematic effects caused by the difference in the SPS model can be eliminated; and (b) we get the opportunity to compare the results obtained using the two SPS models. As we can see from Figs.~\ref{fig:comp_simfits_GLE}, ~\ref{fig:comp_simfits_LE}, and ~\ref{fig:comp_simfits_E} (Sect.~\ref{sec:recov_spatial_resolved_SP}), there is no significant difference between the results obtained with the FSPS and BC16 models, whether or not the mass-$Z$-age prior is applied. The influence of the SPS model is weak on the derived \mass, $Z$, and age in our SED fitting tests. Therefore, using two of the most commonly adopted SPS models in the literature, we demonstrate that the uncertainty in our description of stellar populations in nature has a negligible effect on our SED fitting process.  

\subsection{The influence of wavelength coverage of the imaging data}

\euclid\ covers the NIR and red optical wavelengths, but it is well known that data at shorter wavelengths, preferably covering the Balmer break and the UV, hold sensitive complementary information on stellar populations. This is particularly important when dealing with spatially resolved data where $\rm{S}/\rm{N}$ is typically relatively low. The combination of wavelengths is helpful to break degeneracies among age, $Z$, and dust attenuation \citep[e.g.][]{Worthey1994, Kodama_Arimoto1997, Conroy2013}, although more passbands are not necessarily sufficient for this purpose \citep[e.g.][]{Pforr2012}. Therefore, it is essential to test to what extent \euclid\ imaging data alone can provide constraints on the main stellar population properties and what improvements can be gained by adding more imaging data in the optical and UV. 

From Figs.~\ref{fig:comp_simfits_GLE}, ~\ref{fig:comp_simfits_LE}, and ~\ref{fig:comp_simfits_E}, we can see that \mass\ measurement is very stable against the variation in the wavelength coverage of photometric data. \mass\ is recovered very well using any of the three data cubes as \euclid\ covers the NIR wavelength. On the other hand, age and $Z$ are affected more significantly by the presence or absence of short wavelength data. Age is poorly constrained when using the E data cube with the classical `flat' priors: the estimated ages cluster tightly around $4-5$ Gyr, corresponding to the peak of the mass-weighted age distribution of the SPS models for these priors. However, even with the implementation of informative priors (i.e.~mass-$Z$-age), ages remain more poorly constrained compared to metallicities.  
In the fitting that implements the informative priors, the systematic offset (i.e.~underestimation) in $Z$ measurement is enhanced with the exclusion of \galex\ ($\mu \sim 0.11$ dex) and \galex+\lsst\ ($\mu \sim 0.15$ dex), while the strong proportionality between the inferred and true values is stable against the data variation.              

\subsection{The influence of the mass-$Z$-age prior}

The prior is one of the important components in the Bayesian SED fitting, providing additional information and constraints in the fitting process. In the absence of highly constraining data (e.g.~spectroscopy or multi-wavelength photometry over a wide wavelength range), the role of priors can be significant in getting reliable fitting results. There are multiple forms of priors in SED fitting, and they can be assumed explicitly (e.g.~ranges in the grid of free parameters) or implicitly (e.g.~dust attenuation law, and the form of analytic SFH). There have been only a few studies that adopted empirical scaling relations for a prior in the Bayesian SED fitting. \citet{Leja2019} implemented a prior in the form of mass-$Z$ relation in their integrated SED fitting of high-$z$ galaxies. They adopted the empirical mass-$Z$ relation observed by \citet{Gallazzi2005}. Similar scaling relations also exist among stellar population properties on kpc scales in the local galaxies \citep[see review by][]{Sanchez2020}. This provides an opportunity to use them for informative priors in spatially resolved SED fitting. 

Applying the classical flat priors has resulted in very poor constraints on stellar population parameters, underestimating metallicity and age, and overestimating dust attenuation.

The implementation of the mass-$Z$-age in our work has showcased, instead, its potential. As we can see from Figs.~\ref{fig:comp_simfits_GLE}, ~\ref{fig:comp_simfits_LE}, and ~\ref{fig:comp_simfits_E}, the role of the informative priors is significant in recovering age and $Z$. The most significant effect can be seen in the measurement of $Z$ with the \gledc\ data cube, where fitting with priors can significantly improve the fitting results.

To further test the performance of our SED fitting pipeline, we perform additional fitting tests using simple mock SEDs generated using FSPS without preprocessing, unlike the synthetic images that are generated using the radiative transfer process. We present the results in Appendix~\ref{sec:mock_sedfit_test}. Overall, we obtain trends similar to those seen in the tests using the synthetic images. \mass\ is recovered very well with the three types of data cubes and seems to be independent of priors, while $Z$ and age are recovered relatively well with the three data cubes only if the mass-$Z$-age prior is applied. 

\section{Conclusions and outlooks}
\label{sec:conclusion}

We tested a pipeline for spatially resolved SED fitting of local galaxies as part of \euclid\ science missions on the local Universe. We aimed to test to what extent \euclid-data only can be used to measure spatially resolved stellar populations of local galaxies, and what improvements can be obtained by including imaging data in the optical and UV from other telescopes. We used synthetic images representative of \euclid, \galex, and \lsst\ data based on the TNG50-SKIRT Atlas \citep{Baes2024_a,Baes2024_b}, which were generated using a 3D Monte Carlo radiative transfer process. 

Our pipeline uses functionalities provided by \pixedfit\ \citep{Abdurrouf2021,Abdurrouf2022b_pixedfitcode} for processing the imaging data (i.e.~matching the spatial resolution and sampling of the various passbands), binning neighbouring pixels to achieve chosen $\rm{S}/\rm{N}$ thresholds in multiple bands, and fitting the spatially resolved SEDs to infer the local properties of the stellar populations. In this work, we ran our pipeline on TNG50-SKIRT galaxies with modest levels of current star formation to test its robustness in recovering three main properties on spatially resolved scales: stellar mass surface density (\massd), age, and metallicity ($Z$). We constructed three types of data cubes: \galex\ + \lsst\ + \euclid, \lsst\ + \euclid, and \euclid-only (\gledc, \ledc, and E, respectively), and ran SED fitting on each of them with varying SPS models (FSPS and BC16) and priors. We implemented a new informative prior in the form of the mass-age and mass-$Z$ relations (i.e.~mass-$Z$-age prior), adopted from a combination of empirical and simulated data (see Appendix~\ref{sec:appendix_priors}). 

The mass surface density can be recovered very well using the three data cubes, regardless of the variations in the SPS model and priors (i.e.~whether or not the mass-$Z$-age prior is implemented). The systematic offset ($\mu$) and scatter ($\sigma$) of the logarithmic ratio between the inferred and true values are minimal, $\lesssim 0.05$ (absolute value) and $\sim 0.1$ dex, respectively. This suggests that the \euclid-only data are sufficient to robustly map \massd\ of mildly star-forming local galaxies.  
A slight overestimation of \massd\ seems to occur in the regions that are more actively forming stars or contain younger stellar populations than the rest of the galaxy's region, such as in the spiral arms. 

Not surprisingly, the added prior can improve the $Z$ and age retrieved from the simulated data. In the analysis with the \gledc\ data cube, $Z$ can be recovered reasonably well ($\mu \sim 0.03$ and $\sigma \sim 0.1$ dex) when the fitting implements these priors, while the inferred $Z$ is significantly offset by around $0.6$ dex when these priors are not implemented. In the analysis with the \ledc\ and E data cubes, fitting that implements the added priors results in an offset of around $-0.12$ dex, and $-0.15$ dex, respectively. Despite the offset, there is a good proportionality between the values inferred from SED fitting and the true ones. Without the priors, $Z$ is offset by around $0.5$ dex. For the age, the new priors can significantly improve the fitting result for the simulated \gledc\ data cube but less so for the other two data cubes. In the analysis with the E data cube, age is poorly recovered. This is mainly due to the lack of coverage in the optical and UV, which is essential for constraining the stellar population age.

In future analyses, we plan to further test our ability to recover stellar population parameters by exploring a wider wavelength range, including medium and far-infrared data. With these data, we will further assess our capacity to constrain dust attenuation and the extinction law. Additionally, we will investigate the importance of using single-metallicity models in the analysis of more complex populations, and if needed, allow for more flexibility in star-formation histories with non-parametric descriptions. Spectroscopic measurements will be incorporated where available and help assess the retrieved properties in real, rather than simulated, galaxies.

Using our spatially resolved SED fitting pipeline, we aim to map the stellar population properties of a large number (100\,000s) of galaxies in the local Universe using imaging data from \euclid, combined with optical images from other telescopes, especially \lsst\ and UNIONS surveys. The depth and high spatial resolution of the imaging data, combined with the wide-area coverage of the surveys, make it possible to perform our analysis for many galaxies that have a wide range of global properties and reside in various local density environments, from a low-density field to a dense galaxy cluster. We will produce comprehensive and detailed sets of stellar mass profiles and stellar population gradients across various galactic scales, from centres to peripheries. This will provide constraints on in-situ and ex-situ stellar populations, and the role of feedback processes and mergers as a function of galaxy mass (e.g. \citealt{Tortora+10CG, Tortora+11MtoLgrad}). \euclid\ is poised to uncover thousands of previously undiscovered dwarf low-surface brightness galaxies and the ultra-faint peripheral regions of large galaxies. Consequently, we will enhance and validate our pipeline designed for determining the stellar populations of these faint objects. This will allow us to comprehensively study the internal properties of galaxies and the influencing factors, both internally and externally, for a deeper understanding of galaxy evolution.          

\begin{acknowledgements}
\AckEC
Co-funded by the European Union. Views and opinions expressed are
however, those of the author(s) only and do not necessarily reflect those
of the European Union. Neither the European Union nor the granting
authority can be held responsible for them. CT acknowledges the INAF grant 2022 LEMON. JHK acknowledges grant PID2022-136505NB-I00 funded by MCIN/AEI/10.13039/501100011033 and EU, ERDF.
This project makes use of the MaNGA-Pipe3D data products. We thank the IA-UNAM MaNGA team for creating this catalogue, and the Conacyt Project CB-285080 for supporting them.
\end{acknowledgements}

\bibliographystyle{aa} 
\bibliography{aanda} 

\begin{thebibliography}{125}
\expandafter\ifx\csname natexlab\endcsname\relax\def\natexlab#1{#1}\fi

\bibitem[{{Abdurro'uf} {et~al.}(2022{\natexlab{a}}){Abdurro'uf}, {Accetta}, {Aerts}, {Silva Aguirre}, {Ahumada}, {Ajgaonkar}, {Filiz Ak}, {Alam}, {Allende Prieto}, {Almeida}, {Anders}, {Anderson}, {Andrews}, {Anguiano}, {Aquino-Ort{\'\i}z}, {Arag{\'o}n-Salamanca}, {Argudo-Fern{\'a}ndez}, {Ata}, {Aubert}, {Avila-Reese}, {Badenes}, {Barb{\'a}}, {Barger}, {Barrera-Ballesteros}, {Beaton}, {Beers}, {Belfiore}, {Bender}, {Bernardi}, {Bershady}, {Beutler}, {Bidin}, {Bird}, {Bizyaev}, {Blanc}, {Blanton}, {Boardman}, {Bolton}, {Boquien}, {Borissova}, {Bovy}, {Brandt}, {Brown}, {Brownstein}, {Brusa}, {Buchner}, {Bundy}, {Burchett}, {Bureau}, {Burgasser}, {Cabang}, {Campbell}, {Cappellari}, {Carlberg}, {Wanderley}, {Carrera}, {Cash}, {Chen}, {Chen}, {Cherinka}, {Chiappini}, {Choi}, {Chojnowski}, {Chung}, {Clerc}, {Cohen}, {Comerford}, {Comparat}, {da Costa}, {Covey}, {Crane}, {Cruz-Gonzalez}, {Culhane}, {Cunha}, {Dai}, {Damke}, {Darling}, {Davidson}, {Davies}, {Dawson}, {De Lee}, {Diamond-Stanic}, {Cano-D{\'\i}az},
  {S{\'a}nchez}, {Donor}, {Duckworth}, {Dwelly}, {Eisenstein}, {Elsworth}, {Emsellem}, {Eracleous}, {Escoffier}, {Fan}, {Farr}, {Feng}, {Fern{\'a}ndez-Trincado}, {Feuillet}, {Filipp}, {Fillingham}, {Frinchaboy}, {Fromenteau}, {Galbany}, {Garc{\'\i}a}, {Garc{\'\i}a-Hern{\'a}ndez}, {Ge}, {Geisler}, {Gelfand}, {G{\'e}ron}, {Gibson}, {Goddy}, {Godoy-Rivera}, {Grabowski}, {Green}, {Greener}, {Grier}, {Griffith}, {Guo}, {Guy}, {Hadjara}, {Harding}, {Hasselquist}, {Hayes}, {Hearty}, {Hern{\'a}ndez}, {Hill}, {Hogg}, {Holtzman}, {Horta}, {Hsieh}, {Hsu}, {Hsu}, {Huber}, {Huertas-Company}, {Hutchinson}, {Hwang}, {Ibarra-Medel}, {Chitham}, {Ilha}, {Imig}, {Jaekle}, {Jayasinghe}, {Ji}, {Johnson}, {Jones}, {J{\"o}nsson}, {Katkov}, {Khalatyan}, {Kinemuchi}, {Kisku}, {Knapen}, {Kneib}, {Kollmeier}, {Kong}, {Kounkel}, {Kreckel}, {Krishnarao}, {Lacerna}, {Lane}, {Langgin}, {Lavender}, {Law}, {Lazarz}, {Leung}, {Leung}, {Lewis}, {Li}, {Li}, {Lian}, {Liang}, {Lin}, {Lin}, {Lin}, {Lintott}, {Long}, {Longa-Pe{\~n}a},
  {L{\'o}pez-Cob{\'a}}, {Lu}, {Lundgren}, {Luo}, {Mackereth}, {de la Macorra}, {Mahadevan}, {Majewski}, {Manchado}, {Mandeville}, {Maraston}, {Margalef-Bentabol}, {Masseron}, {Masters}, {Mathur}, {McDermid}, {Mckay}, {Merloni}, {Merrifield}, {Meszaros}, {Miglio}, {Di Mille}, {Minniti}, {Minsley}, {Monachesi}, {Moon}, {Mosser}, {Mulchaey}, {Muna}, {Mu{\~n}oz}, {Myers}, {Myers}, {Nadathur}, {Nair}, {Nandra}, {Neumann}, {Newman}, {Nidever}, {Nikakhtar}, {Nitschelm}, {O'Connell}, {Garma-Oehmichen}, {Luan Souza de Oliveira}, {Olney}, {Oravetz}, {Ortigoza-Urdaneta}, {Osorio}, {Otter}, {Pace}, {Padilla}, {Pan}, {Pan}, {Parikh}, {Parker}, {Peirani}, {Pe{\~n}a Ram{\'\i}rez}, {Penny}, {Percival}, {Perez-Fournon}, {Pinsonneault}, {Poidevin}, {Poovelil}, {Price-Whelan}, {B{\'a}rbara de Andrade Queiroz}, {Raddick}, {Ray}, {Rembold}, {Riddle}, {Riffel}, {Riffel}, {Rix}, {Robin}, {Rodr{\'\i}guez-Puebla}, {Roman-Lopes}, {Rom{\'a}n-Z{\'u}{\~n}iga}, {Rose}, {Ross}, {Rossi}, {Rubin}, {Salvato}, {S{\'a}nchez},
  {S{\'a}nchez-Gallego}, {Sanderson}, {Santana Rojas}, {Sarceno}, {Sarmiento}, {Sayres}, {Sazonova}, {Schaefer}, {Schiavon}, {Schlegel}, {Schneider}, {Schultheis}, {Schwope}, {Serenelli}, {Serna}, {Shao}, {Shapiro}, {Sharma}, {Shen}, {Shetrone}, {Shu}, {Simon}, {Skrutskie}, {Smethurst}, {Smith}, {Sobeck}, {Spoo}, {Sprague}, {Stark}, {Stassun}, {Steinmetz}, {Stello}, {Stone-Martinez}, {Storchi-Bergmann}, {Stringfellow}, {Stutz}, {Su}, {Taghizadeh-Popp}, {Talbot}, {Tayar}, {Telles}, {Teske}, {Thakar}, {Theissen}, {Tkachenko}, {Thomas}, {Tojeiro}, {Hernandez Toledo}, {Troup}, {Trump}, {Trussler}, {Turner}, {Tuttle}, {Unda-Sanzana}, {V{\'a}zquez-Mata}, {Valentini}, {Valenzuela}, {Vargas-Gonz{\'a}lez}, {Vargas-Maga{\~n}a}, {Alfaro}, {Villanova}, {Vincenzo}, {Wake}, {Warfield}, {Washington}, {Weaver}, {Weijmans}, {Weinberg}, {Weiss}, {Westfall}, {Wild}, {Wilde}, {Wilson}, {Wilson}, {Wilson}, {Wolf}, {Wood-Vasey}, {Yan}, {Zamora}, {Zasowski}, {Zhang}, {Zhao}, {Zheng}, {Zheng}, \& {Zhu}}]{Abdurrouf2022manga}
{Abdurro'uf}, {Accetta}, K., {Aerts}, C., {et~al.} 2022{\natexlab{a}}, \apjs, 259, 35

\bibitem[{{Abdurro'uf} \& {Akiyama}(2017)}]{Abdurrouf2017}
{Abdurro'uf} \& {Akiyama}, M. 2017, \mnras, 469, 2806

\bibitem[{{Abdurro'uf} \& {Akiyama}(2018)}]{Abdurrouf2018}
{Abdurro'uf} \& {Akiyama}, M. 2018, \mnras, 479, 5083

\bibitem[{{Abdurro'uf} {et~al.}(2023){Abdurro'uf}, {Coe}, {Jung}, {Ferguson}, {Brammer}, {Iyer}, {Bradley}, {Dayal}, {Windhorst}, {Zitrin}, {Meena}, {Oguri}, {Diego}, {Kokorev}, {Dimauro}, {Adamo}, {Conselice}, {Welch}, {Vanzella}, {Hsiao}, {Xu}, {Roy}, \& {Mulcahey}}]{Abdurrouf2023}
{Abdurro'uf}, {Coe}, D., {Jung}, I., {et~al.} 2023, \apj, 945, 117

\bibitem[{{Abdurro'uf} {et~al.}(2022{\natexlab{b}}){Abdurro'uf}, {Lin}, {Hirashita}, {Morishita}, {Tacchella}, {Akiyama}, {Takeuchi}, \& {Wu}}]{Abdurrouf2022a}
{Abdurro'uf}, {Lin}, Y.-T., {Hirashita}, H., {et~al.} 2022{\natexlab{b}}, \apj, 926, 81

\bibitem[{{Abdurro'uf} {et~al.}(2022{\natexlab{c}}){Abdurro'uf}, {Lin}, {Hirashita}, {Morishita}, {Tacchella}, {Wu}, {Akiyama}, \& {Takeuchi}}]{Abdurrouf2022c}
{Abdurro'uf}, {Lin}, Y.-T., {Hirashita}, H., {et~al.} 2022{\natexlab{c}}, \apj, 935, 98

\bibitem[{{Abdurro'uf} {et~al.}(2021){Abdurro'uf}, {Lin}, {Wu}, \& {Akiyama}}]{Abdurrouf2021}
{Abdurro'uf}, {Lin}, Y.-T., {Wu}, P.-F., \& {Akiyama}, M. 2021, \apjs, 254, 15

\bibitem[{{Abdurro'uf} {et~al.}(2022{\natexlab{d}}){Abdurro'uf}, {Lin}, {Wu}, \& {Akiyama}}]{Abdurrouf2022b_pixedfitcode}
{Abdurro'uf}, {Lin}, Y.-T., {Wu}, P.-F., \& {Akiyama}, M. 2022{\natexlab{d}}, {piXedfit: Analyze spatially resolved SEDs of galaxies}, Astrophysics Source Code Library, record ascl:2207.033

\bibitem[{{Acquaviva} {et~al.}(2011){Acquaviva}, {Gawiser}, \& {Guaita}}]{Acquaviva2011}
{Acquaviva}, V., {Gawiser}, E., \& {Guaita}, L. 2011, \apj, 737, 47

\bibitem[{{Aihara} {et~al.}(2018){Aihara}, {Arimoto}, {Armstrong}, {Arnouts}, {Bahcall}, {Bickerton}, {Bosch}, {Bundy}, {Capak}, {Chan}, {Chiba}, {Coupon}, {Egami}, {Enoki}, {Finet}, {Fujimori}, {Fujimoto}, {Furusawa}, {Furusawa}, {Goto}, {Goulding}, {Greco}, {Greene}, {Gunn}, {Hamana}, {Harikane}, {Hashimoto}, {Hattori}, {Hayashi}, {Hayashi}, {He{\l}miniak}, {Higuchi}, {Hikage}, {Ho}, {Hsieh}, {Huang}, {Huang}, {Ikeda}, {Imanishi}, {Inoue}, {Iwasawa}, {Iwata}, {Jaelani}, {Jian}, {Kamata}, {Karoji}, {Kashikawa}, {Katayama}, {Kawanomoto}, {Kayo}, {Koda}, {Koike}, {Kojima}, {Komiyama}, {Konno}, {Koshida}, {Koyama}, {Kusakabe}, {Leauthaud}, {Lee}, {Lin}, {Lin}, {Lupton}, {Mandelbaum}, {Matsuoka}, {Medezinski}, {Mineo}, {Miyama}, {Miyatake}, {Miyazaki}, {Momose}, {More}, {More}, {Moritani}, {Moriya}, {Morokuma}, {Mukae}, {Murata}, {Murayama}, {Nagao}, {Nakata}, {Niida}, {Niikura}, {Nishizawa}, {Obuchi}, {Oguri}, {Oishi}, {Okabe}, {Okamoto}, {Okura}, {Ono}, {Onodera}, {Onoue}, {Osato}, {Ouchi}, {Price}, {Pyo},
  {Sako}, {Sawicki}, {Shibuya}, {Shimasaku}, {Shimono}, {Shirasaki}, {Silverman}, {Simet}, {Speagle}, {Spergel}, {Strauss}, {Sugahara}, {Sugiyama}, {Suto}, {Suyu}, {Suzuki}, {Tait}, {Takada}, {Takata}, {Tamura}, {Tanaka}, {Tanaka}, {Tanaka}, {Tanaka}, {Terai}, {Terashima}, {Toba}, {Tominaga}, {Toshikawa}, {Turner}, {Uchida}, {Uchiyama}, {Umetsu}, {Uraguchi}, {Urata}, {Usuda}, {Utsumi}, {Wang}, {Wang}, {Wong}, {Yabe}, {Yamada}, {Yamanoi}, {Yasuda}, {Yeh}, {Yonehara}, \& {Yuma}}]{Aihara2018}
{Aihara}, H., {Arimoto}, N., {Armstrong}, R., {et~al.} 2018, \pasj, 70, S4

\bibitem[{{Baes} {et~al.}(2024{\natexlab{a}}){Baes}, {Gebek}, {Tr{\v{c}}ka}, {Camps}, {van der Wel}, {Abdurro'uf}, {Andreadis}, {Tulu}, {Emana}, {Fritz}, {Kelly}, {Kova{\v{c}}i{\'c}}, {La Marca}, {Martorano}, {Mosenkov}, {Nersesian}, {Rodriguez-Gomez}, {Tortora}, {Vander Meulen}, \& {Wang}}]{Baes2024_a}
{Baes}, M., {Gebek}, A., {Tr{\v{c}}ka}, A., {et~al.} 2024{\natexlab{a}}, \aap, 683, A181

\bibitem[{{Baes} {et~al.}(2024{\natexlab{b}}){Baes}, {Mosenkov}, {Kelly}, {Abdurro'uf}, {Andreadis}, {Bokona Tulu}, {Camps}, {Tassama Emana}, {Fritz}, {Gebek}, {Kova{\v{c}}i{\'c}}, {La Marca}, {Martorano}, {Nersesian}, {Rodriguez-Gomez}, {Tortora}, {Tr{\v{c}}ka}, {Vander Meulen}, {van der Wel}, \& {Wang}}]{Baes2024_b}
{Baes}, M., {Mosenkov}, A., {Kelly}, R., {et~al.} 2024{\natexlab{b}}, \aap, 683, A182

\bibitem[{{Baldry} {et~al.}(2010){Baldry}, {Robotham}, {Hill}, {Driver}, {Liske}, {Norberg}, {Bamford}, {Hopkins}, {Loveday}, {Peacock}, {Cameron}, {Croom}, {Cross}, {Doyle}, {Dye}, {Frenk}, {Jones}, {van Kampen}, {Kelvin}, {Nichol}, {Parkinson}, {Popescu}, {Prescott}, {Sharp}, {Sutherland}, {Thomas}, \& {Tuffs}}]{Baldry2010}
{Baldry}, I.~K., {Robotham}, A.~S.~G., {Hill}, D.~T., {et~al.} 2010, \mnras, 404, 86

\bibitem[{Barbary(2016)}]{Barbary2016_sep}
Barbary, K. 2016, Journal of Open Source Software, 1, 58

\bibitem[{{Bellstedt} {et~al.}(2024){Bellstedt}, {Robotham}, {Driver}, {Lagos}, {Davies}, \& {Cook}}]{2024Bellstedt}
{Bellstedt}, S., {Robotham}, A. S.~G., {Driver}, S.~P., {et~al.} 2024, \mnras, 528, 5452

\bibitem[{{Bertin} \& {Arnouts}(1996)}]{Bertin1996}
{Bertin}, E. \& {Arnouts}, S. 1996, \aaps, 117, 393

\bibitem[{{Bianchi} {et~al.}(2014){Bianchi}, {Conti}, \& {Shiao}}]{Bianchi2014}
{Bianchi}, L., {Conti}, A., \& {Shiao}, B. 2014, Advances in Space Research, 53, 900

\bibitem[{{Blanton} \& {Moustakas}(2009)}]{Blanton2009}
{Blanton}, M.~R. \& {Moustakas}, J. 2009, \araa, 47, 159

\bibitem[{{Bolzonella} {et~al.}(2000){Bolzonella}, {Miralles}, \& {Pell{\'o}}}]{Bolzonella2000}
{Bolzonella}, M., {Miralles}, J.~M., \& {Pell{\'o}}, R. 2000, \aap, 363, 476

\bibitem[{{Boquien} {et~al.}(2019){Boquien}, {Burgarella}, {Roehlly}, {Buat}, {Ciesla}, {Corre}, {Inoue}, \& {Salas}}]{Boquien2019}
{Boquien}, M., {Burgarella}, D., {Roehlly}, Y., {et~al.} 2019, \aap, 622, A103

\bibitem[{{Bradley} {et~al.}(2022){Bradley}, {Sip{\H{o}}cz}, {Robitaille}, {Tollerud}, {Vin{\'\i}cius}, {Deil}, {Barbary}, {Wilson}, {Busko}, {Donath}, {G{\"u}nther}, {Cara}, {Lim}, {Me{\ss}linger}, {Conseil}, {Bostroem}, {Droettboom}, {Bray}, {Andersen Bratholm}, {Barentsen}, {Craig}, {Rathi}, {Pascual}, {Perren}, {Georgiev}, {De Val-Borro}, {Kerzendorf}, {Bach}, {Quint}, \& {Souchereau}}]{Bradley2022_photutils}
{Bradley}, L., {Sip{\H{o}}cz}, B., {Robitaille}, T., {et~al.} 2022, {astropy/photutils: 1.5.0}, Zenodo

\bibitem[{{Brinchmann} {et~al.}(2004){Brinchmann}, {Charlot}, {White}, {Tremonti}, {Kauffmann}, {Heckman}, \& {Brinkmann}}]{Brinchmann2004}
{Brinchmann}, J., {Charlot}, S., {White}, S.~D.~M., {et~al.} 2004, \mnras, 351, 1151

\bibitem[{{Bruzual} \& {Charlot}(2003)}]{Bruzual2003}
{Bruzual}, G. \& {Charlot}, S. 2003, \mnras, 344, 1000

\bibitem[{{Bundy} {et~al.}(2015){Bundy}, {Bershady}, {Law}, {Yan}, {Drory}, {MacDonald}, {Wake}, {Cherinka}, {S{\'a}nchez-Gallego}, {Weijmans}, {Thomas}, {Tremonti}, {Masters}, {Coccato}, {Diamond-Stanic}, {Arag{\'o}n-Salamanca}, {Avila-Reese}, {Badenes}, {Falc{\'o}n-Barroso}, {Belfiore}, {Bizyaev}, {Blanc}, {Bland-Hawthorn}, {Blanton}, {Brownstein}, {Byler}, {Cappellari}, {Conroy}, {Dutton}, {Emsellem}, {Etherington}, {Frinchaboy}, {Fu}, {Gunn}, {Harding}, {Johnston}, {Kauffmann}, {Kinemuchi}, {Klaene}, {Knapen}, {Leauthaud}, {Li}, {Lin}, {Maiolino}, {Malanushenko}, {Malanushenko}, {Mao}, {Maraston}, {McDermid}, {Merrifield}, {Nichol}, {Oravetz}, {Pan}, {Parejko}, {Sanchez}, {Schlegel}, {Simmons}, {Steele}, {Steinmetz}, {Thanjavur}, {Thompson}, {Tinker}, {van den Bosch}, {Westfall}, {Wilkinson}, {Wright}, {Xiao}, \& {Zhang}}]{Bundy2015}
{Bundy}, K., {Bershady}, M.~A., {Law}, D.~R., {et~al.} 2015, \apj, 798, 7

\bibitem[{{Calzetti} {et~al.}(2000){Calzetti}, {Armus}, {Bohlin}, {Kinney}, {Koornneef}, \& {Storchi-Bergmann}}]{Calzetti2000}
{Calzetti}, D., {Armus}, L., {Bohlin}, R.~C., {et~al.} 2000, \apj, 533, 682

\bibitem[{{Camps} \& {Baes}(2015)}]{Camps2015}
{Camps}, P. \& {Baes}, M. 2015, Astronomy and Computing, 9, 20

\bibitem[{{Camps} \& {Baes}(2020)}]{Camps2020}
{Camps}, P. \& {Baes}, M. 2020, Astronomy and Computing, 31, 100381

\bibitem[{{Cappellari} {et~al.}(2011){Cappellari}, {Emsellem}, {Krajnovi{\'c}}, {McDermid}, {Scott}, {Verdoes Kleijn}, {Young}, {Alatalo}, {Bacon}, {Blitz}, {Bois}, {Bournaud}, {Bureau}, {Davies}, {Davis}, {de Zeeuw}, {Duc}, {Khochfar}, {Kuntschner}, {Lablanche}, {Morganti}, {Naab}, {Oosterloo}, {Sarzi}, {Serra}, \& {Weijmans}}]{Cappellari2011}
{Cappellari}, M., {Emsellem}, E., {Krajnovi{\'c}}, D., {et~al.} 2011, \mnras, 413, 813

\bibitem[{{Carnall} {et~al.}(2018){Carnall}, {McLure}, {Dunlop}, \& {Dav{\'e}}}]{Carnall2018}
{Carnall}, A.~C., {McLure}, R.~J., {Dunlop}, J.~S., \& {Dav{\'e}}, R. 2018, \mnras, 480, 4379

\bibitem[{{Chabrier}(2003)}]{Chabrier2003}
{Chabrier}, G. 2003, \pasp, 115, 763

\bibitem[{{Colless} {et~al.}(2001){Colless}, {Dalton}, {Maddox}, {Sutherland}, {Norberg}, {Cole}, {Bland-Hawthorn}, {Bridges}, {Cannon}, {Collins}, {Couch}, {Cross}, {Deeley}, {De Propris}, {Driver}, {Efstathiou}, {Ellis}, {Frenk}, {Glazebrook}, {Jackson}, {Lahav}, {Lewis}, {Lumsden}, {Madgwick}, {Peacock}, {Peterson}, {Price}, {Seaborne}, \& {Taylor}}]{2001Colless}
{Colless}, M., {Dalton}, G., {Maddox}, S., {et~al.} 2001, \mnras, 328, 1039

\bibitem[{{Conroy}(2013)}]{Conroy2013}
{Conroy}, C. 2013, \araa, 51, 393

\bibitem[{{Conroy} \& {Gunn}(2010)}]{Conroy2010}
{Conroy}, C. \& {Gunn}, J.~E. 2010, \apj, 712, 833

\bibitem[{{Conroy} {et~al.}(2009){Conroy}, {Gunn}, \& {White}}]{Conroy2009}
{Conroy}, C., {Gunn}, J.~E., \& {White}, M. 2009, \apj, 699, 486

\bibitem[{{Conti} {et~al.}(2003){Conti}, {Connolly}, {Hopkins}, {Budav{\'a}ri}, {Szalay}, {Csabai}, {Schmidt}, {Adams}, \& {Petrovic}}]{Conti2003}
{Conti}, A., {Connolly}, A.~J., {Hopkins}, A.~M., {et~al.} 2003, \aj, 126, 2330

\bibitem[{{Croom} {et~al.}(2012){Croom}, {Lawrence}, {Bland-Hawthorn}, {Bryant}, {Fogarty}, {Richards}, {Goodwin}, {Farrell}, {Miziarski}, {Heald}, {Jones}, {Lee}, {Colless}, {Brough}, {Hopkins}, {Bauer}, {Birchall}, {Ellis}, {Horton}, {Leon-Saval}, {Lewis}, {L{\'o}pez-S{\'a}nchez}, {Min}, {Trinh}, \& {Trowland}}]{Croom2012}
{Croom}, S.~M., {Lawrence}, J.~S., {Bland-Hawthorn}, J., {et~al.} 2012, \mnras, 421, 872

\bibitem[{{Cropper} {et~al.}(2016){Cropper}, {Pottinger}, {Niemi}, {Azzollini}, {Denniston}, {Szafraniec}, {Awan}, {Mellier}, {Berthe}, {Martignac}, {Cara}, {Di Giorgio}, {Sciortino}, {Bozzo}, {Genolet}, {Cole}, {Philippon}, {Hailey}, {Hunt}, {Swindells}, {Holland}, {Gow}, {Murray}, {Hall}, {Skottfelt}, {Amiaux}, {Laureijs}, {Racca}, {Salvignol}, {Short}, {Lorenzo Alvarez}, {Kitching}, {Hoekstra}, {Massey}, \& {Israel}}]{2016ropper}
{Cropper}, M., {Pottinger}, S., {Niemi}, S., {et~al.} 2016, in Society of Photo-Optical Instrumentation Engineers (SPIE) Conference Series, Vol. 9904, Space Telescopes and Instrumentation 2016: Optical, Infrared, and Millimeter Wave, ed. H.~A. {MacEwen}, G.~G. {Fazio}, M.~{Lystrup}, N.~{Batalha}, N.~{Siegler}, \& E.~C. {Tong}, 99040Q

\bibitem[{{Cuillandre} {et~al.}(2025){Cuillandre}, {Bolzonella}, {Boselli}, {Marleau}, {Mondelin}, {Sorce}, {Stone}, {Buitrago}, {Cantiello}, {George}, {Hatch}, {Quilley}, {Mannucci}, {Saifollahi}, {S{\'a}nchez-Janssen}, {Tarsitano}, {Tortora}, {Xu}, {Bouy}, {Gwyn}, {Kluge}, {Lan{\c{c}}on}, {Laureijs}, {Schirmer}, {Abdurro'uf}, {Awad}, {Baes}, {Bournaud}, {Carollo}, {Codis}, {Conselice}, {De Lapparent}, {Duc}, {Ferr{\'e}-Mateu}, {Gillard}, {Golden-Marx}, {Jablonka}, {Habas}, {Hunt}, {Mei}, {Miville-Desch{\^e}nes}, {Montes}, {Nersesian}, {Peletier}, {Poulain}, {Scaramella}, {Scialpi}, {Sola}, {Stephan}, {Ulivi}, {Urbano}, {Z{\"o}ller}, {Aghanim}, {Altieri}, {Amara}, {Andreon}, {Auricchio}, {Baldi}, {Balestra}, {Bardelli}, {Bender}, {Biviano}, {Bodendorf}, {Bonino}, {Branchini}, {Brescia}, {Brinchmann}, {Camera}, {Capobianco}, {Carbone}, {Carretero}, {Casas}, {Castander}, {Castellano}, {Castignani}, {Cavuoti}, {Cimatti}, {Congedo}, {Conversi}, {Copin}, {Courbin}, {Courtois}, {Cropper}, {Da Silva}, {Degaudenzi},
  {De Lucia}, {Di Giorgio}, {Dinis}, {Douspis}, {Dubath}, {Duncan}, {Dupac}, {Dusini}, {Farina}, {Farrens}, {Ferriol}, {Fotopoulou}, {Frailis}, {Franceschi}, {Galeotta}, {Gillis}, {Giocoli}, {G{\'o}mez-Alvarez}, {Grazian}, {Grupp}, {Guzzo}, {Haugan}, {Hoar}, {Hoekstra}, {Holmes}, {Hook}, {Hormuth}, {Hornstrup}, {Hudelot}, {Jahnke}, {Jhabvala}, {Keih{\"a}nen}, {Kermiche}, {Kiessling}, {Kilbinger}, {Kitching}, {Kohley}, {Kubik}, {Kuijken}, {K{\"u}mmel}, {Kunz}, {Kurki-Suonio}, {Lahav}, {Le Mignant}, {Ligori}, {Lilje}, {Lindholm}, {Lloro}, {Maino}, {Maiorano}, {Mansutti}, {Marggraf}, {Markovic}, {Martinet}, {Marulli}, {Massey}, {Maurogordato}, {McCracken}, {Medinaceli}, {Melchior}, {Mellier}, {Meneghetti}, {Merlin}, {Meylan}, {Mohr}, {Mora}, {Moresco}, {Moscardini}, {Nakajima}, {Nichol}, {Niemi}, {Padilla}, {Paltani}, {Pasian}, {Pedersen}, {Percival}, {Pettorino}, {Pires}, {Polenta}, {Poncet}, {Popa}, {Pozzetti}, {Raison}, {Renzi}, {Rhodes}, {Riccio}, {Romelli}, {Roncarelli}, {Saglia}, {Sapone}, {Schneider},
  {Schrabback}, {Secroun}, {Seidel}, {Serrano}, {Simon}, {Sirignano}, {Sirri}, {Skottfelt}, {Stanco}, {Tallada-Cresp{\'\i}}, {Taylor}, {Teplitz}, {Tereno}, {Toledo-Moreo}, {Tutusaus}, {Valentijn}, {Valenziano}, {Vassallo}, {Verdoes Kleijn}, {Wang}, {Weller}, {Zucca}, {Burigana}, \& {Scottez}}]{Cuillandre+24_Perseus-LF}
{Cuillandre}, J.~C., {Bolzonella}, M., {Boselli}, A., {et~al.} 2025, \aap, 697, A11

\bibitem[{{da Cunha} {et~al.}(2008){da Cunha}, {Charlot}, \& {Elbaz}}]{daCunha2008}
{da Cunha}, E., {Charlot}, S., \& {Elbaz}, D. 2008, \mnras, 388, 1595

\bibitem[{{Dark Energy Survey Collaboration} {et~al.}(2016){Dark Energy Survey Collaboration}, {Abbott}, {Abdalla}, {Aleksi{\'c}}, {Allam}, {Amara}, {Bacon}, {Balbinot}, {Banerji}, {Bechtol}, {Benoit-L{\'e}vy}, {Bernstein}, {Bertin}, {Blazek}, {Bonnett}, {Bridle}, {Brooks}, {Brunner}, {Buckley-Geer}, {Burke}, {Caminha}, {Capozzi}, {Carlsen}, {Carnero-Rosell}, {Carollo}, {Carrasco-Kind}, {Carretero}, {Castander}, {Clerkin}, {Collett}, {Conselice}, {Crocce}, {Cunha}, {D'Andrea}, {da Costa}, {Davis}, {Desai}, {Diehl}, {Dietrich}, {Dodelson}, {Doel}, {Drlica-Wagner}, {Estrada}, {Etherington}, {Evrard}, {Fabbri}, {Finley}, {Flaugher}, {Foley}, {Fosalba}, {Frieman}, {Garc{\'\i}a-Bellido}, {Gaztanaga}, {Gerdes}, {Giannantonio}, {Goldstein}, {Gruen}, {Gruendl}, {Guarnieri}, {Gutierrez}, {Hartley}, {Honscheid}, {Jain}, {James}, {Jeltema}, {Jouvel}, {Kessler}, {King}, {Kirk}, {Kron}, {Kuehn}, {Kuropatkin}, {Lahav}, {Li}, {Lima}, {Lin}, {Maia}, {Makler}, {Manera}, {Maraston}, {Marshall}, {Martini}, {McMahon},
  {Melchior}, {Merson}, {Miller}, {Miquel}, {Mohr}, {Morice-Atkinson}, {Naidoo}, {Neilsen}, {Nichol}, {Nord}, {Ogando}, {Ostrovski}, {Palmese}, {Papadopoulos}, {Peiris}, {Peoples}, {Percival}, {Plazas}, {Reed}, {Refregier}, {Romer}, {Roodman}, {Ross}, {Rozo}, {Rykoff}, {Sadeh}, {Sako}, {S{\'a}nchez}, {Sanchez}, {Santiago}, {Scarpine}, {Schubnell}, {Sevilla-Noarbe}, {Sheldon}, {Smith}, {Smith}, {Soares-Santos}, {Sobreira}, {Soumagnac}, {Suchyta}, {Sullivan}, {Swanson}, {Tarle}, {Thaler}, {Thomas}, {Thomas}, {Tucker}, {Vieira}, {Vikram}, {Walker}, {Wechsler}, {Weller}, {Wester}, {Whiteway}, {Wilcox}, {Yanny}, {Zhang}, \& {Zuntz}}]{DES2016}
{Dark Energy Survey Collaboration}, {Abbott}, T., {Abdalla}, F.~B., {et~al.} 2016, \mnras, 460, 1270

\bibitem[{{Davies} {et~al.}(2017){Davies}, {Baes}, {Bianchi}, {Jones}, {Madden}, {Xilouris}, {Bocchio}, {Casasola}, {Cassara}, {Clark}, {De Looze}, {Evans}, {Fritz}, {Galametz}, {Galliano}, {Lianou}, {Mosenkov}, {Smith}, {Verstocken}, {Viaene}, {Vika}, {Wagle}, \& {Ysard}}]{Davies2017}
{Davies}, J.~I., {Baes}, M., {Bianchi}, S., {et~al.} 2017, \pasp, 129, 044102

\bibitem[{{De Vis} {et~al.}(2019){De Vis}, {Jones}, {Viaene}, {Casasola}, {Clark}, {Baes}, {Bianchi}, {Cassara}, {Davies}, {De Looze}, {Galametz}, {Galliano}, {Lianou}, {Madden}, {Manilla-Robles}, {Mosenkov}, {Nersesian}, {Roychowdhury}, {Xilouris}, \& {Ysard}}]{DeVis2019}
{De Vis}, P., {Jones}, A., {Viaene}, S., {et~al.} 2019, \aap, 623, A5

\bibitem[{{de Zeeuw} {et~al.}(2002){de Zeeuw}, {Bureau}, {Emsellem}, {Bacon}, {Carollo}, {Copin}, {Davies}, {Kuntschner}, {Miller}, {Monnet}, {Peletier}, \& {Verolme}}]{deZeeuw2002}
{de Zeeuw}, P.~T., {Bureau}, M., {Emsellem}, E., {et~al.} 2002, \mnras, 329, 513

\bibitem[{{DESI Collaboration} {et~al.}(2016){DESI Collaboration}, {Aghamousa}, {Aguilar}, {Ahlen}, {Alam}, {Allen}, {Allende Prieto}, {Annis}, {Bailey}, {Balland}, {Ballester}, {Baltay}, {Beaufore}, {Bebek}, {Beers}, {Bell}, {Bernal}, {Besuner}, {Beutler}, {Blake}, {Bleuler}, {Blomqvist}, {Blum}, {Bolton}, {Briceno}, {Brooks}, {Brownstein}, {Buckley-Geer}, {Burden}, {Burtin}, {Busca}, {Cahn}, {Cai}, {Cardiel-Sas}, {Carlberg}, {Carton}, {Casas}, {Castander}, {Cervantes-Cota}, {Claybaugh}, {Close}, {Coker}, {Cole}, {Comparat}, {Cooper}, {Cousinou}, {Crocce}, {Cuby}, {Cunningham}, {Davis}, {Dawson}, {de la Macorra}, {De Vicente}, {Delubac}, {Derwent}, {Dey}, {Dhungana}, {Ding}, {Doel}, {Duan}, {Ealet}, {Edelstein}, {Eftekharzadeh}, {Eisenstein}, {Elliott}, {Escoffier}, {Evatt}, {Fagrelius}, {Fan}, {Fanning}, {Farahi}, {Farihi}, {Favole}, {Feng}, {Fernandez}, {Findlay}, {Finkbeiner}, {Fitzpatrick}, {Flaugher}, {Flender}, {Font-Ribera}, {Forero-Romero}, {Fosalba}, {Frenk}, {Fumagalli}, {Gaensicke}, {Gallo},
  {Garcia-Bellido}, {Gaztanaga}, {Pietro Gentile Fusillo}, {Gerard}, {Gershkovich}, {Giannantonio}, {Gillet}, {Gonzalez-de-Rivera}, {Gonzalez-Perez}, {Gott}, {Graur}, {Gutierrez}, {Guy}, {Habib}, {Heetderks}, {Heetderks}, {Heitmann}, {Hellwing}, {Herrera}, {Ho}, {Holland}, {Honscheid}, {Huff}, {Hutchinson}, {Huterer}, {Hwang}, {Illa Laguna}, {Ishikawa}, {Jacobs}, {Jeffrey}, {Jelinsky}, {Jennings}, {Jiang}, {Jimenez}, {Johnson}, {Joyce}, {Jullo}, {Juneau}, {Kama}, {Karcher}, {Karkar}, {Kehoe}, {Kennamer}, {Kent}, {Kilbinger}, {Kim}, {Kirkby}, {Kisner}, {Kitanidis}, {Kneib}, {Koposov}, {Kovacs}, {Koyama}, {Kremin}, {Kron}, {Kronig}, {Kueter-Young}, {Lacey}, {Lafever}, {Lahav}, {Lambert}, {Lampton}, {Landriau}, {Lang}, {Lauer}, {Le Goff}, {Le Guillou}, {Le Van Suu}, {Lee}, {Lee}, {Leitner}, {Lesser}, {Levi}, {L'Huillier}, {Li}, {Liang}, {Lin}, {Linder}, {Loebman}, {Luki{\'c}}, {Ma}, {MacCrann}, {Magneville}, {Makarem}, {Manera}, {Manser}, {Marshall}, {Martini}, {Massey}, {Matheson}, {McCauley}, {McDonald},
  {McGreer}, {Meisner}, {Metcalfe}, {Miller}, {Miquel}, {Moustakas}, {Myers}, {Naik}, {Newman}, {Nichol}, {Nicola}, {Nicolati da Costa}, {Nie}, {Niz}, {Norberg}, {Nord}, {Norman}, {Nugent}, {O'Brien}, {Oh}, {Olsen}, {Padilla}, {Padmanabhan}, {Padmanabhan}, {Palanque-Delabrouille}, {Palmese}, {Pappalardo}, {P{\^a}ris}, {Park}, {Patej}, {Peacock}, {Peiris}, {Peng}, {Percival}, {Perruchot}, {Pieri}, {Pogge}, {Pollack}, {Poppett}, {Prada}, {Prakash}, {Probst}, {Rabinowitz}, {Raichoor}, {Ree}, {Refregier}, {Regal}, {Reid}, {Reil}, {Rezaie}, {Rockosi}, {Roe}, {Ronayette}, {Roodman}, {Ross}, {Ross}, {Rossi}, {Rozo}, {Ruhlmann-Kleider}, {Rykoff}, {Sabiu}, {Samushia}, {Sanchez}, {Sanchez}, {Schlegel}, {Schneider}, {Schubnell}, {Secroun}, {Seljak}, {Seo}, {Serrano}, {Shafieloo}, {Shan}, {Sharples}, {Sholl}, {Shourt}, {Silber}, {Silva}, {Sirk}, {Slosar}, {Smith}, {Smoot}, {Som}, {Song}, {Sprayberry}, {Staten}, {Stefanik}, {Tarle}, {Sien Tie}, {Tinker}, {Tojeiro}, {Valdes}, {Valenzuela}, {Valluri}, {Vargas-Magana},
  {Verde}, {Walker}, {Wang}, {Wang}, {Weaver}, {Weaverdyck}, {Wechsler}, {Weinberg}, {White}, {Yang}, {Yeche}, {Zhang}, {Zhao}, {Zheng}, {Zhou}, {Zhou}, {Zhu}, {Zou}, \& {Zu}}]{2016DESI}
{DESI Collaboration}, {Aghamousa}, A., {Aguilar}, J., {et~al.} 2016, arXiv:1611.00036

\bibitem[{{Dobbels} {et~al.}(2020){Dobbels}, {Baes}, {Viaene}, {Bianchi}, {Davies}, {Casasola}, {Clark}, {Fritz}, {Galametz}, {Galliano}, {Mosenkov}, {Nersesian}, \& {Tr{\v{c}}ka}}]{Dobbels2020}
{Dobbels}, W., {Baes}, M., {Viaene}, S., {et~al.} 2020, \aap, 634, A57

\bibitem[{{Driver} {et~al.}(2022){Driver}, {Bellstedt}, {Robotham}, {Baldry}, {Davies}, {Liske}, {Obreschkow}, {Taylor}, {Wright}, {Alpaslan}, {Bamford}, {Bauer}, {Bland-Hawthorn}, {Bilicki}, {Bravo}, {Brough}, {Casura}, {Cluver}, {Colless}, {Conselice}, {Croom}, {de Jong}, {D'Eugenio}, {De Propris}, {Dogruel}, {Drinkwater}, {Dvornik}, {Farrow}, {Frenk}, {Giblin}, {Graham}, {Grootes}, {Gunawardhana}, {Hashemizadeh}, {H{\"a}u{\ss}ler}, {Heymans}, {Hildebrandt}, {Holwerda}, {Hopkins}, {Jarrett}, {Heath Jones}, {Kelvin}, {Koushan}, {Kuijken}, {Lara-L{\'o}pez}, {Lange}, {L{\'o}pez-S{\'a}nchez}, {Loveday}, {Mahajan}, {Meyer}, {Moffett}, {Napolitano}, {Norberg}, {Owers}, {Radovich}, {Raouf}, {Peacock}, {Phillipps}, {Pimbblet}, {Popescu}, {Said}, {Sansom}, {Seibert}, {Sutherland}, {Thorne}, {Tuffs}, {Turner}, {van der Wel}, {van Kampen}, \& {Wilkins}}]{2022Driver}
{Driver}, S.~P., {Bellstedt}, S., {Robotham}, A. S.~G., {et~al.} 2022, \mnras, 513, 439

\bibitem[{{Euclid Collaboration: Bisigello} {et~al.}(2023){Euclid Collaboration: Bisigello}, {Conselice}, {Baes}, {et~al.}}]{Bisigello2023}
{Euclid Collaboration: Bisigello}, L., {Conselice}, C.~J., {Baes}, M., {et~al.} 2023, \mnras, 520, 3529

\bibitem[{{Euclid Collaboration: Cropper} {et~al.}(2025){Euclid Collaboration: Cropper}, {Al-Bahlawan}, {Amiaux}, {et~al.}}]{2024Cropper_euclid}
{Euclid Collaboration: Cropper}, M., {Al-Bahlawan}, A., {Amiaux}, J., {et~al.} 2025, A\&A, 697, A2

\bibitem[{{Euclid Collaboration: Enia} {et~al.}(2025){Euclid Collaboration: Enia}, {Pozzetti}, {Bolzonella}, {et~al.}}]{Q1-SP031-Enia}
{Euclid Collaboration: Enia}, A., {Pozzetti}, L., {Bolzonella}, M., {et~al.} 2025, A\&A, in press (Euclid Q1 SI), \url{https://doi.org/10.1051/0004-6361/202554576}, arXiv:2503.15314

\bibitem[{{Euclid Collaboration: Hormuth} {et~al.}(2025){Euclid Collaboration: Hormuth}, {Jahnke}, {Schirmer}, {et~al.}}]{2024Hormuth_euclid}
{Euclid Collaboration: Hormuth}, F., {Jahnke}, K., {Schirmer}, M., {et~al.} 2025, A\&A, 697, A4

\bibitem[{{Euclid Collaboration: Jahnke} {et~al.}(2025){Euclid Collaboration: Jahnke}, {Gillard}, {Schirmer}, {et~al.}}]{2024Jahnke_euclid}
{Euclid Collaboration: Jahnke}, K., {Gillard}, W., {Schirmer}, M., {et~al.} 2025, A\&A, 697, A3

\bibitem[{{Euclid Collaboration: Kova{\v{c}}i{\'c}} {et~al.}(2025){Euclid Collaboration: Kova{\v{c}}i{\'c}}, {Baes}, {Nersesian}, {et~al.}}]{2025Kovacic_euclid}
{Euclid Collaboration: Kova{\v{c}}i{\'c}}, I., {Baes}, M., {Nersesian}, A., {et~al.} 2025, \aap, 695, A284

\bibitem[{{Euclid Collaboration: Martinet} {et~al.}(2019){Euclid Collaboration: Martinet}, {Schrabback}, {Hoekstra}, {et~al.}}]{Martinet2019}
{Euclid Collaboration: Martinet}, N., {Schrabback}, T., {Hoekstra}, H., {et~al.} 2019, \aap, 627, A59

\bibitem[{{Euclid Collaboration: Mellier} {et~al.}(2025){Euclid Collaboration: Mellier}, {Abdurro'uf}, {Acevedo~Barroso}, {et~al.}}]{Mellier+24}
{Euclid Collaboration: Mellier}, Y., {Abdurro'uf}, {Acevedo~Barroso}, J., {et~al.} 2025, A\&A, 697, A1

\bibitem[{{Euclid Collaboration: Merlin} {et~al.}(2023){Euclid Collaboration: Merlin}, {Castellano}, {Bretonni{\`e}re}, {et~al.}}]{Merlin2023}
{Euclid Collaboration: Merlin}, E., {Castellano}, M., {Bretonni{\`e}re}, H., {et~al.} 2023, \aap, 671, A101

\bibitem[{{Euclid Collaboration: Scaramella} {et~al.}(2022){Euclid Collaboration: Scaramella}, {Amiaux}, {Mellier}, {et~al.}}]{Scaramella2022}
{Euclid Collaboration: Scaramella}, R., {Amiaux}, J., {Mellier}, Y., {et~al.} 2022, \aap, 662, A112

\bibitem[{{Euclid Collaboration: Schirmer} {et~al.}(2022){Euclid Collaboration: Schirmer}, {Jahnke}, {Seidel}, {et~al.}}]{Schirmer2022}
{Euclid Collaboration: Schirmer}, M., {Jahnke}, K., {Seidel}, G., {et~al.} 2022, \aap, 662, A92

\bibitem[{{Falc{\'o}n-Barroso} {et~al.}(2011){Falc{\'o}n-Barroso}, {S{\'a}nchez-Bl{\'a}zquez}, {Vazdekis}, {Ricciardelli}, {Cardiel}, {Cenarro}, {Gorgas}, \& {Peletier}}]{Falcon-Barroso2011}
{Falc{\'o}n-Barroso}, J., {S{\'a}nchez-Bl{\'a}zquez}, P., {Vazdekis}, A., {et~al.} 2011, \aap, 532, A95

\bibitem[{{Gallazzi} {et~al.}(2005){Gallazzi}, {Charlot}, {Brinchmann}, {White}, \& {Tremonti}}]{Gallazzi2005}
{Gallazzi}, A., {Charlot}, S., {Brinchmann}, J., {White}, S. D.~M., \& {Tremonti}, C.~A. 2005, \mnras, 362, 41

\bibitem[{{Galliano} {et~al.}(2021){Galliano}, {Nersesian}, {Bianchi}, {De Looze}, {Roychowdhury}, {Baes}, {Casasola}, {Cassar{\'a}}, {Dobbels}, {Fritz}, {Galametz}, {Jones}, {Madden}, {Mosenkov}, {Xilouris}, \& {Ysard}}]{Galliano2021}
{Galliano}, F., {Nersesian}, A., {Bianchi}, S., {et~al.} 2021, \aap, 649, A18

\bibitem[{{Gardner} {et~al.}(2023){Gardner}, {Mather}, {Abbott}, {Abell}, {Abernathy}, {Abney}, {Abraham}, {Abraham}, {Abul-Huda}, {Acton}, {Adams}, {Adams}, {Adler}, {Adriaensen}, {Aguilar}, {Ahmed}, {Ahmed}, {Ahmed}, {Albat}, {Albert}, {Alberts}, {Aldridge}, {Allen}, {Allen}, {Altenburg}, {Altunc}, {Alvarez}, {{\'A}lvarez-M{\'a}rquez}, {Alves de Oliveira}, {Ambrose}, {Anandakrishnan}, {Andersen}, {Anderson}, {Anderson}, {Anderson}, {Anderson}, {Aprea}, {Archer}, {Arenberg}, {Argyriou}, {Arribas}, {Artigau}, {Arvai}, {Atcheson}, {Atkinson}, {Averbukh}, {Aymergen}, {Bacinski}, {Baggett}, {Bagnasco}, {Baker}, {Balzano}, {Banks}, {Baran}, {Barker}, {Barrett}, {Barringer}, {Barto}, {Bast}, {Baudoz}, {Baum}, {Beatty}, {Beaulieu}, {Bechtold}, {Beck}, {Beddard}, {Beichman}, {Bellagama}, {Bely}, {Berger}, {Bergeron}, {Bernier}, {Bertch}, {Beskow}, {Betz}, {Biagetti}, {Birkmann}, {Bjorklund}, {Blackwood}, {Blazek}, {Blossfeld}, {Bluth}, {Boccaletti}, {Boegner}, {Bohlin}, {Boia}, {B{\"o}ker}, {Bonaventura}, {Bond},
  {Bosley}, {Boucarut}, {Bouchet}, {Bouwman}, {Bower}, {Bowers}, {Bowers}, {Boyce}, {Boyer}, {Boyer}, {Boyer}, {Boyer}, {Bradley}, {Brady}, {Brandl}, {Brannen}, {Breda}, {Bremmer}, {Brennan}, {Bresnahan}, {Bright}, {Broiles}, {Bromenschenkel}, {Brooks}, {Brooks}, {Brown}, {Brown}, {Brown}, {Bruce}, {Bryson}, {Bujanda}, {Bullock}, {Bunker}, {Bureo}, {Burt}, {Bush}, {Bushouse}, {Bussman}, {Cabaud}, {Cale}, {Calhoon}, {Calvani}, {Canipe}, {Caputo}, {Cara}, {Carey}, {Case}, {Cesari}, {Cetorelli}, {Chance}, {Chandler}, {Chaney}, {Chapman}, {Charlot}, {Chayer}, {Cheezum}, {Chen}, {Chen}, {Cherinka}, {Chichester}, {Chilton}, {Chittiraibalan}, {Clampin}, {Clark}, {Clark}, {Clark}, {Claybrooks}, {Cleveland}, {Cohen}, {Cohen}, {Col{\'o}n}, {Coleman}, {Colina}, {Comber}, {Comeau}, {Comer}, {Conde Reis}, {Connolly}, {Conroy}, {Contos}, {Contreras}, {Cook}, {Cooper}, {Cooper}, {Correia}, {Correnti}, {Cossou}, {Costanza}, {Coulais}, {Cox}, {Coyle}, {Cracraft}, {Crew}, {Curtis}, {Cusveller}, {Da Costa Maciel}, {Dailey},
  {Daugeron}, {Davidson}, {Davies}, {Davis}, {Davis}, {Day}, {de Chambure}, {de Jong}, {De Marchi}, {Dean}, {Decker}, {Delisa}, {Dell}, {Dellagatta}, {Dembinska}, {Demosthenes}, {Dencheva}, {Deneu}, {DePriest}, {Deschenes}, {Dethienne}, {Detre}, {Diaz}, {Dicken}, {DiFelice}, {Dillman}, {Disharoon}, {Dixon}, {Doggett}, {Dominguez}, {Donaldson}, {Doria-Warner}, {Santos}, {Doty}, {Douglas}, {Doyon}, {Dressler}, {Driggers}, {Driggers}, {Dunn}, {DuPrie}, {Dupuis}, {Durning}, {Dutta}, {Earl}, {Eccleston}, {Ecobichon}, {Egami}, {Ehrenwinkler}, {Eisenhamer}, {Eisenhower}, {Eisenstein}, {El Hamel}, {Elie}, {Elliott}, {Elliott}, {Engesser}, {Espinoza}, {Etienne}, {Etxaluze}, {Evans}, {Fabreguettes}, {Falcolini}, {Falini}, {Fatig}, {Feeney}, {Feinberg}, {Fels}, {Ferdous}, {Ferguson}, {Ferrarese}, {Ferreira}, {Ferruit}, {Ferry}, {Filippazzo}, {Firre}, {Fix}, {Flagey}, {Flanagan}, {Fleming}, {Florian}, {Flynn}, {Foiadelli}, {Fontaine}, {Fontanella}, {Forshay}, {Fortner}, {Fox}, {Framarini}, {Francisco}, {Franck}, {Franx},
  {Franz}, {Friedman}, {Friend}, {Frost}, {Fu}, {Fullerton}, {Gaillard}, {Galkin}, {Gallagher}, {Galyer}, {Garc{\'\i}a Mar{\'\i}n}, {Gardner}, {Garland}, {Garrett}, {Gasman}, {G{\'a}sp{\'a}r}, {Gastaud}, {Gaudreau}, {Gauthier}, {Geers}, {Geithner}, {Gennaro}, {Gerber}, {Gereau}, {Giampaoli}, {Giardino}, {Gibbons}, {Gilbert}, {Gilman}, {Girard}, {Giuliano}, {Gkountis}, {Glasse}, {Glassmire}, {Glauser}, {Glazer}, {Goldberg}, {Golimowski}, {Gonzaga}, {Gordon}, {Gordon}, {Goudfrooij}, {Gough}, {Graham}, {Grau}, {Green}, {Greene}, {Greene}, {Greenfield}, {Greenhouse}, {Greve}, {Greville}, {Grimaldi}, {Groe}, {Groebner}, {Grumm}, {Grundy}, {G{\"u}del}, {Guillard}, {Guldalian}, {Gunn}, {Gurule}, {Gutman}, {Guy}, {Guyot}, {Hack}, {Haderlein}, {Hagan}, {Hagedorn}, {Hainline}, {Haley}, {Hami}, {Hamilton}, {Hammann}, {Hammel}, {Hanley}, {Hansen}, {Hardy}, {Harnisch}, {Harr}, {Harris}, {Hart}, {Hartig}, {Hasan}, {Hashim}, {Hashimoto}, {Haskins}, {Hawkins}, {Hayden}, {Hayden}, {Healy}, {Hecht}, {Heeg}, {Hejal}, {Helm},
  {Hengemihle}, {Henning}, {Henry}, {Henry}, {Henshaw}, {Hernandez}, {Herrington}, {Heske}, {Hesman}, {Hickey}, {Hilbert}, {Hines}, {Hinz}, {Hirsch}, {Hitcho}, {Hodapp}, {Hodge}, {Hoffman}, {Holfeltz}, {Holler}, {Hoppa}, {Horner}, {Howard}, {Howard}, {Huber}, {Hunkeler}, {Hunter}, {Hunter}, {Hurd}, {Hurst}, {Hutchings}, {Hylan}, {Ignat}, {Illingworth}, {Irish}, {Isaacs}, {Jackson}, {Jaffe}, {Jahic}, {Jahromi}, {Jakobsen}, {James}, {James}, {James}, {Jamieson}, {Jandra}, {Jayawardhana}, {Jedrzejewski}, {Jeffers}, {Jensen}, {Joanne}, {Johns}, {Johnson}, {Johnson}, {Johnson}, {Johnson}, {Johnson}, {Johnson}, {Johnstone}, {Jollet}, {Jones}, {Jones}, {Jones}, {Jones}, {Jones}, {Jordan}, {Jordan}, {Jue}, {Jurkowski}, {Justis}, {Justtanont}, {Kaleida}, {Kalirai}, {Kalmanson}, {Kaltenegger}, {Kammerer}, {Kan}, {Kanarek}, {Kao}, {Karakla}, {Karl}, {Kassin}, {Kauffman}, {Kavanagh}, {Kelley}, {Kelly}, {Kendrew}, {Kennedy}, {Kenny}, {Keski-Kuha}, {Keyes}, {Khan}, {Kidwell}, {Kimble}, {King}, {King}, {Kinzel}, {Kirk},
  {Kirkpatrick}, {Klaassen}, {Klingemann}, {Klintworth}, {Knapp}, {Knight}, {Knollenberg}, {Knutsen}, {Koehler}, {Koekemoer}, {Kofler}, {Kontson}, {Kovacs}, {Kozhurina-Platais}, {Krause}, {Kriss}, {Krist}, {Kristoffersen}, {Krogel}, {Krueger}, {Kulp}, {Kumari}, {Kwan}, {Kyprianou}, {Labador}, {Labiano}, {Lafreni{\`e}re}, {Lagage}, {Laidler}, {Laine}, {Laird}, {Lajoie}, {Lallo}, {Lam}, {LaMassa}, {Lambros}, {Lampenfield}, {Lander}, {Langston}, {Larson}, {Larson}, {LaVerghetta}, {Law}, {Lawrence}, {Lee}, {Lee}, {Lee}, {Leisenring}, {Leveille}, {Levenson}, {Levi}, {Levine}, {Lewis}, {Lewis}, {Lewis}, {Libralato}, {Lidon}, {Liebrecht}, {Lightsey}, {Lilly}, {Lim}, {Lim}, {Ling}, {Link}, {Link}, {Lipinski}, {Liu}, {Lo}, {Lobmeyer}, {Logue}, {Long}, {Long}, {Long}, {Long}, {L{\'o}pez-Caniego}, {Lotz}, {Love-Pruitt}, {Lubskiy}, {Luers}, {Luetgens}, {Luevano}, {Lui}, {Lund}, {Lundquist}, {Lunine}, {L{\"u}tzgendorf}, {Lynch}, {MacDonald}, {MacDonald}, {Macias}, {Macklis}, {Maghami}, {Maharaja}, {Maiolino},
  {Makrygiannis}, {Malla}, {Malumuth}, {Manjavacas}, {Marini}, {Marrione}, {Marston}, {Martel}, {Martin}, {Martin}, {Martinez}, {Maschmann}, {Masci}, {Masetti}, {Maszkiewicz}, {Matthews}, {Matuskey}, {McBrayer}, {McCarthy}, {McCaughrean}, {McClare}, {McClare}, {McCloskey}, {McClurg}, {McCoy}, {McElwain}, {McGregor}, {McGuffey}, {McKay}, {McKenzie}, {McLean}, {McMaster}, {McNeil}, {De Meester}, {Mehalick}, {Meixner}, {Mel{\'e}ndez}, {Menzel}, {Menzel}, {Merz}, {Mesterharm}, {Meyer}, {Meyett}, {Meza}, {Midwinter}, {Milam}, {Miller}, {Miller}, {Miskey}, {Misselt}, {Mitchell}, {Mohan}, {Montoya}, {Moran}, {Morishita}, {Moro-Mart{\'\i}n}, {Morrison}, {Morrison}, {Morse}, {Moschos}, {Moseley}, {Mosier}, {Mosner}, {Mountain}, {Muckenthaler}, {Mueller}, {Mueller}, {Muhiem}, {M{\"u}hlmann}, {Mullally}, {Mullen}, {Munger}, {Murphy}, {Murray}, {Muzerolle}, {Mycroft}, {Myers}, {Myers}, {Myers}, {Myers}, {Myrick}, {Nagle}, {Nayak}, {Naylor}, {Neff}, {Nelan}, {Nella}, {Nguyen}, {Nguyen}, {Nickson}, {Nidhiry}, {Niedner},
  {Nieto-Santisteban}, {Nikolov}, {Nishisaka}, {Noriega-Crespo}, {Nota}, {O'Mara}, {Oboryshko}, {O'Brien}, {Ochs}, {Offenberg}, {Ogle}, {Ohl}, {Olmsted}, {Osborne}, {O'Shaughnessy}, {{\"O}stlin}, {O'Sullivan}, {Otor}, {Ottens}, {Ouellette}, {Outlaw}, {Owens}, {Pacifici}, {Page}, {Paranilam}, {Park}, {Parrish}, {Paschal}, {Patapis}, {Patel}, {Patrick}, {Pattishall}, {Paul}, {Paul}, {Pauly}, {Pavlovsky}, {Pe{\~n}a-Guerrero}, {Pedder}, {Peek}, {Pelham}, {Penanen}, {Perriello}, {Perrin}, {Perrine}, {Perrygo}, {Peslier}, {Petach}, {Peterson}, {Pfarr}, {Pierson}, {Pietraszkiewicz}, {Pilchen}, {Pipher}, {Pirzkal}, {Pitman}, {Player}, {Plesha}, {Plitzke}, {Pohner}, {Poletis}, {Pollizzi}, {Polster}, {Pontius}, {Pontoppidan}, {Porges}, {Potter}, {Prescott}, {Proffitt}, {Pueyo}, {Quispe Neira}, {Radich}, {Rager}, {Rameau}, {Ramey}, {Ramos Alarcon}, {Rampini}, {Rapp}, {Rashford}, {Rauscher}, {Ravindranath}, {Rawle}, {Rawlings}, {Ray}, {Regan}, {Rehm}, {Rehm}, {Reid}, {Reis}, {Renk}, {Reoch}, {Ressler}, {Rest},
  {Reynolds}, {Richon}, {Richon}, {Ridgaway}, {Riedel}, {Rieke}, {Rieke}, {Rifelli}, {Rigby}, {Riggs}, {Ringel}, {Ritchie}, {Rix}, {Robberto}, {Robinson}, {Robinson}, {Robinson}, {Rock}, {Rodriguez}, {Rodr{\'\i}guez del Pino}, {Roellig}, {Rohrbach}, {Roman}, {Romelfanger}, {Romo}, {Rosales}, {Rose}, {Roteliuk}, {Roth}, {Rothwell}, {Rouzaud}, {Rowe}, {Rowlands}, {Roy}, {Royer}, {Rui}, {Rumler}, {Rumpl}, {Russ}, {Ryan}, {Ryan}, {Saad}, {Sabata}, {Sabatino}, {Sabbi}, {Sabelhaus}, {Sabia}, {Sahu}, {Saif}, {Salvignol}, {Samara-Ratna}, {Samuelson}, {Sanders}, {Sappington}, {Sargent}, {Sauer}, {Savadkin}, {Sawicki}, {Schappell}, {Scheffer}, {Scheithauer}, {Scherer}, {Schiff}, {Schlawin}, {Schmeitzky}, {Schmitz}, {Schmude}, {Schneider}, {Schreiber}, {Schroeven-Deceuninck}, {Schultz}, {Schwab}, {Schwartz}, {Scoccimarro}, {Scott}, {Scott}, {Seaton}, {Seely}, {Seery}, {Seidleck}, {Sembach}, {Shanahan}, {Shaughnessy}, {Shaw}, {Shay}, {Sheehan}, {Sheth}, {Shih}, {Shivaei}, {Siegel}, {Sienkiewicz}, {Simmons}, {Simon},
  {Sirianni}, {Sivaramakrishnan}, {Slade}, {Sloan}, {Slocum}, {Slowinski}, {Smith}, {Smith}, {Smith}, {Smith}, {Smith}, {Smith}, {Smolik}, {Soderblom}, {Sohn}, {Sokol}, {Sonneborn}, {Sontag}, {Sooy}, {Soummer}, {Southwood}, {Spain}, {Sparmo}, {Speer}, {Spencer}, {Sprofera}, {Stallcup}, {Stanley}, {Stansberry}, {Stark}, {Starr}, {Stassi}, {Steck}, {Steeley}, {Stephens}, {Stephenson}, {Stewart}, {Stiavelli}, {}, {Strada}, {Straughn}, {Streetman}, {Strickland}, {Strobele}, {Stuhlinger}, {Stys}, {Such}, {Sukhatme}, {Sullivan}, {Sullivan}, {Sumner}, {Sun}, {Sunnquist}, {Swade}, {Swam}, {Swenton}, {Swoish}, {Tam Litten}, {Tamas}, {Tao}, {Taylor}, {Taylor}, {te Plate}, {Van Tea}, {Teague}, {Telfer}, {Temim}, {Texter}, {Thatte}, {Thompson}, {Thompson}, {Thomson}, {Thronson}, {Tierney}, {Tikkanen}, {Tinnin}, {Tippet}, {Todd}, {Tran}, {Trauger}, {Trejo}, {Vinh Truong}, {Tsukamoto}, {Tufail}, {Tumlinson}, {Tustain}, {Tyra}, {Ubeda}, {Underwood}, {Uzzo}, {Vaclavik}, {Valenduc}, {Valenti}, {Van Campen}, {van de Wetering},
  {Van Der Marel}, {van Haarlem}, {Vandenbussche}, {van Dishoeck}, {Vanterpool}, {Vernoy}, {Vila Costas}, {Volk}, {Voorzaat}, {Voyton}, {Vydra}, {Waddy}, {Waelkens}, {Wahlgren}, {Walker}, {Wander}, {Warfield}, {Warner}, {Wasiak}, {Wasiak}, {Wehner}, {Weiler}, {Weilert}, {Weiss}, {Wells}, {Welty}, {Wheate}, {Wheeler}, {White}, {Whitehouse}, {Whiteleather}, {Whitman}, {Williams}, {Willmer}, {Willott}, {Willoughby}, {Wilson}, {Wilson}, {Wilson}, {Windhorst}, {Wislowski}, {Wolfe}, {Wolfe}, {Wolff}, {Wondel}, {Woo}, {Woods}, {Worden}, {Workman}, {Wright}, {Wu}, {Wu}, {Wun}, {Wymer}, {Yadetie}, {Yan}, {Yang}, {Yates}, {Yeager}, {Yerger}, {Young}, {Young}, {Yu}, {Yu}, {Zak}, {Zeidler}, {Zepp}, {Zhou}, {Zincke}, {Zonak}, \& {Zondag}}]{Gardner2023}
{Gardner}, J.~P., {Mather}, J.~C., {Abbott}, R., {et~al.} 2023, \pasp, 135, 068001

\bibitem[{{Gilda} {et~al.}(2021){Gilda}, {Lower}, \& {Narayanan}}]{Gilda2021}
{Gilda}, S., {Lower}, S., \& {Narayanan}, D. 2021, \apj, 916, 43

\bibitem[{{Gim{\'e}nez-Arteaga} {et~al.}(2023){Gim{\'e}nez-Arteaga}, {Oesch}, {Brammer}, {Valentino}, {Mason}, {Weibel}, {Barrufet}, {Fujimoto}, {Heintz}, {Nelson}, {Strait}, {Suess}, \& {Gibson}}]{Gimenez-Arteaga+23}
{Gim{\'e}nez-Arteaga}, C., {Oesch}, P.~A., {Brammer}, G.~B., {et~al.} 2023, \apj, 948, 126

\bibitem[{{Girardi} {et~al.}(2000){Girardi}, {Bressan}, {Bertelli}, \& {Chiosi}}]{Girardi2000}
{Girardi}, L., {Bressan}, A., {Bertelli}, G., \& {Chiosi}, C. 2000, \aaps, 141, 371

\bibitem[{{Groves} {et~al.}(2008){Groves}, {Dopita}, {Sutherland}, {Kewley}, {Fischera}, {Leitherer}, {Brandl}, \& {van Breugel}}]{2008Groves}
{Groves}, B., {Dopita}, M.~A., {Sutherland}, R.~S., {et~al.} 2008, \apjs, 176, 438

\bibitem[{{Hahn} \& {Melchior}(2022)}]{Hahn2022}
{Hahn}, C. \& {Melchior}, P. 2022, \apj, 938, 11

\bibitem[{{Han} \& {Han}(2014)}]{Han2014}
{Han}, Y. \& {Han}, Z. 2014, \apjs, 215, 2

\bibitem[{{Hunt} {et~al.}(2025){Hunt}, {Annibali}, {Cuillandre}, {Ferguson}, {Jablonka}, {Larsen}, {Marleau}, {Schinnerer}, {Schirmer}, {Stone}, {Tortora}, {Saifollahi}, {Lan{\c{c}}on}, {Bolzonella}, {Gwyn}, {Kluge}, {Laureijs}, {Carollo}, {Collins}, {Dimauro}, {Duc}, {Erkal}, {Howell}, {Nally}, {Saremi}, {Scaramella}, {Belokurov}, {Conselice}, {Knapen}, {McConnachie}, {McDonald}, {Miro Carretero}, {Roman}, {Sauvage}, {Sola}, {Aghanim}, {Altieri}, {Andreon}, {Auricchio}, {Awan}, {Azzollini}, {Baldi}, {Balestra}, {Bardelli}, {Basset}, {Bender}, {Bonino}, {Branchini}, {Brescia}, {Brinchmann}, {Camera}, {Candini}, {Capobianco}, {Carbone}, {Carretero}, {Casas}, {Castellano}, {Cavuoti}, {Cimatti}, {Congedo}, {Conversi}, {Copin}, {Corcione}, {Courbin}, {Courtois}, {Cropper}, {Da Silva}, {Degaudenzi}, {De Lucia}, {Di Giorgio}, {Dinis}, {Dubath}, {Dupac}, {Dusini}, {Farina}, {Farrens}, {Ferriol}, {Fosalba}, {Frailis}, {Franceschi}, {Fumana}, {Galeotta}, {Garilli}, {George}, {Gillard}, {Gillis}, {Giocoli},
  {G{\'o}mez-Alvarez}, {Granett}, {Grazian}, {Grupp}, {Guzzo}, {Haugan}, {Hoar}, {Hoekstra}, {Holliman}, {Holmes}, {Hook}, {Hormuth}, {Hornstrup}, {Hudelot}, {Jahnke}, {Keih{\"a}nen}, {Kermiche}, {Kiessling}, {Kilbinger}, {Kitching}, {Kohley}, {Kubik}, {Kuijken}, {K{\"u}mmel}, {Kunz}, {Kurki-Suonio}, {Lahav}, {Le Mignant}, {Lilje}, {Lindholm}, {Lloro}, {Maiorano}, {Mansutti}, {Marggraf}, {Markovic}, {Martinet}, {Marulli}, {Massey}, {Maurogordato}, {McCracken}, {Medinaceli}, {Mei}, {Mellier}, {Meneghetti}, {Merlin}, {Meylan}, {Moresco}, {Moscardini}, {Munari}, {Nakajima}, {Nichol}, {Niemi}, {Nightingale}, {Padilla}, {Paltani}, {Pasian}, {Pedersen}, {Percival}, {Pettorino}, {Pires}, {Polenta}, {Poncet}, {Popa}, {Pozzetti}, {Racca}, {Raison}, {Rebolo}, {Refregier}, {Renzi}, {Rhodes}, {Riccio}, {Romelli}, {Roncarelli}, {Rossetti}, {Saglia}, {Sapone}, {Sartoris}, {Schneider}, {Schrabback}, {Scodeggio}, {Secroun}, {Seidel}, {Serrano}, {Sirignano}, {Sirri}, {Skottfelt}, {Stanco}, {Tallada-Cresp{\'\i}}, {Tavagnacco},
  {Taylor}, {Teplitz}, {Tereno}, {Toledo-Moreo}, {Torradeflot}, {Tutusaus}, {Valentijn}, {Valenziano}, {Vassallo}, {Verdoes Kleijn}, {Veropalumbo}, {Wang}, {Weller}, {Williams}, {Zamorani}, {Zucca}, {Burigana}, {Scottez}, {Miluzio}, {Simon}, {Mora}, {Mart{\'\i}n-Fleitas}, \& {Scott}}]{Hunt+24_showcase}
{Hunt}, L.~K., {Annibali}, F., {Cuillandre}, J.~C., {et~al.} 2025, \aap, 697, A9

\bibitem[{{Ibata} {et~al.}(2017){Ibata}, {McConnachie}, {Cuillandre}, {Fantin}, {Haywood}, {Martin}, {Bergeron}, {Beckmann}, {Bernard}, {Bonifacio}, {Caffau}, {Carlberg}, {C{\^o}t{\'e}}, {Cabanac}, {Chapman}, {Duc}, {Durret}, {Famaey}, {Fabbro}, {Gwyn}, {Hammer}, {Hill}, {Hudson}, {Lan{\c{c}}on}, {Lewis}, {Malhan}, {di Matteo}, {McCracken}, {Mei}, {Mellier}, {Navarro}, {Pires}, {Pritchet}, {Reyl{\'e}}, {Richer}, {Robin}, {S{\'a}nchez-Janssen}, {Sawicki}, {Scott}, {Scottez}, {Spekkens}, {Starkenburg}, {Thomas}, \& {Venn}}]{Ibata2017}
{Ibata}, R.~A., {McConnachie}, A., {Cuillandre}, J.-C., {et~al.} 2017, \apj, 848, 128

\bibitem[{{Ivezi{\'c}} {et~al.}(2019){Ivezi{\'c}}, {Kahn}, {Tyson}, {Abel}, {Acosta}, {Allsman}, {Alonso}, {AlSayyad}, {Anderson}, {Andrew}, {Angel}, {Angeli}, {Ansari}, {Antilogus}, {Araujo}, {Armstrong}, {Arndt}, {Astier}, {Aubourg}, {Auza}, {Axelrod}, {Bard}, {Barr}, {Barrau}, {Bartlett}, {Bauer}, {Bauman}, {Baumont}, {Bechtol}, {Bechtol}, {Becker}, {Becla}, {Beldica}, {Bellavia}, {Bianco}, {Biswas}, {Blanc}, {Blazek}, {Blandford}, {Bloom}, {Bogart}, {Bond}, {Booth}, {Borgland}, {Borne}, {Bosch}, {Boutigny}, {Brackett}, {Bradshaw}, {Brandt}, {Brown}, {Bullock}, {Burchat}, {Burke}, {Cagnoli}, {Calabrese}, {Callahan}, {Callen}, {Carlin}, {Carlson}, {Chandrasekharan}, {Charles-Emerson}, {Chesley}, {Cheu}, {Chiang}, {Chiang}, {Chirino}, {Chow}, {Ciardi}, {Claver}, {Cohen-Tanugi}, {Cockrum}, {Coles}, {Connolly}, {Cook}, {Cooray}, {Covey}, {Cribbs}, {Cui}, {Cutri}, {Daly}, {Daniel}, {Daruich}, {Daubard}, {Daues}, {Dawson}, {Delgado}, {Dellapenna}, {de Peyster}, {de Val-Borro}, {Digel}, {Doherty}, {Dubois},
  {Dubois-Felsmann}, {Durech}, {Economou}, {Eifler}, {Eracleous}, {Emmons}, {Fausti Neto}, {Ferguson}, {Figueroa}, {Fisher-Levine}, {Focke}, {Foss}, {Frank}, {Freemon}, {Gangler}, {Gawiser}, {Geary}, {Gee}, {Geha}, {Gessner}, {Gibson}, {Gilmore}, {Glanzman}, {Glick}, {Goldina}, {Goldstein}, {Goodenow}, {Graham}, {Gressler}, {Gris}, {Guy}, {Guyonnet}, {Haller}, {Harris}, {Hascall}, {Haupt}, {Hernandez}, {Herrmann}, {Hileman}, {Hoblitt}, {Hodgson}, {Hogan}, {Howard}, {Huang}, {Huffer}, {Ingraham}, {Innes}, {Jacoby}, {Jain}, {Jammes}, {Jee}, {Jenness}, {Jernigan}, {Jevremovi{\'c}}, {Johns}, {Johnson}, {Johnson}, {Jones}, {Juramy-Gilles}, {Juri{\'c}}, {Kalirai}, {Kallivayalil}, {Kalmbach}, {Kantor}, {Karst}, {Kasliwal}, {Kelly}, {Kessler}, {Kinnison}, {Kirkby}, {Knox}, {Kotov}, {Krabbendam}, {Krughoff}, {Kub{\'a}nek}, {Kuczewski}, {Kulkarni}, {Ku}, {Kurita}, {Lage}, {Lambert}, {Lange}, {Langton}, {Le Guillou}, {Levine}, {Liang}, {Lim}, {Lintott}, {Long}, {Lopez}, {Lotz}, {Lupton}, {Lust}, {MacArthur}, {Mahabal},
  {Mandelbaum}, {Markiewicz}, {Marsh}, {Marshall}, {Marshall}, {May}, {McKercher}, {McQueen}, {Meyers}, {Migliore}, {Miller}, {Mills}, {Miraval}, {Moeyens}, {Moolekamp}, {Monet}, {Moniez}, {Monkewitz}, {Montgomery}, {Morrison}, {Mueller}, {Muller}, {Mu{\~n}oz Arancibia}, {Neill}, {Newbry}, {Nief}, {Nomerotski}, {Nordby}, {O'Connor}, {Oliver}, {Olivier}, {Olsen}, {O'Mullane}, {Ortiz}, {Osier}, {Owen}, {Pain}, {Palecek}, {Parejko}, {Parsons}, {Pease}, {Peterson}, {Peterson}, {Petravick}, {Libby Petrick}, {Petry}, {Pierfederici}, {Pietrowicz}, {Pike}, {Pinto}, {Plante}, {Plate}, {Plutchak}, {Price}, {Prouza}, {Radeka}, {Rajagopal}, {Rasmussen}, {Regnault}, {Reil}, {Reiss}, {Reuter}, {Ridgway}, {Riot}, {Ritz}, {Robinson}, {Roby}, {Roodman}, {Rosing}, {Roucelle}, {Rumore}, {Russo}, {Saha}, {Sassolas}, {Schalk}, {Schellart}, {Schindler}, {Schmidt}, {Schneider}, {Schneider}, {Schoening}, {Schumacher}, {Schwamb}, {Sebag}, {Selvy}, {Sembroski}, {Seppala}, {Serio}, {Serrano}, {Shaw}, {Shipsey}, {Sick}, {Silvestri},
  {Slater}, {Smith}, {Smith}, {Sobhani}, {Soldahl}, {Storrie-Lombardi}, {Stover}, {Strauss}, {Street}, {Stubbs}, {Sullivan}, {Sweeney}, {Swinbank}, {Szalay}, {Takacs}, {Tether}, {Thaler}, {Thayer}, {Thomas}, {Thornton}, {Thukral}, {Tice}, {Trilling}, {Turri}, {Van Berg}, {Vanden Berk}, {Vetter}, {Virieux}, {Vucina}, {Wahl}, {Walkowicz}, {Walsh}, {Walter}, {Wang}, {Wang}, {Warner}, {Wiecha}, {Willman}, {Winters}, {Wittman}, {Wolff}, {Wood-Vasey}, {Wu}, {Xin}, {Yoachim}, \& {Zhan}}]{Ivezic2019}
{Ivezi{\'c}}, {\v{Z}}., {Kahn}, S.~M., {Tyson}, J.~A., {et~al.} 2019, \apj, 873, 111

\bibitem[{{Jones} {et~al.}(2017){Jones}, {K{\"o}hler}, {Ysard}, {Bocchio}, \& {Verstraete}}]{Jones2017}
{Jones}, A.~P., {K{\"o}hler}, M., {Ysard}, N., {Bocchio}, M., \& {Verstraete}, L. 2017, \aap, 602, A46

\bibitem[{{Kodama} \& {Arimoto}(1997)}]{Kodama_Arimoto1997}
{Kodama}, T. \& {Arimoto}, N. 1997, \aap, 320, 41

\bibitem[{{Kroupa} \& {Boily}(2002)}]{Kroupa2002}
{Kroupa}, P. \& {Boily}, C.~M. 2002, \mnras, 336, 1188

\bibitem[{{Lacerda} {et~al.}(2022){Lacerda}, {S{\'a}nchez}, {Mej{\'\i}a-Narv{\'a}ez}, {Camps-Fari{\~n}a}, {Espinosa-Ponce}, {Barrera-Ballesteros}, {Ibarra-Medel}, \& {Lugo-Aranda}}]{2022Lacerda}
{Lacerda}, E. A.~D., {S{\'a}nchez}, S.~F., {Mej{\'\i}a-Narv{\'a}ez}, A., {et~al.} 2022, \na, 97, 101895

\bibitem[{{Laureijs} {et~al.}(2011){Laureijs}, {Amiaux}, {Arduini}, {Augu{\`e}res}, {Brinchmann}, {Cole}, {Cropper}, {Dabin}, {Duvet}, {Ealet}, {Garilli}, {Gondoin}, {Guzzo}, {Hoar}, {Hoekstra}, {Holmes}, {Kitching}, {Maciaszek}, {Mellier}, {Pasian}, {Percival}, {Rhodes}, {Saavedra Criado}, {Sauvage}, {Scaramella}, {Valenziano}, {Warren}, {Bender}, {Castander}, {Cimatti}, {Le F{\`e}vre}, {Kurki-Suonio}, {Levi}, {Lilje}, {Meylan}, {Nichol}, {Pedersen}, {Popa}, {Rebolo Lopez}, {Rix}, {Rottgering}, {Zeilinger}, {Grupp}, {Hudelot}, {Massey}, {Meneghetti}, {Miller}, {Paltani}, {Paulin-Henriksson}, {Pires}, {Saxton}, {Schrabback}, {Seidel}, {Walsh}, {Aghanim}, {Amendola}, {Bartlett}, {Baccigalupi}, {Beaulieu}, {Benabed}, {Cuby}, {Elbaz}, {Fosalba}, {Gavazzi}, {Helmi}, {Hook}, {Irwin}, {Kneib}, {Kunz}, {Mannucci}, {Moscardini}, {Tao}, {Teyssier}, {Weller}, {Zamorani}, {Zapatero Osorio}, {Boulade}, {Foumond}, {Di Giorgio}, {Guttridge}, {James}, {Kemp}, {Martignac}, {Spencer}, {Walton}, {Bl{\"u}mchen}, {Bonoli},
  {Bortoletto}, {Cerna}, {Corcione}, {Fabron}, {Jahnke}, {Ligori}, {Madrid}, {Martin}, {Morgante}, {Pamplona}, {Prieto}, {Riva}, {Toledo}, {Trifoglio}, {Zerbi}, {Abdalla}, {Douspis}, {Grenet}, {Borgani}, {Bouwens}, {Courbin}, {Delouis}, {Dubath}, {Fontana}, {Frailis}, {Grazian}, {Koppenh{\"o}fer}, {Mansutti}, {Melchior}, {Mignoli}, {Mohr}, {Neissner}, {Noddle}, {Poncet}, {Scodeggio}, {Serrano}, {Shane}, {Starck}, {Surace}, {Taylor}, {Verdoes-Kleijn}, {Vuerli}, {Williams}, {Zacchei}, {Altieri}, {Escudero Sanz}, {Kohley}, {Oosterbroek}, {Astier}, {Bacon}, {Bardelli}, {Baugh}, {Bellagamba}, {Benoist}, {Bianchi}, {Biviano}, {Branchini}, {Carbone}, {Cardone}, {Clements}, {Colombi}, {Conselice}, {Cresci}, {Deacon}, {Dunlop}, {Fedeli}, {Fontanot}, {Franzetti}, {Giocoli}, {Garcia-Bellido}, {Gow}, {Heavens}, {Hewett}, {Heymans}, {Holland}, {Huang}, {Ilbert}, {Joachimi}, {Jennins}, {Kerins}, {Kiessling}, {Kirk}, {Kotak}, {Krause}, {Lahav}, {van Leeuwen}, {Lesgourgues}, {Lombardi}, {Magliocchetti}, {Maguire},
  {Majerotto}, {Maoli}, {Marulli}, {Maurogordato}, {McCracken}, {McLure}, {Melchiorri}, {Merson}, {Moresco}, {Nonino}, {Norberg}, {Peacock}, {Pello}, {Penny}, {Pettorino}, {Di Porto}, {Pozzetti}, {Quercellini}, {Radovich}, {Rassat}, {Roche}, {Ronayette}, {Rossetti}, {Sartoris}, {Schneider}, {Semboloni}, {Serjeant}, {Simpson}, {Skordis}, {Smadja}, {Smartt}, {Spano}, {Spiro}, {Sullivan}, {Tilquin}, {Trotta}, {Verde}, {Wang}, {Williger}, {Zhao}, {Zoubian}, \& {Zucca}}]{Laureijs2011}
{Laureijs}, R., {Amiaux}, J., {Arduini}, S., {et~al.} 2011, arXiv:1110.3193

\bibitem[{{Leja} {et~al.}(2019){Leja}, {Johnson}, {Conroy}, {van Dokkum}, {Speagle}, {Brammer}, {Momcheva}, {Skelton}, {Whitaker}, {Franx}, \& {Nelson}}]{Leja2019}
{Leja}, J., {Johnson}, B.~D., {Conroy}, C., {et~al.} 2019, \apj, 877, 140

\bibitem[{{Leja} {et~al.}(2017){Leja}, {Johnson}, {Conroy}, {van Dokkum}, \& {Byler}}]{Leja2017}
{Leja}, J., {Johnson}, B.~D., {Conroy}, C., {van Dokkum}, P.~G., \& {Byler}, N. 2017, \apj, 837, 170

\bibitem[{{Lovell} {et~al.}(2019){Lovell}, {Acquaviva}, {Thomas}, {Iyer}, {Gawiser}, \& {Wilkins}}]{Lovell2019}
{Lovell}, C.~C., {Acquaviva}, V., {Thomas}, P.~A., {et~al.} 2019, \mnras, 490, 5503

\bibitem[{{Maciaszek} {et~al.}(2022){Maciaszek}, {Ealet}, {Gillard}, {Jahnke}, {Barbier}, {Prieto}, {Bon}, {Bonnefoi}, {Caillat}, {Carle}, {Costille}, {Ducret}, {Fabron}, {Foulon}, {Gimenez}, {Grassi}, {Jaquet}, {Le Mignant}, {Martin}, {Pamplona}, {Sanchez}, {Cl{\'e}mens}, {Caillat}, {Niclas}, {Secroun}, {Kubik}, {Ferriol}, {Berthe}, {Barri{\`e}re}, {Fontignie}, {Valenziano}, {Auricchio}, {Battaglia}, {De Rosa}, {Farinelli}, {Franceschi}, {Medinaceli}, {Morgante}, {Sortino}, {Trifoglio}, {Corcione}, {Capobianco}, {Ligori}, {Dusini}, {Borsato}, {Dal Corso}, {Laudisio}, {Sirignano}, {Stanco}, {Ventura}, {Patrizii}, {Chiarusi}, {Fornari}, {Giacomini}, {Margiotta}, {Mauri}, {Pasqualini}, {Sirri}, {Spurio}, {Tenti}, {Travaglini}, {Bonoli}, {Bortoletto}, {Balestra}, {Dalessandro}, {Grupp}, {Penka}, {Steinwagner}, {Hormuth}, {Schirmer}, {Seidel}, {Padilla}, {Casas}, {Lloro}, {Toledo-Moreo}, {Gomez}, {Colodro-Conde}, {Liz{\'a}n}, {Diaz}, {Lilje}, {Andersen}, {Andersen}, {S{\o}rensen}, {Hornstrup}, {Jessen}, {Thizy},
  {Holmes}, {Pniel}, {Jhabvala}, {Pravdo}, {Seiffert}, {Waczynski}, {Laureij}, {Racca}, {Salvignol}, {Boenke}, {Strada}, \& {Mellier}}]{2022Maciaszek}
{Maciaszek}, T., {Ealet}, A., {Gillard}, W., {et~al.} 2022, in Society of Photo-Optical Instrumentation Engineers (SPIE) Conference Series, Vol. 12180, Space Telescopes and Instrumentation 2022: Optical, Infrared, and Millimeter Wave, ed. L.~E. {Coyle}, S.~{Matsuura}, \& M.~D. {Perrin}, 121801K

\bibitem[{{Marigo} \& {Girardi}(2007)}]{Marigo2007}
{Marigo}, P. \& {Girardi}, L. 2007, \aap, 469, 239

\bibitem[{{Marigo} {et~al.}(2008){Marigo}, {Girardi}, {Bressan}, {Groenewegen}, {Silva}, \& {Granato}}]{Marigo2008}
{Marigo}, P., {Girardi}, L., {Bressan}, A., {et~al.} 2008, \aap, 482, 883

\bibitem[{{Marinacci} {et~al.}(2018){Marinacci}, {Vogelsberger}, {Pakmor}, {Torrey}, {Springel}, {Hernquist}, {Nelson}, {Weinberger}, {Pillepich}, {Naiman}, \& {Genel}}]{Marinacci2018}
{Marinacci}, F., {Vogelsberger}, M., {Pakmor}, R., {et~al.} 2018, \mnras, 480, 5113

\bibitem[{{Marleau} {et~al.}(2025){Marleau}, {Cuillandre}, {Cantiello}, {Carollo}, {Duc}, {Habas}, {Hunt}, {Jablonka}, {Mirabile}, {Mondelin}, {Poulain}, {Saifollahi}, {S{\'a}nchez-Janssen}, {Sola}, {Urbano}, {Z{\"o}ller}, {Bolzonella}, {Lan{\c{c}}on}, {Laureijs}, {Marchal}, {Schirmer}, {Stone}, {Boselli}, {Ferr{\'e}-Mateu}, {Hatch}, {Kluge}, {Montes}, {Sorce}, {Tortora}, {Venhola}, {Golden-Marx}, {Aghanim}, {Amara}, {Andreon}, {Auricchio}, {Baccigalupi}, {Baldi}, {Balestra}, {Bardelli}, {Battaglia}, {Bender}, {Bodendorf}, {Branchini}, {Brescia}, {Brinchmann}, {Camera}, {Candini}, {Capobianco}, {Carbone}, {Carretero}, {Casas}, {Castellano}, {Cavuoti}, {Cimatti}, {Congedo}, {Conselice}, {Conversi}, {Copin}, {Courbin}, {Courtois}, {Cropper}, {Da Silva}, {Degaudenzi}, {De Lucia}, {Di Giorgio}, {Dinis}, {Douspis}, {Duncan}, {Dupac}, {Dusini}, {Ealet}, {Farina}, {Farrens}, {Ferriol}, {Fosalba}, {Fotopoulou}, {Frailis}, {Franceschi}, {Fumana}, {Galeotta}, {Garilli}, {George}, {Gillard}, {Gillis}, {Giocoli},
  {G{\'o}mez-Alvarez}, {Grazian}, {Grupp}, {Guzzo}, {Hailey}, {Haugan}, {Hoar}, {Hoekstra}, {Holmes}, {Hook}, {Hormuth}, {Hornstrup}, {Hu}, {Hudelot}, {Jahnke}, {Jhabvala}, {Keih{\"a}nen}, {Kermiche}, {Kiessling}, {Kitching}, {Kohley}, {Kubik}, {Kuijken}, {K{\"u}mmel}, {Kunz}, {Kurki-Suonio}, {Lahav}, {Le Mignant}, {Ligori}, {Lilje}, {Lindholm}, {Lloro}, {Maino}, {Maiorano}, {Mansutti}, {Marggraf}, {Markovic}, {Martinet}, {Marulli}, {Massey}, {Maurogordato}, {McCracken}, {Medinaceli}, {Mei}, {Mellier}, {Meneghetti}, {Merlin}, {Meylan}, {Moresco}, {Moscardini}, {Munari}, {Nakajima}, {Nichol}, {Niemi}, {Padilla}, {Paltani}, {Pasian}, {Pedersen}, {Percival}, {Pettorino}, {Pires}, {Polenta}, {Poncet}, {Popa}, {Pozzetti}, {Raison}, {Rebolo}, {Refregier}, {Renzi}, {Rhodes}, {Riccio}, {Rix}, {Romelli}, {Roncarelli}, {Rossetti}, {Saglia}, {Sapone}, {Scaramella}, {Schneider}, {Secroun}, {Seidel}, {Seiffert}, {Serrano}, {Sirignano}, {Sirri}, {Stanco}, {Tallada-Cresp{\'\i}}, {Taylor}, {Teplitz}, {Tereno},
  {Toledo-Moreo}, {Tsyganov}, {Tutusaus}, {Valentijn}, {Valenziano}, {Vassallo}, {Verdoes Kleijn}, {Veropalumbo}, {Wang}, {Weller}, {Williams}, {Zamorani}, {Zucca}, {Biviano}, {Burigana}, {Scottez}, {Viel}, {Simon}, {Mora}, {Mart{\'\i}n-Fleitas}, \& {Scott}}]{Marleau+24}
{Marleau}, F.~R., {Cuillandre}, J.~C., {Cantiello}, M., {et~al.} 2025, \aap, 697, A12

\bibitem[{{Naiman} {et~al.}(2018){Naiman}, {Pillepich}, {Springel}, {Ramirez-Ruiz}, {Torrey}, {Vogelsberger}, {Pakmor}, {Nelson}, {Marinacci}, {Hernquist}, {Weinberger}, \& {Genel}}]{Naiman2018}
{Naiman}, J.~P., {Pillepich}, A., {Springel}, V., {et~al.} 2018, \mnras, 477, 1206

\bibitem[{{Nelson} {et~al.}(2019){Nelson}, {Pillepich}, {Springel}, {Pakmor}, {Weinberger}, {Genel}, {Torrey}, {Vogelsberger}, {Marinacci}, \& {Hernquist}}]{Nelson2019}
{Nelson}, D., {Pillepich}, A., {Springel}, V., {et~al.} 2019, \mnras, 490, 3234

\bibitem[{{Nelson} {et~al.}(2018){Nelson}, {Pillepich}, {Springel}, {Weinberger}, {Hernquist}, {Pakmor}, {Genel}, {Torrey}, {Vogelsberger}, {Kauffmann}, {Marinacci}, \& {Naiman}}]{Nelson2018}
{Nelson}, D., {Pillepich}, A., {Springel}, V., {et~al.} 2018, \mnras, 475, 624

\bibitem[{{Noeske} {et~al.}(2007){Noeske}, {Weiner}, {Faber}, {Papovich}, {Koo}, {Somerville}, {Bundy}, {Conselice}, {Newman}, {Schiminovich}, {Le Floc'h}, {Coil}, {Rieke}, {Lotz}, {Primack}, {Barmby}, {Cooper}, {Davis}, {Ellis}, {Fazio}, {Guhathakurta}, {Huang}, {Kassin}, {Martin}, {Phillips}, {Rich}, {Small}, {Willmer}, \& {Wilson}}]{Noeske2007}
{Noeske}, K.~G., {Weiner}, B.~J., {Faber}, S.~M., {et~al.} 2007, \apjl, 660, L43

\bibitem[{{O'Connor}(2019)}]{OConnor2019}
{O'Connor}, P. 2019, Journal of Astronomical Telescopes, Instruments, and Systems, 5, 041508

\bibitem[{{Pacifici} {et~al.}(2023){Pacifici}, {Iyer}, {Mobasher}, {da Cunha}, {Acquaviva}, {Burgarella}, {Calistro Rivera}, {Carnall}, {Chang}, {Chartab}, {Cooke}, {Fairhurst}, {Kartaltepe}, {Leja}, {Ma{\l}ek}, {Salmon}, {Torelli}, {Vidal-Garc{\'\i}a}, {Boquien}, {Brammer}, {Brown}, {Capak}, {Chevallard}, {Circosta}, {Croton}, {Davidzon}, {Dickinson}, {Duncan}, {Faber}, {Ferguson}, {Fontana}, {Guo}, {Haeussler}, {Hemmati}, {Jafariyazani}, {Kassin}, {Larson}, {Lee}, {Mantha}, {Marchi}, {Nayyeri}, {Newman}, {Pandya}, {Pforr}, {Reddy}, {Sanders}, {Shah}, {Shahidi}, {Stevans}, {Triani}, {Tyler}, {Vanderhoof}, {de la Vega}, {Wang}, \& {Weston}}]{Pacifici2023}
{Pacifici}, C., {Iyer}, K.~G., {Mobasher}, B., {et~al.} 2023, \apj, 944, 141

\bibitem[{{Peterson} {et~al.}(2015){Peterson}, {Jernigan}, {Kahn}, {Rasmussen}, {Peng}, {Ahmad}, {Bankert}, {Chang}, {Claver}, {Gilmore}, {Grace}, {Hannel}, {Hodge}, {Lorenz}, {Lupu}, {Meert}, {Nagarajan}, {Todd}, {Winans}, \& {Young}}]{Peterson2015}
{Peterson}, J.~R., {Jernigan}, J.~G., {Kahn}, S.~M., {et~al.} 2015, \apjs, 218, 14

\bibitem[{{Pforr} {et~al.}(2012){Pforr}, {Maraston}, \& {Tonini}}]{Pforr2012}
{Pforr}, J., {Maraston}, C., \& {Tonini}, C. 2012, \mnras, 422, 3285

\bibitem[{{Pillepich} {et~al.}(2019){Pillepich}, {Nelson}, {Springel}, {Pakmor}, {Torrey}, {Weinberger}, {Vogelsberger}, {Marinacci}, {Genel}, {van der Wel}, \& {Hernquist}}]{Pillepich2019}
{Pillepich}, A., {Nelson}, D., {Springel}, V., {et~al.} 2019, \mnras, 490, 3196

\bibitem[{{Popesso} {et~al.}(2023){Popesso}, {Concas}, {Cresci}, {Belli}, {Rodighiero}, {Inami}, {Dickinson}, {Ilbert}, {Pannella}, \& {Elbaz}}]{2023Popesso}
{Popesso}, P., {Concas}, A., {Cresci}, G., {et~al.} 2023, \mnras, 519, 1526

\bibitem[{{Robotham} {et~al.}(2022){Robotham}, {Bellstedt}, \& {Driver}}]{2022Robotham}
{Robotham}, A.~S.~G., {Bellstedt}, S., \& {Driver}, S.~P. 2022, \mnras, 513, 2985

\bibitem[{{R{\"o}ck} {et~al.}(2016){R{\"o}ck}, {Vazdekis}, {Ricciardelli}, {Peletier}, {Knapen}, \& {Falc{\'o}n-Barroso}}]{Rock+16}
{R{\"o}ck}, B., {Vazdekis}, A., {Ricciardelli}, E., {et~al.} 2016, \aap, 589, A73

\bibitem[{{Saifollahi} {et~al.}(2025){Saifollahi}, {Voggel}, {Lan{\c{c}}on}, {Cantiello}, {Raj}, {Cuillandre}, {Larsen}, {Marleau}, {Venhola}, {Schirmer}, {Carollo}, {Duc}, {Ferguson}, {Hunt}, {K{\"u}mmel}, {Laureijs}, {Marchal}, {Nucita}, {Peletier}, {Poulain}, {Rejkuba}, {S{\'a}nchez-Janssen}, {Urbano}, {Abdurro'uf}, {Altieri}, {Baes}, {Bolzonella}, {Conselice}, {Cote}, {Dimauro}, {Gonzalez}, {Habas}, {Hudelot}, {Kluge}, {Lonare}, {Massari}, {Romelli}, {Scaramella}, {Sola}, {Stone}, {Tortora}, {van Mierlo}, {Knapen}, {Mart{\'\i}n-Fleitas}, {Mora}, {Rom{\'a}n}, {Aghanim}, {Amara}, {Andreon}, {Auricchio}, {Baldi}, {Balestra}, {Bardelli}, {Basset}, {Bender}, {Bonino}, {Branchini}, {Brescia}, {Brinchmann}, {Camera}, {Capobianco}, {Carbone}, {Carretero}, {Casas}, {Castellano}, {Cavuoti}, {Cimatti}, {Congedo}, {Conversi}, {Copin}, {Courbin}, {Courtois}, {Cropper}, {Da Silva}, {Degaudenzi}, {Di Giorgio}, {Dinis}, {Dubath}, {Dupac}, {Dusini}, {Fabricius}, {Farina}, {Farrens}, {Ferriol}, {Fosalba}, {Frailis},
  {Franceschi}, {Fumana}, {Galeotta}, {Garilli}, {Gillard}, {Gillis}, {Giocoli}, {G{\'o}mez-Alvarez}, {Granett}, {Grazian}, {Grupp}, {Guzzo}, {Haugan}, {Hoar}, {Hoekstra}, {Holmes}, {Hook}, {Hormuth}, {Hornstrup}, {Jahnke}, {Jhabvala}, {Keih{\"a}nen}, {Kermiche}, {Kiessling}, {Kitching}, {Kohley}, {Kubik}, {Kuijken}, {Kunz}, {Kurki-Suonio}, {Lahav}, {Le Mignant}, {Ligori}, {Lilje}, {Lindholm}, {Lloro}, {Maino}, {Maiorano}, {Mansutti}, {Marggraf}, {Markovic}, {Martinet}, {Marulli}, {Massey}, {Maurogordato}, {McCracken}, {Medinaceli}, {Mei}, {Melchior}, {Mellier}, {Meneghetti}, {Meylan}, {Moresco}, {Moscardini}, {Munari}, {Nakajima}, {Nichol}, {Niemi}, {Padilla}, {Paltani}, {Pasian}, {Pedersen}, {Percival}, {Pettorino}, {Pires}, {Polenta}, {Poncet}, {Popa}, {Pozzetti}, {Racca}, {Raison}, {Rebolo}, {Refregier}, {Renzi}, {Rhodes}, {Riccio}, {Roncarelli}, {Rossetti}, {Saglia}, {Sapone}, {Sartoris}, {Schneider}, {Schrabback}, {Secroun}, {Seidel}, {Serrano}, {Sirignano}, {Sirri}, {Stanco}, {Tallada-Cresp{\'\i}},
  {Taylor}, {Teplitz}, {Tereno}, {Toledo-Moreo}, {Torradeflot}, {Tsyganov}, {Tutusaus}, {Valentijn}, {Valenziano}, {Vassallo}, {Verdoes Kleijn}, {Veropalumbo}, {Wang}, {Weller}, {Williams}, {Zamorani}, {Zucca}, {Biviano}, {Burigana}, {Scottez}, {Simon}, {Balogh}, \& {Scott}}]{Saifollahi+24}
{Saifollahi}, T., {Voggel}, K., {Lan{\c{c}}on}, A., {et~al.} 2025, \aap, 697, A10

\bibitem[{{Salucci} {et~al.}(2008){Salucci}, {Yegorova}, \& {Drory}}]{Salucci+08}
{Salucci}, P., {Yegorova}, I.~A., \& {Drory}, N. 2008, \mnras, 388, 159

\bibitem[{{S{\'a}nchez}(2020)}]{Sanchez2020}
{S{\'a}nchez}, S.~F. 2020, \araa, 58, 99

\bibitem[{{S{\'a}nchez} {et~al.}(2022){S{\'a}nchez}, {Barrera-Ballesteros}, {Lacerda}, {Mej{\'\i}a-Narvaez}, {Camps-Fari{\~n}a}, {Bruzual}, {Espinosa-Ponce}, {Rodr{\'\i}guez-Puebla}, {Calette}, {Ibarra-Medel}, {Avila-Reese}, {Hernandez-Toledo}, {Bershady}, {Cano-Diaz}, \& {Munguia-Cordova}}]{Sanchez2022}
{S{\'a}nchez}, S.~F., {Barrera-Ballesteros}, J.~K., {Lacerda}, E., {et~al.} 2022, \apjs, 262, 36

\bibitem[{{S{\'a}nchez} {et~al.}(2012){S{\'a}nchez}, {Kennicutt}, {Gil de Paz}, {van de Ven}, {V{\'\i}lchez}, {Wisotzki}, {Walcher}, {Mast}, {Aguerri}, {Albiol-P{\'e}rez}, {Alonso-Herrero}, {Alves}, {Bakos}, {Bart{\'a}kov{\'a}}, {Bland-Hawthorn}, {Boselli}, {Bomans}, {Castillo-Morales}, {Cortijo-Ferrero}, {de Lorenzo-C{\'a}ceres}, {Del Olmo}, {Dettmar}, {D{\'\i}az}, {Ellis}, {Falc{\'o}n-Barroso}, {Flores}, {Gallazzi}, {Garc{\'\i}a-Lorenzo}, {Gonz{\'a}lez Delgado}, {Gruel}, {Haines}, {Hao}, {Husemann}, {Igl{\'e}sias-P{\'a}ramo}, {Jahnke}, {Johnson}, {Jungwiert}, {Kalinova}, {Kehrig}, {Kupko}, {L{\'o}pez-S{\'a}nchez}, {Lyubenova}, {Marino}, {M{\'a}rmol-Queralt{\'o}}, {M{\'a}rquez}, {Masegosa}, {Meidt}, {Mendez-Abreu}, {Monreal-Ibero}, {Montijo}, {Mour{\~a}o}, {Palacios-Navarro}, {Papaderos}, {Pasquali}, {Peletier}, {P{\'e}rez}, {P{\'e}rez}, {Quirrenbach}, {Rela{\~n}o}, {Rosales-Ortega}, {Roth}, {Ruiz-Lara}, {S{\'a}nchez-Bl{\'a}zquez}, {Sengupta}, {Singh}, {Stanishev}, {Trager}, {Vazdekis}, {Viironen}, {Wild},
  {Zibetti}, \& {Ziegler}}]{Sanchez2012}
{S{\'a}nchez}, S.~F., {Kennicutt}, R.~C., {Gil de Paz}, A., {et~al.} 2012, \aap, 538, A8

\bibitem[{{S{\'a}nchez-Bl{\'a}zquez} {et~al.}(2006){S{\'a}nchez-Bl{\'a}zquez}, {Peletier}, {Jim{\'e}nez-Vicente}, {Cardiel}, {Cenarro}, {Falc{\'o}n-Barroso}, {Gorgas}, {Selam}, \& {Vazdekis}}]{Sanchez-Blazquez2006}
{S{\'a}nchez-Bl{\'a}zquez}, P., {Peletier}, R.~F., {Jim{\'e}nez-Vicente}, J., {et~al.} 2006, \mnras, 371, 703

\bibitem[{{Sawicki}(2012)}]{Sawicki2012}
{Sawicki}, M. 2012, \pasp, 124, 1208

\bibitem[{{Sawicki} \& {Yee}(1998)}]{Sawicki1998}
{Sawicki}, M. \& {Yee}, H.~K.~C. 1998, \aj, 115, 1329

\bibitem[{{Secroun} {et~al.}(2018){Secroun}, {Barbier}, {Buton}, {Cl{\'e}mens}, {Conversi}, {Ealet}, {Ferriol}, {Fornari}, {Gillard}, {Kohley}, {Kubik}, {Rosset}, {Serra}, {Smadja}, \& {Zoubian}}]{Secroun2018}
{Secroun}, A., {Barbier}, R., {Buton}, C., {et~al.} 2018, in Society of Photo-Optical Instrumentation Engineers (SPIE) Conference Series, Vol. 10709, High Energy, Optical, and Infrared Detectors for Astronomy VIII, ed. A.~D. {Holland} \& J.~{Beletic}, 1070921

\bibitem[{{Sorba} \& {Sawicki}(2015)}]{2015Sorba}
{Sorba}, R. \& {Sawicki}, M. 2015, \mnras, 452, 235

\bibitem[{{Sorba} \& {Sawicki}(2018)}]{2018Sorba}
{Sorba}, R. \& {Sawicki}, M. 2018, \mnras, 476, 1532

\bibitem[{{Speagle} {et~al.}(2014){Speagle}, {Steinhardt}, {Capak}, \& {Silverman}}]{Speagle2014}
{Speagle}, J.~S., {Steinhardt}, C.~L., {Capak}, P.~L., \& {Silverman}, J.~D. 2014, \apjs, 214, 15

\bibitem[{{Spergel} {et~al.}(2015){Spergel}, {Gehrels}, {Baltay}, {Bennett}, {Breckinridge}, {Donahue}, {Dressler}, {Gaudi}, {Greene}, {Guyon}, {Hirata}, {Kalirai}, {Kasdin}, {Macintosh}, {Moos}, {Perlmutter}, {Postman}, {Rauscher}, {Rhodes}, {Wang}, {Weinberg}, {Benford}, {Hudson}, {Jeong}, {Mellier}, {Traub}, {Yamada}, {Capak}, {Colbert}, {Masters}, {Penny}, {Savransky}, {Stern}, {Zimmerman}, {Barry}, {Bartusek}, {Carpenter}, {Cheng}, {Content}, {Dekens}, {Demers}, {Grady}, {Jackson}, {Kuan}, {Kruk}, {Melton}, {Nemati}, {Parvin}, {Poberezhskiy}, {Peddie}, {Ruffa}, {Wallace}, {Whipple}, {Wollack}, \& {Zhao}}]{2015Spergel_roman}
{Spergel}, D., {Gehrels}, N., {Baltay}, C., {et~al.} 2015, arXiv e-prints, arXiv:1503.03757

\bibitem[{{Springel}(2010)}]{Springel2010}
{Springel}, V. 2010, \mnras, 401, 791

\bibitem[{{Springel} {et~al.}(2018){Springel}, {Pakmor}, {Pillepich}, {Weinberger}, {Nelson}, {Hernquist}, {Vogelsberger}, {Genel}, {Torrey}, {Marinacci}, \& {Naiman}}]{Springel2018}
{Springel}, V., {Pakmor}, R., {Pillepich}, A., {et~al.} 2018, \mnras, 475, 676

\bibitem[{{Tortora} {et~al.}(2010){Tortora}, {Napolitano}, {Cardone}, {Capaccioli}, {Jetzer}, \& {Molinaro}}]{Tortora+10CG}
{Tortora}, C., {Napolitano}, N.~R., {Cardone}, V.~F., {et~al.} 2010, \mnras, 407, 144

\bibitem[{{Tortora} {et~al.}(2011){Tortora}, {Napolitano}, {Romanowsky}, {Jetzer}, {Cardone}, \& {Capaccioli}}]{Tortora+11MtoLgrad}
{Tortora}, C., {Napolitano}, N.~R., {Romanowsky}, A.~J., {et~al.} 2011, \mnras, 418, 1557

\bibitem[{{Tremonti} {et~al.}(2004){Tremonti}, {Heckman}, {Kauffmann}, {Brinchmann}, {Charlot}, {White}, {Seibert}, {Peng}, {Schlegel}, {Uomoto}, {Fukugita}, \& {Brinkmann}}]{Tremonti2004}
{Tremonti}, C.~A., {Heckman}, T.~M., {Kauffmann}, G., {et~al.} 2004, \apj, 613, 898

\bibitem[{{Tr{\v{c}}ka} {et~al.}(2022){Tr{\v{c}}ka}, {Baes}, {Camps}, {Kapoor}, {Nelson}, {Pillepich}, {Barrientos}, {Hernquist}, {Marinacci}, \& {Vogelsberger}}]{Trcka2022}
{Tr{\v{c}}ka}, A., {Baes}, M., {Camps}, P., {et~al.} 2022, \mnras, 516, 3728

\bibitem[{{Viaene} {et~al.}(2014){Viaene}, {Fritz}, {Baes}, {Bendo}, {Blommaert}, {Boquien}, {Boselli}, {Ciesla}, {Cortese}, {De Looze}, {Gear}, {Gentile}, {Hughes}, {Jarrett}, {Karczewski}, {Smith}, {Spinoglio}, {Tamm}, {Tempel}, {Thilker}, \& {Verstappen}}]{Viaene2014}
{Viaene}, S., {Fritz}, J., {Baes}, M., {et~al.} 2014, \aap, 567, A71

\bibitem[{{Walcher} {et~al.}(2011){Walcher}, {Groves}, {Budav{\'a}ri}, \& {Dale}}]{2011Walcher}
{Walcher}, J., {Groves}, B., {Budav{\'a}ri}, T., \& {Dale}, D. 2011, \apss, 331, 1

\bibitem[{{Welikala} {et~al.}(2011){Welikala}, {Hopkins}, {Robertson}, {Connolly}, {Tasca}, {Koekemoer}, {Ilbert}, {Bardelli}, {Kneib}, \& {Zentner}}]{Welikala2011}
{Welikala}, N., {Hopkins}, A.~M., {Robertson}, B.~E., {et~al.} 2011, arXiv:1112.2657

\bibitem[{{Wisnioski} {et~al.}(2015){Wisnioski}, {F{\"o}rster Schreiber}, {Wuyts}, {Wuyts}, {Bandara}, {Wilman}, {Genzel}, {Bender}, {Davies}, {Fossati}, {Lang}, {Mendel}, {Beifiori}, {Brammer}, {Chan}, {Fabricius}, {Fudamoto}, {Kulkarni}, {Kurk}, {Lutz}, {Nelson}, {Momcheva}, {Rosario}, {Saglia}, {Seitz}, {Tacconi}, \& {van Dokkum}}]{Wisnioski2015}
{Wisnioski}, E., {F{\"o}rster Schreiber}, N.~M., {Wuyts}, S., {et~al.} 2015, \apj, 799, 209

\bibitem[{{Worthey}(1994)}]{Worthey1994}
{Worthey}, G. 1994, \apjs, 95, 107

\bibitem[{{Wuyts} {et~al.}(2012){Wuyts}, {F{\"o}rster Schreiber}, {Genzel}, {Guo}, {Barro}, {Bell}, {Dekel}, {Faber}, {Ferguson}, {Giavalisco}, {Grogin}, {Hathi}, {Huang}, {Kocevski}, {Koekemoer}, {Koo}, {Lotz}, {Lutz}, {McGrath}, {Newman}, {Rosario}, {Saintonge}, {Tacconi}, {Weiner}, \& {van der Wel}}]{2012Wuyts}
{Wuyts}, S., {F{\"o}rster Schreiber}, N.~M., {Genzel}, R., {et~al.} 2012, \apj, 753, 114

\bibitem[{{Wuyts} {et~al.}(2013){Wuyts}, {F{\"o}rster Schreiber}, {Nelson}, {van Dokkum}, {Brammer}, {Chang}, {Faber}, {Ferguson}, {Franx}, {Fumagalli}, {Genzel}, {Grogin}, {Kocevski}, {Koekemoer}, {Lundgren}, {Lutz}, {McGrath}, {Momcheva}, {Rosario}, {Skelton}, {Tacconi}, {van der Wel}, \& {Whitaker}}]{Wuyts2013}
{Wuyts}, S., {F{\"o}rster Schreiber}, N.~M., {Nelson}, E.~J., {et~al.} 2013, \apj, 779, 135

\bibitem[{{Yang} {et~al.}(2022){Yang}, {Boquien}, {Brandt}, {Buat}, {Burgarella}, {Ciesla}, {Lehmer}, {Ma{\l}ek}, {Mountrichas}, {Papovich}, {Pons}, {Stalevski}, {Theul{\'e}}, \& {Zhu}}]{Yang2022}
{Yang}, G., {Boquien}, M., {Brandt}, W.~N., {et~al.} 2022, \apj, 927, 192

\bibitem[{{York} {et~al.}(2000){York}, {Adelman}, {Anderson}, {Anderson}, {Annis}, {Bahcall}, {Bakken}, {Barkhouser}, {Bastian}, {Berman}, {Boroski}, {Bracker}, {Briegel}, {Briggs}, {Brinkmann}, {Brunner}, {Burles}, {Carey}, {Carr}, {Castander}, {Chen}, {Colestock}, {Connolly}, {Crocker}, {Csabai}, {Czarapata}, {Davis}, {Doi}, {Dombeck}, {Eisenstein}, {Ellman}, {Elms}, {Evans}, {Fan}, {Federwitz}, {Fiscelli}, {Friedman}, {Frieman}, {Fukugita}, {Gillespie}, {Gunn}, {Gurbani}, {de Haas}, {Haldeman}, {Harris}, {Hayes}, {Heckman}, {Hennessy}, {Hindsley}, {Holm}, {Holmgren}, {Huang}, {Hull}, {Husby}, {Ichikawa}, {Ichikawa}, {Ivezi{\'c}}, {Kent}, {Kim}, {Kinney}, {Klaene}, {Kleinman}, {Kleinman}, {Knapp}, {Korienek}, {Kron}, {Kunszt}, {Lamb}, {Lee}, {Leger}, {Limmongkol}, {Lindenmeyer}, {Long}, {Loomis}, {Loveday}, {Lucinio}, {Lupton}, {MacKinnon}, {Mannery}, {Mantsch}, {Margon}, {McGehee}, {McKay}, {Meiksin}, {Merelli}, {Monet}, {Munn}, {Narayanan}, {Nash}, {Neilsen}, {Neswold}, {Newberg}, {Nichol}, {Nicinski},
  {Nonino}, {Okada}, {Okamura}, {Ostriker}, {Owen}, {Pauls}, {Peoples}, {Peterson}, {Petravick}, {Pier}, {Pope}, {Pordes}, {Prosapio}, {Rechenmacher}, {Quinn}, {Richards}, {Richmond}, {Rivetta}, {Rockosi}, {Ruthmansdorfer}, {Sandford}, {Schlegel}, {Schneider}, {Sekiguchi}, {Sergey}, {Shimasaku}, {Siegmund}, {Smee}, {Smith}, {Snedden}, {Stone}, {Stoughton}, {Strauss}, {Stubbs}, {SubbaRao}, {Szalay}, {Szapudi}, {Szokoly}, {Thakar}, {Tremonti}, {Tucker}, {Uomoto}, {Vanden Berk}, {Vogeley}, {Waddell}, {Wang}, {Watanabe}, {Weinberg}, {Yanny}, {Yasuda}, \& {SDSS Collaboration}}]{2000York}
{York}, D.~G., {Adelman}, J., {Anderson}, John~E., J., {et~al.} 2000, \aj, 120, 1579

\bibitem[{{Zabel} {et~al.}(2021){Zabel}, {Davis}, {Smith}, {Sarzi}, {Loni}, {Serra}, {Lara-L{\'o}pez}, {Cigan}, {Baes}, {Bendo}, {De Looze}, {Iodice}, {Kleiner}, {Koribalski}, {Peletier}, {Pinna}, \& {de Zeeuw}}]{Zabel2021}
{Zabel}, N., {Davis}, T.~A., {Smith}, M. W.~L., {et~al.} 2021, \mnras, 502, 4723

\bibitem[{{Zibetti} {et~al.}(2009){Zibetti}, {Charlot}, \& {Rix}}]{2009Zibetti}
{Zibetti}, S., {Charlot}, S., \& {Rix}, H.-W. 2009, \mnras, 400, 1181

\end{thebibliography}

\begin{appendix}

\clearpage
\section{Constructing informative priors in the forms of scaling relations on spatially resolved scales}
\label{sec:appendix_priors}
\FloatBarrier 

A prior can be constructed based on empirical data from previous observations that are deemed reliable. In the study of spatially resolved properties of local galaxies, it has been known that \massd\ correlates strongly with other key physical properties, such as SFR surface density, $Z$, stellar population age, and gas-phase metallicity \citep[see review by][]{Sanchez2020}. These empirical scaling relations can be used as a prior in SED fitting. Previous studies have applied this on a global scale. For example, \citet{Leja2019} adopted the mass-$Z$ relation observed by \citet{Gallazzi2005} as a prior in their SED fitting. We extend this attempt to spatially resolved scales.  

To construct a prior in the form of mass-age and mass-$Z$ relations (i.e.~mass-$Z$-age prior), first, we combine observational data taken from the MaNGA survey and simulated data from the TNG50-SKIRT database. The MaNGA data is in the form of 2D maps of stellar population properties of 10\,220 local ($0.01<z<0.15$) galaxies that are produced by the pipe3D pipeline \citep{Sanchez2022,2022Lacerda}.\footnote{\url{https://www.sdss4.org/dr17/manga/manga-data/manga-pipe3d-value-added-catalog/}} We retrieve the maps of all the galaxies and combine their spaxels, which totals 16\,250\,999. For the simulated data, we combine pixels from all maps of 1\,160 galaxies produced by the TNG50-SKIRT Atlas. For simplicity and to limit the number of pixels, we only use the maps from one viewing angle: ``O1''. The total of collected pixels is 488\,235\,803. These are original simulated maps, without post-processing as done for the synthetic imaging data (see Sect.~\ref{sec:gen_noisefree_images}).   

\begin{figure}[H]
\vspace{-10pt}
\centering
\includegraphics[width=0.35\textwidth]{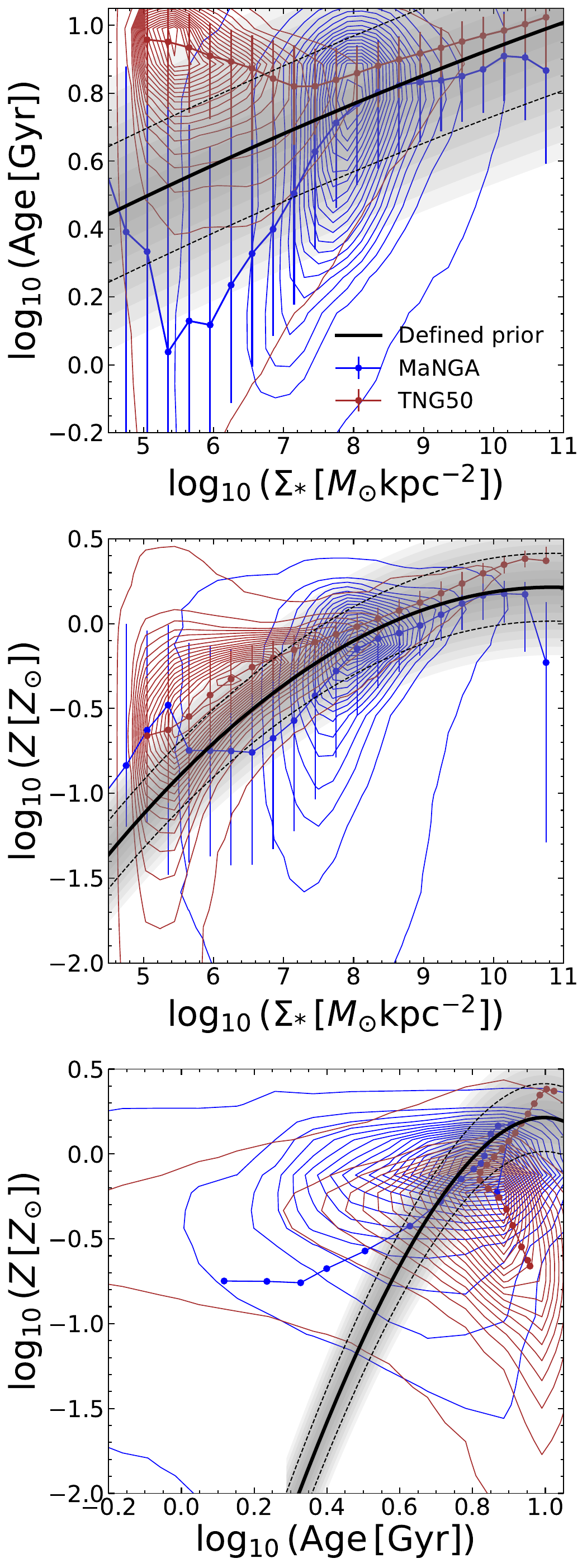}
\caption{Constructing informative priors in the forms of mass-$Z$-age relations based on observational data from the MaNGA survey and simulations from TNG50. The median profiles are shown by solid black lines, which we use for priors. The dashed black lines indicate $\pm 0.2$ around the median profiles. The rightmost panel shows age-$Z$ relation. Its median profiles are derived from combined median profiles of mass-$Z$ and mass-age relations to illustrate how the mass-$Z$ and mass-age priors behave on the $Z$-age plane.} 
\label{fig:priors_age_metals}
\end{figure}

Using these data, we examine the scaling relations between \massd, age, and $Z$ in both observations and simulations. These mass-$Z$-age relations are shown in Fig.~\ref{fig:priors_age_metals}. Tight relationships are evident from the density contours and the median profiles of both simulation and empirical data, especially for the mass-$Z$ and mass-age relations. The age-$Z$ relation exhibits greater scatter than the other two relations. Its median profiles are derived from combined median profiles of mass-$Z$ and mass-age relations rather than from the data directly. This is to illustrate how the mass-$Z$ and mass-age priors behave on the $Z$-age plane. Further analysis of these scaling relations is beyond the scope of this paper and we defer this for future work.  

The deviation in the median profiles of the mass–age and mass–$Z$ relations from MaNGA and TNG50 primarily occurs in low \massd\ regions, below $\sim 10^{7}\,M_{\odot} \, \rm{kpc}^{-2}$. In higher mass density regions, the relations from MaNGA and TNG50 appear consistent, albeit with slightly different normalizations. The contours of MaNGA data reveal that most spaxels lie above this density threshold, and the majority of our sample galaxies exhibit spatially resolved mass densities above this threshold (see Figures C.1–C.3). 

From these relations, we construct priors by further taking the median between the MaNGA and TNG50 median profiles for mass bin above $\Sigma_{*}=10^{6}\,\msun\,{\rm kpc}^{-2}$ and then fit the median profile with a second-degree polynomial function. The prior is defined to be a Gaussian function along the $y$ axis, centred at the median profile (for each point of \massd) and standard deviation of $0.3$ dex. 

\begin{figure*}[t]
\centering
\includegraphics[width=0.75\textwidth]{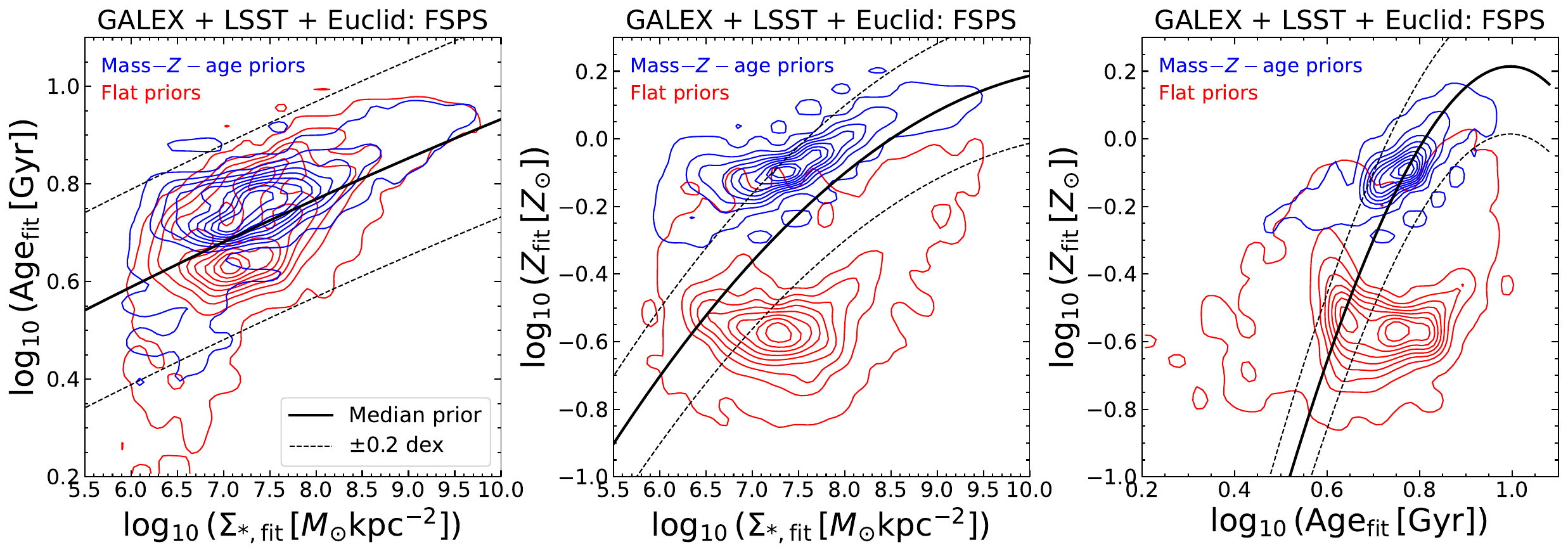}
\includegraphics[width=0.75\textwidth]{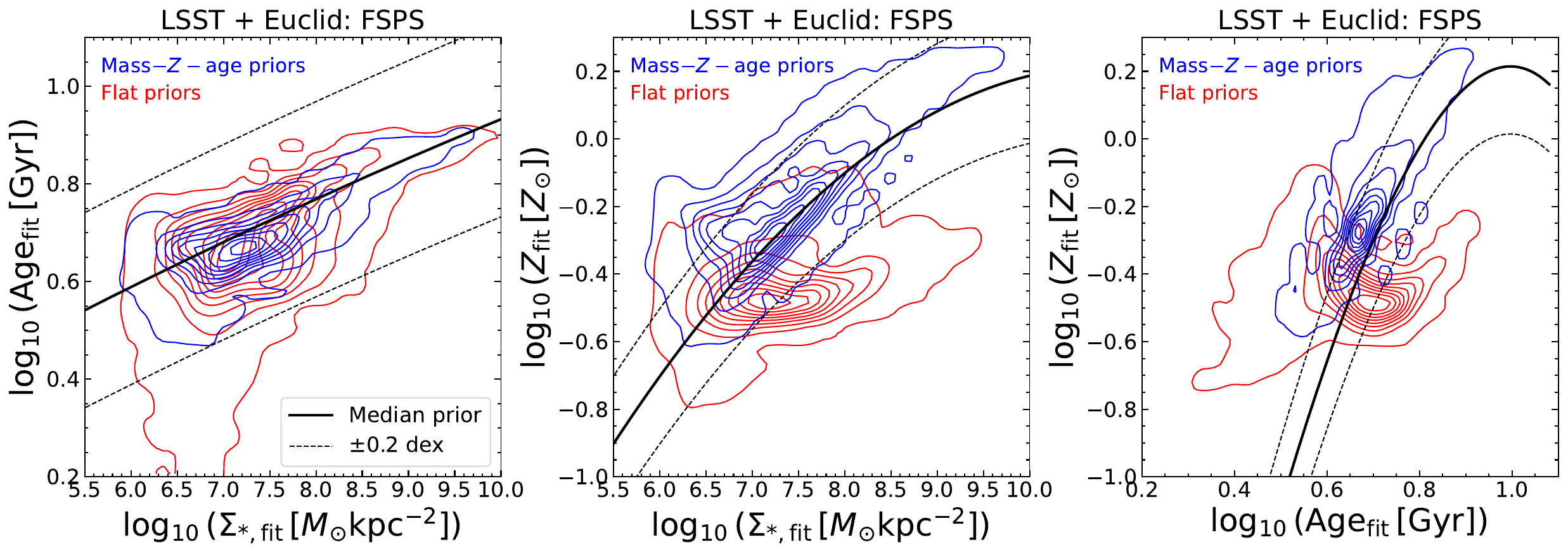}
\includegraphics[width=0.75\textwidth]{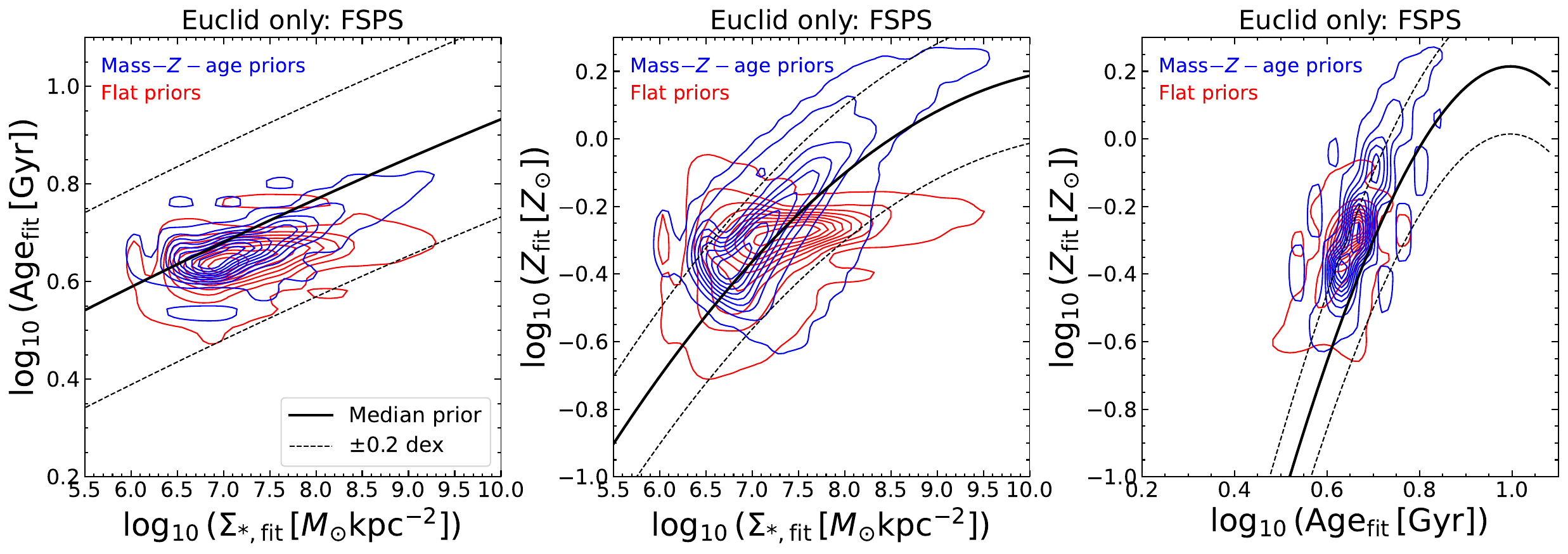}
\caption{Distributions of derived surface mass density, $Z$, and age values, compared with the mass–$Z$–age priors. From top to bottom, results for \gledc, \ledc\, and E-only data cubes are shown. The solid lines show the median prior, while the dashed lines show the $\pm 0.2$ dex range. The colour scheme for SED fitting results is similar to Figs. \ref{fig:comp_simfits_GLE}, \ref{fig:comp_simfits_LE}, and \ref{fig:comp_simfits_E}.}
\label{fig:priors_vs_data}
\end{figure*}

In Fig. \ref{fig:priors_vs_data}, we have also explored the impact of these priors on the results of the paper shown in Figs. \ref{fig:comp_simfits_GLE}, \ref{fig:comp_simfits_LE}, and \ref{fig:comp_simfits_E}. 
In the upper row, the location of the retrieved properties relative to the prior, shown in blue when the mass-$Z$-age prior is implemented, is comparable to the location seen for the TNG50 models in Fig.\,\ref{fig:priors_age_metals}, and suggests that the prior is efficient without being the sole driver of the results when G+L+E data are used (the retrieved parameters are not exactly located along the ridge-line of the prior). In the lower two panels, the results are closer to the prior, indicating the role of the prior relative to other constraints is becoming predominant.

\section{SED fitting tests using simple mock SEDs}
\label{sec:mock_sedfit_test}


\begin{figure*}[h]
\centering
\includegraphics[width=0.85\textwidth]{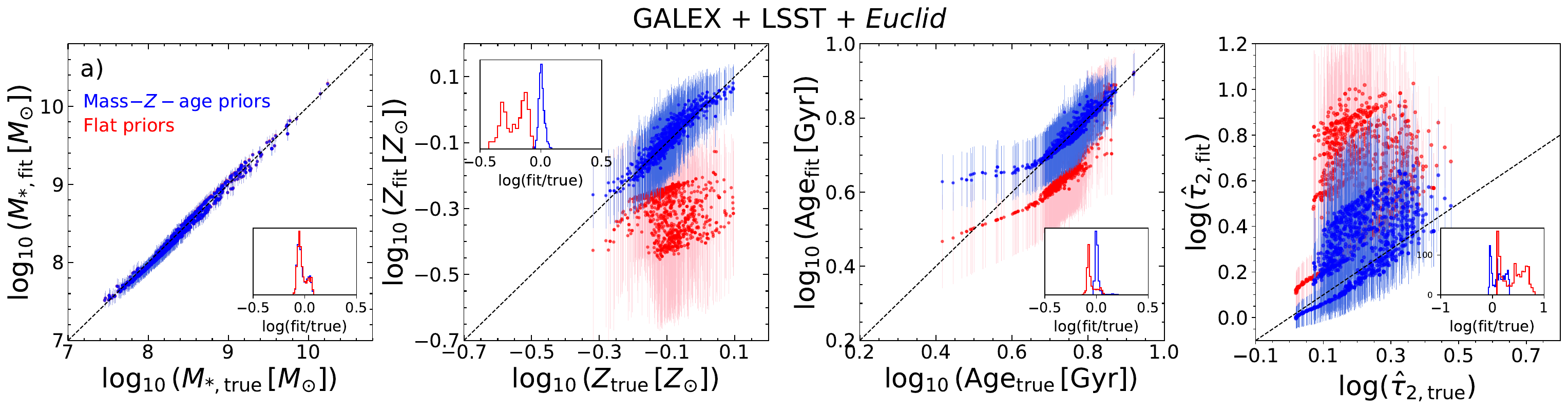}
\includegraphics[width=0.85\textwidth]{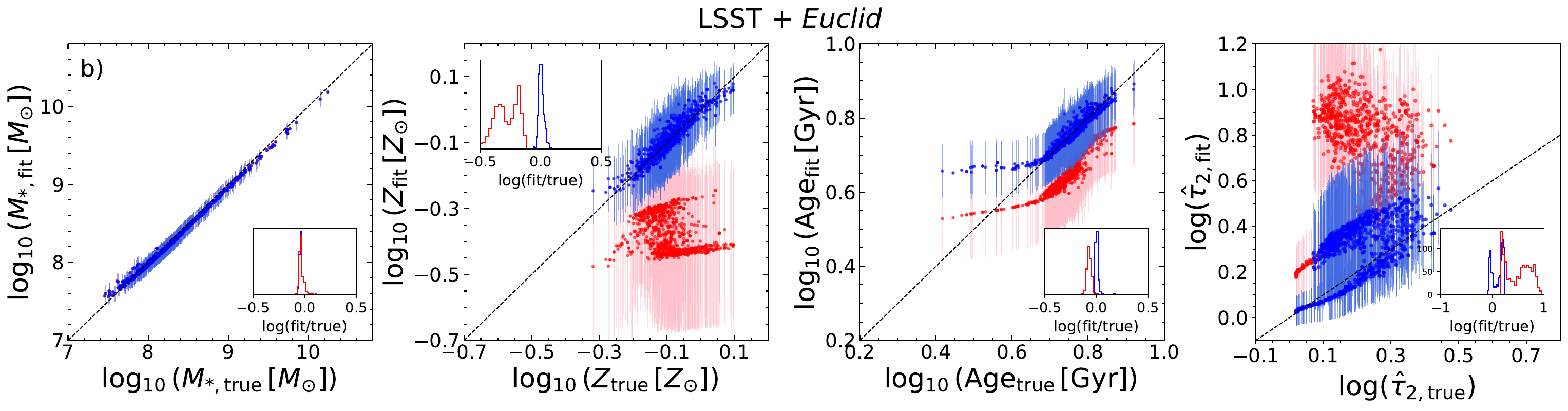}
\includegraphics[width=0.85\textwidth]{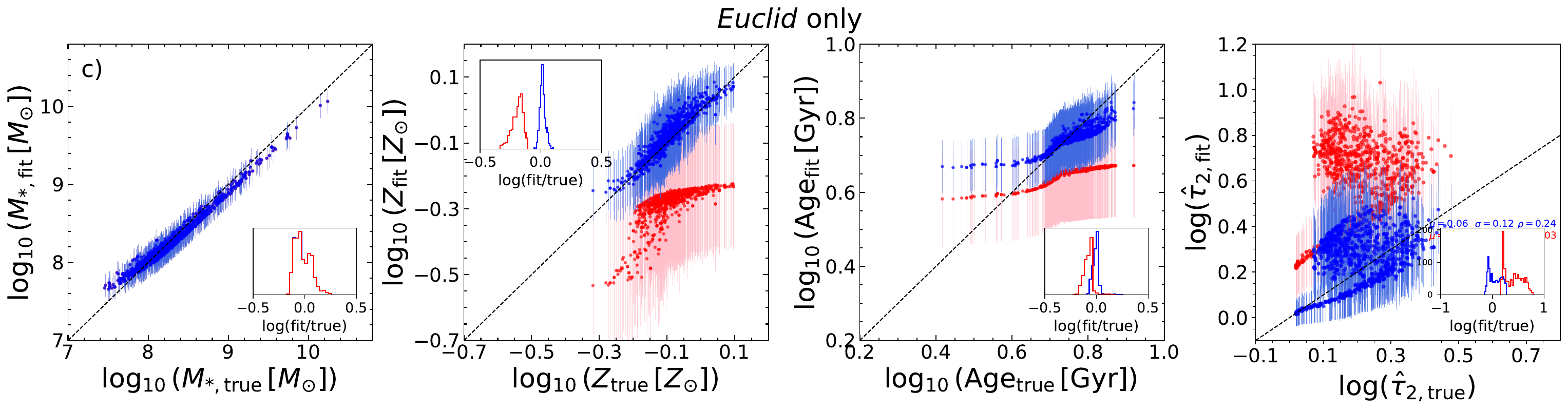}
\caption{Results of SED fitting tests using simple mock SEDs generated with FSPS. Three sets of mock SEDs are made: \gledc, \ledc, and E, and the results of fitting on them are shown in the top, middle, and bottom rows, respectively. We also fit the mock data with and without applying the mass-age and mass-$Z$ priors and the results are shown in blue and red colours, respectively. The histograms in the insets show ratios (in logarithmic scale) between the best-fit and true values.}
\label{fig:mock_sedfit_tests}
\end{figure*}

\begin{table*}[h]
\caption{The robustness of the recovery of the stellar population properties from the SED fitting test on mock SEDs that apply the mass-$Z$-age priors. \label{tab:apdx_robustness_fits_with_priors}}
\centering
\begin{tabular}{lSSS}
\hline \hline \\ [-5pt]
 & \multicolumn{3}{c}{Data cube} \\
\cline{2-4} \\ [-5pt] 
 & \multicolumn{1}{c}{\galex\ + \lsst\ + \euclid} & \multicolumn{1}{c}{\lsst\ + \euclid} & \multicolumn{1}{c}{\euclid-only} \\
\hline \\ [-5pt]
Stellar mass density & & & \\
\hline \\ [-5pt]
$\mu$ & -0.03 & -0.03 & -0.02 \\ 
$\sigma$ & 0.04 & 0.03 & 0.08 \\
$\rho$ & 1.00 & 1.00 & 0.98 \\
\hline \\ [-5pt]
Age & & & \\
\hline \\ [-5pt]
$\mu$ & 0.01 & 0.01 & -0.01 \\ 
$\sigma$ & 0.02 & 0.03 & 0.04 \\
$\rho$ & 0.97 & 0.97 & 0.93 \\
\hline \\ [-5pt]
Metallicity & & & \\
\hline \\ [-5pt]
$\mu$ & 0.01 & 0.01 & 0.01 \\ 
$\sigma$ & 0.02 & 0.02 & 0.02 \\
$\rho$ & 0.93 & 0.93 & 0.93 \\
\hline \\ [-5pt]
Dust optical depth & & & \\
\hline \\ [-5pt]
$\mu$ & 0.11 & 0.10 & 0.06 \\ 
$\sigma$ & 0.12 & 0.11 & 0.12 \\
$\rho$ & 0.56 & 0.49 & 0.24 \\
\hline
\end{tabular}
\end{table*}

\begin{table*}[h]
\caption{The robustness of the recovery of the stellar population properties from the SED fitting test on mock SEDs that do not apply the mass-$Z$-age priors. \label{tab:apdx_robustness_fits_no_priors}}
\centering
\begin{tabular}{lSSS}
\hline \hline \\ [-5pt]
 & \multicolumn{3}{c}{Data cube} \\
\cline{2-4} \\ [-5pt] 
 & \multicolumn{1}{c}{\galex\ + \lsst\ + \euclid} & \multicolumn{1}{c}{\lsst\ + \euclid} & \multicolumn{1}{c}{\euclid-only} \\
\hline \\ [-5pt]
Stellar mass density & & &\\
\hline \\ [-5pt]
$\mu$ & -0.03 & -0.03 & -0.02 \\ 
$\sigma$ & 0.04 & 0.03 & 0.08 \\
$\rho$ & 1.00 & 1.00 & 0.98 \\
\hline \\ [-5pt]
Age & & &\\
\hline \\ [-5pt]
$\mu$ & -0.05 & -0.07 & -0.08 \\ 
$\sigma$ & 0.04 & 0.03 & 0.04 \\
$\rho$ & 0.97 & 0.96 & 0.97 \\
\hline \\ [-5pt]
Metallicity & & & \\
\hline \\ [-5pt]
$\mu$ & -0.22 & -0.28 & -0.19 \\ 
$\sigma$ & 0.10 & 0.10 & 0.05 \\
$\rho$ & 0.13 & 0.30 & 0.81 \\
\hline \\ [-5pt]
Dust optical depth & & & \\
\hline \\ [-5pt]
$\mu$ & 0.40 & 0.50 & 0.41 \\ 
$\sigma$ & 0.22 & 0.23 & 0.17 \\
$\rho$ & 0.17 & -0.04 & 0.03 \\
\hline
\end{tabular}
\end{table*}

The complexity of the radiative transfer process in generating synthetic images might distort the SEDs of individual pixels and change them from being pure spectra of the model of a composite stellar population (CSP). This complexity arises from the intricate spatial configuration of stars and dust within galaxies, which shapes the composite SED of each pixel in the synthetic images produced through radiative transfer modelling. In a dust-free environment, the pixel's SED is simply the sum of the spectra from the stellar populations within that pixel. However, when dust is present, the situation becomes more complex: the amount of attenuation affecting each spectrum varies depending on the distribution of dust particles (and the resulting cumulative optical depth) along the line of sight to each source. 

This can make our fitting tests on the synthetic images non-trivial. To further test our SED fitting pipeline, especially our new priors, we perform additional SED fitting tests using simple mock SEDs generated from CSP models. We generate mock SEDs using FSPS. Instead of generating random parameters, the properties of mock data are adopted from the inferred properties of spatial bins that are randomly chosen from our simulated galaxy sample. For this, we use results from the analysis on the \gledc\ data, in which the fitting uses the FSPS model and applies the mass-$Z$-age prior. We randomly chose 40 spatial bins from each galaxy to take their properties and produce 1\,000 mock SEDs. To check the effect of wavelength coverage (by the photometric data), we make three data types (\gledc, \ledc, and E) by simply removing the photometry associated with the filters. We assign $\rm{S}/\rm{N}$ of 5 in all filters. The flux is perturbed by adding Gaussian noise with a standard deviation equal to Flux/5. The flux uncertainty is then computed based on the perturbed flux value.  

We run fitting to the mock SEDs using the same set-ups as those for the tests on synthetic images, described in Sect.~\ref{sec:sed_fitting}. To check the effect of the mass-$Z$-age prior, we fit each of the mock SEDs twice, one applies the priors and the other does not. The results, obtained with \gledc, \ledc, and E data cubes, are shown in Fig.~\ref{fig:mock_sedfit_tests}. 
As we can see from this figure, \mass\ is recovered very well, and the \mass\ estimate does not seem to be affected by whether or not the priors are applied. This is in part due to the coverage of NIR from \euclid. 

Without the aid of the mass-$Z$-age priors, age and $Z$ are barely recovered. Similar to the case with synthetic TNG images, both parameters are underestimated. The worst-case scenario occurs when only \euclid\ bands are used. In this case, ages cluster around 4-5 Gyr (i.e.~centred around the peak of the mass-weighted age distribution in the SPS models), and metallicities are concentrated around $0.5-0.6 \, \rm Z_{\rm \odot}$. For data cubes with a wider wavelength coverage, age estimation improves, while metallicities show no further improvement. In all data cubes, to counterbalance the underestimated ages and metallicities, larger dust attenuations (not shown) are generated in the SED fitting procedure.  

Instead, age and $Z$ are recovered relatively well with the three datasets and by applying the priors. $Z$ is recovered equally well with the three datasets, in contrast to the results of fitting tests to the synthetic images, where \ledc\ and E data cubes give $\sim 0.1$ dex offset (see Figs.~\ref{fig:comp_simfits_LE} and~\ref{fig:comp_simfits_E}). On the other hand, using the priors, age is recovered well above $\sim 5$ Gyr with the 3 data cubes, although a slight offset is present in the results with E data. This trend is also different from the results obtained with synthetic images, where age is barely recovered with \ledc\ and E data cube. 

\FloatBarrier 

\section{Maps of the spatially resolved stellar population properties}
\label{sec:maps_properties_all}

\begin{figure*}[h]
\centering 
\includegraphics[width=0.9\textwidth]{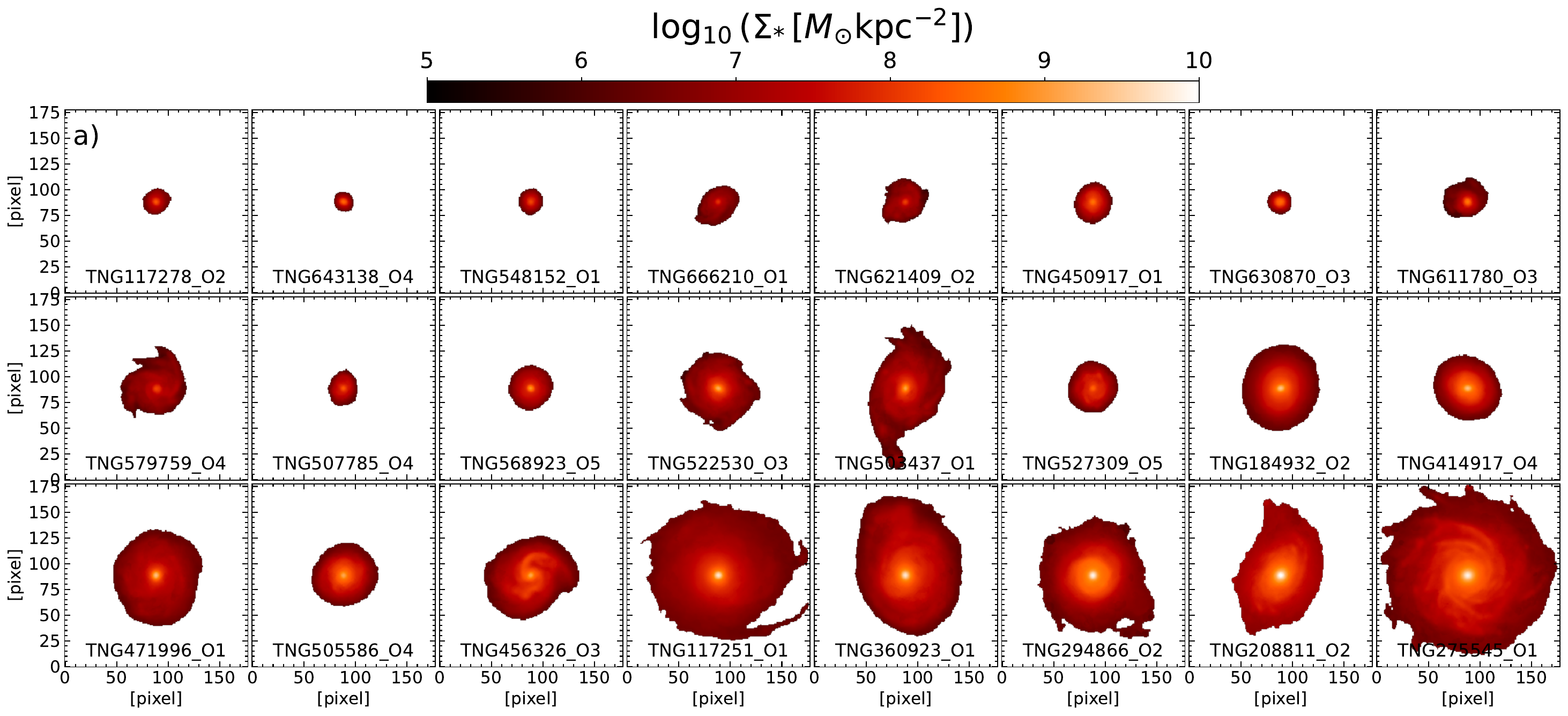}
\includegraphics[width=0.9\textwidth]{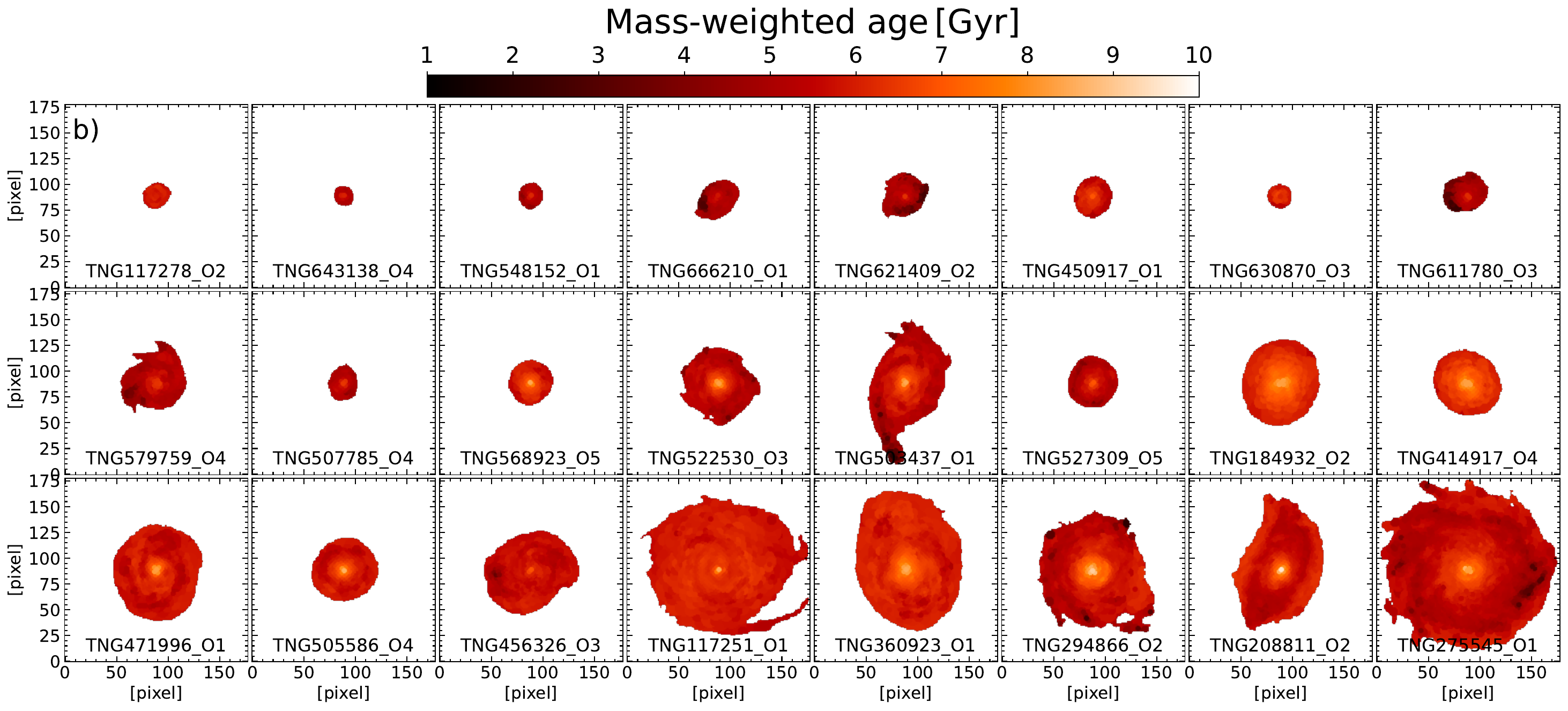}
\includegraphics[width=0.9\textwidth]{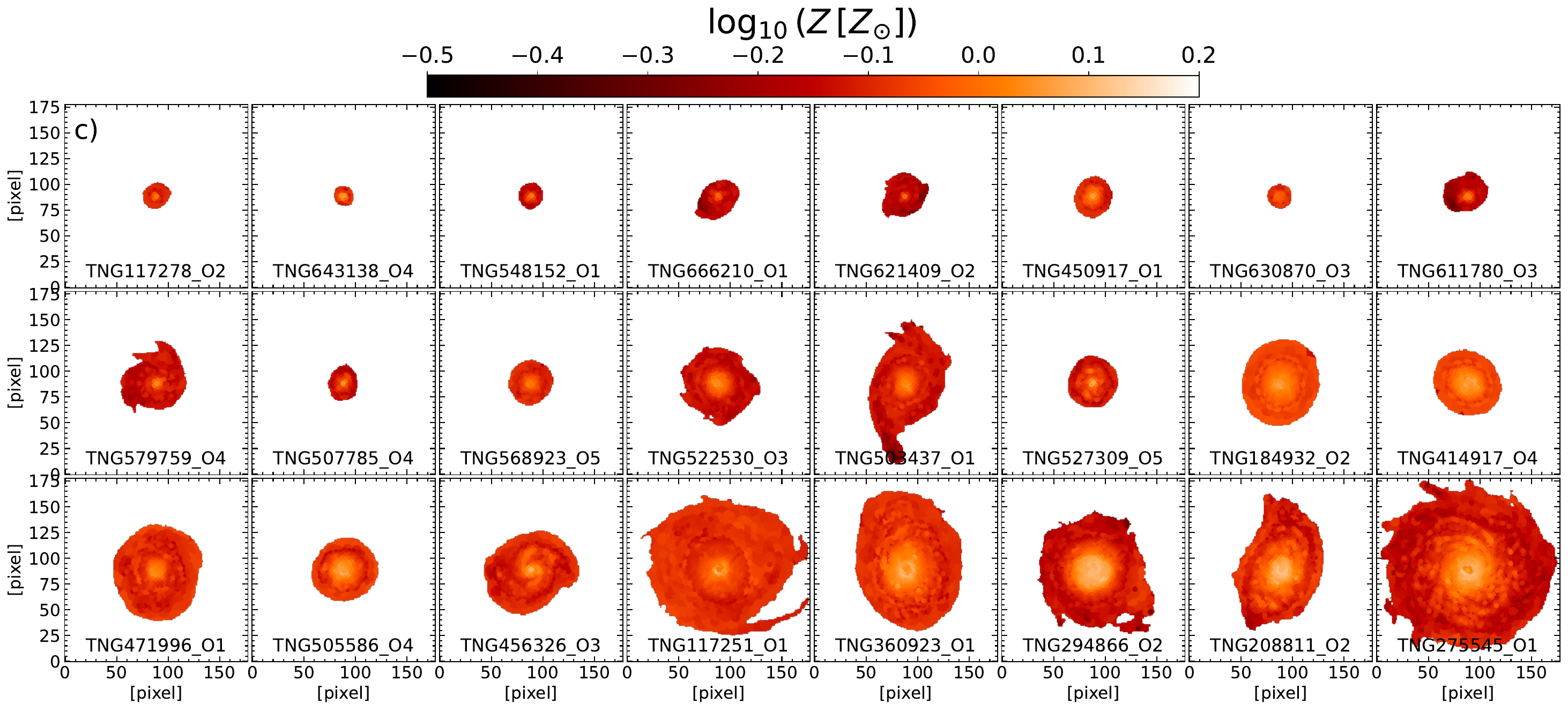}
\caption{Maps of the spatially resolved stellar population properties of 24 galaxies in our sample obtained from analysis on \gledc\ data cubes. The maps include stellar mass surface density (\massd; a), age (b), and metallicity (c).}
\label{fig:comb_maps_props_gle}
\end{figure*}
\begin{figure*}[h]
\centering 
\includegraphics[width=0.9\textwidth]{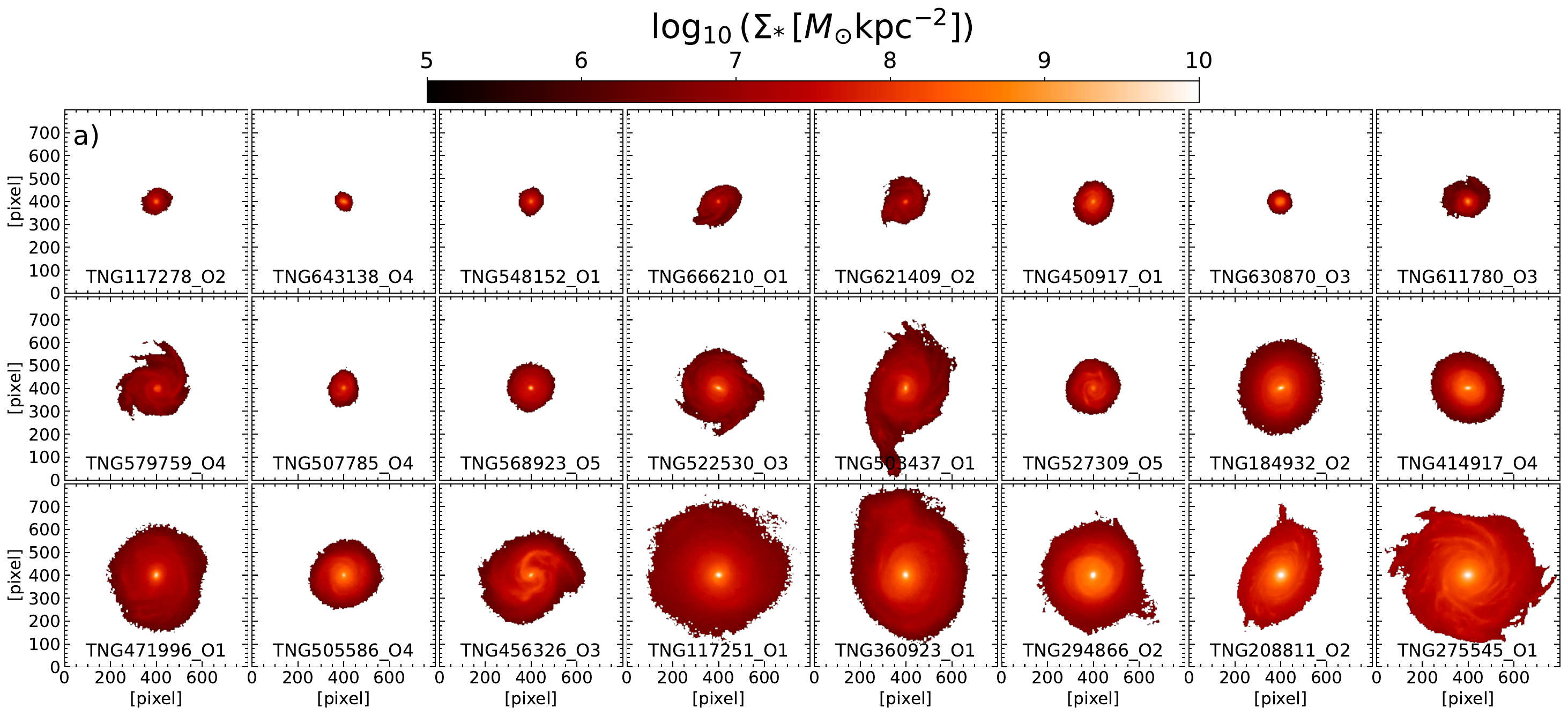}
\includegraphics[width=0.9\textwidth]{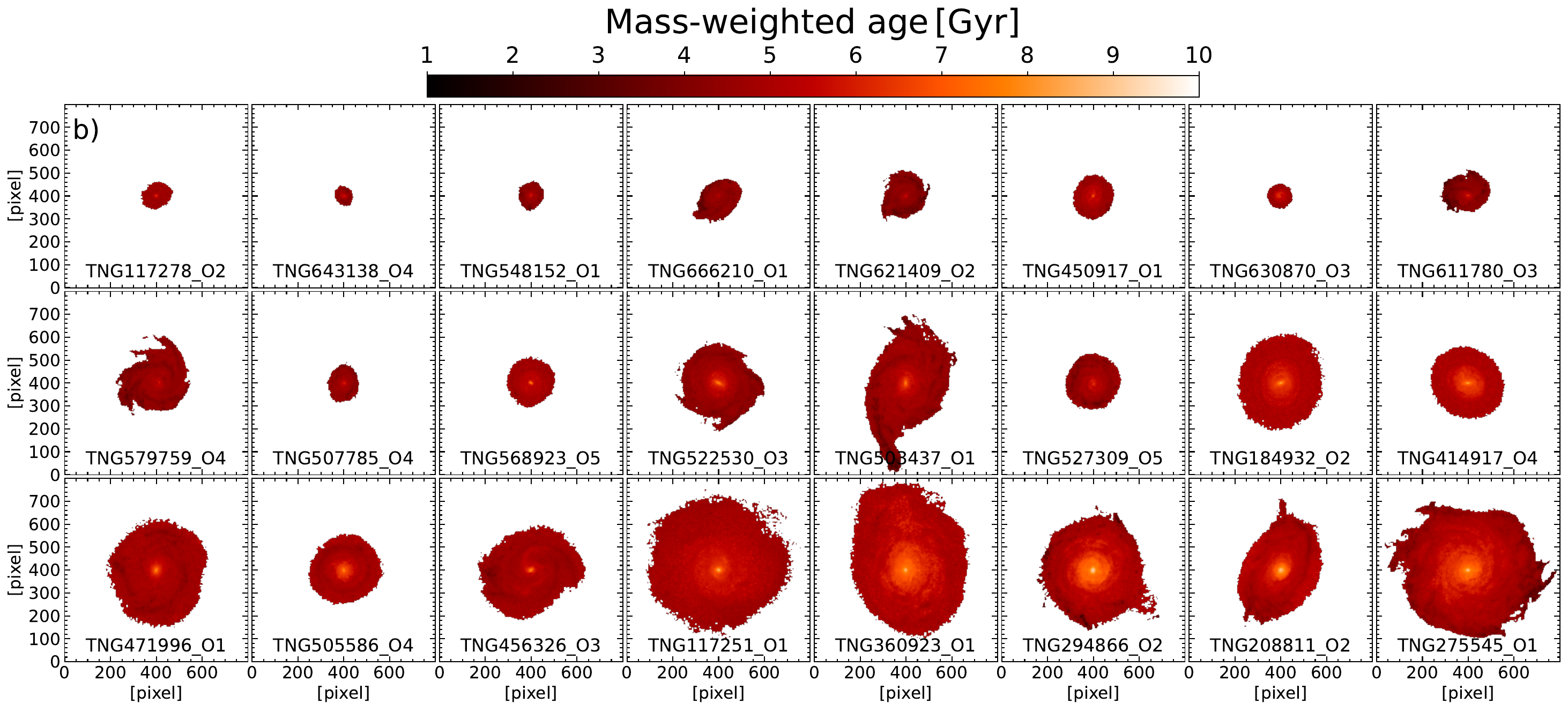}
\includegraphics[width=0.9\textwidth]{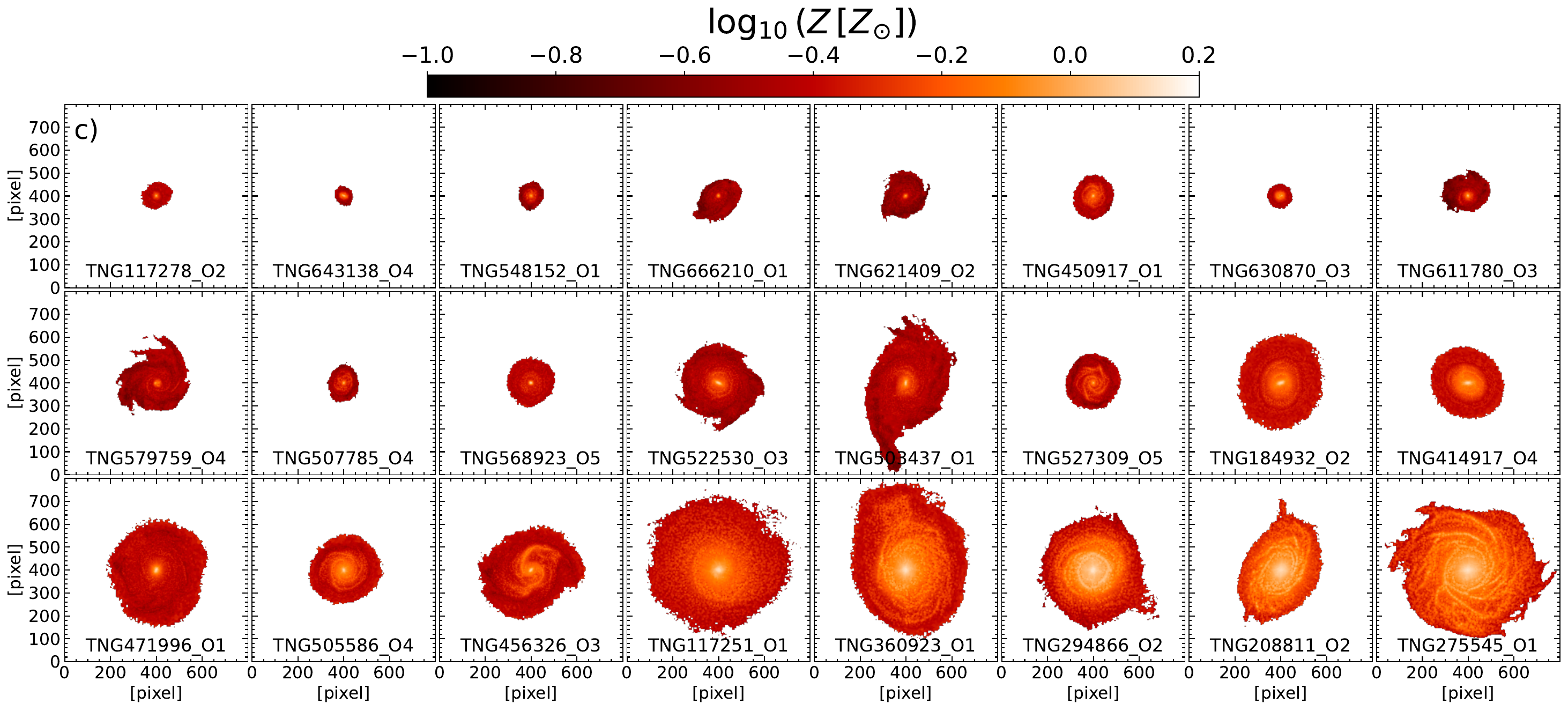}
\caption{Same as Fig.~\ref{fig:comb_maps_props_gle} but for the results obtained from the analysis on \ledc\ data cubes.}
\label{fig:comb_maps_props_le}
\end{figure*}
\begin{figure*}[h]
\centering 
\includegraphics[width=0.9\textwidth]{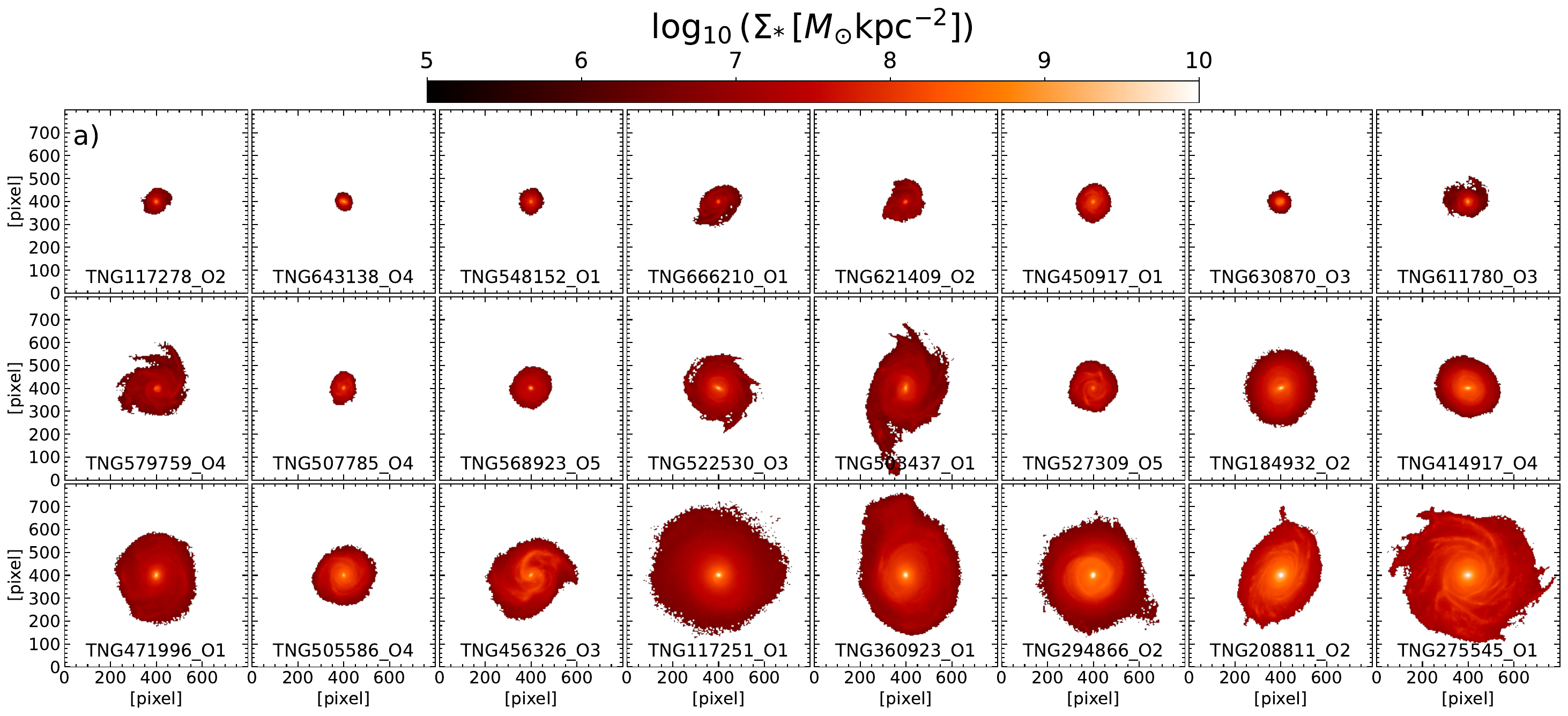}
\includegraphics[width=0.9\textwidth]{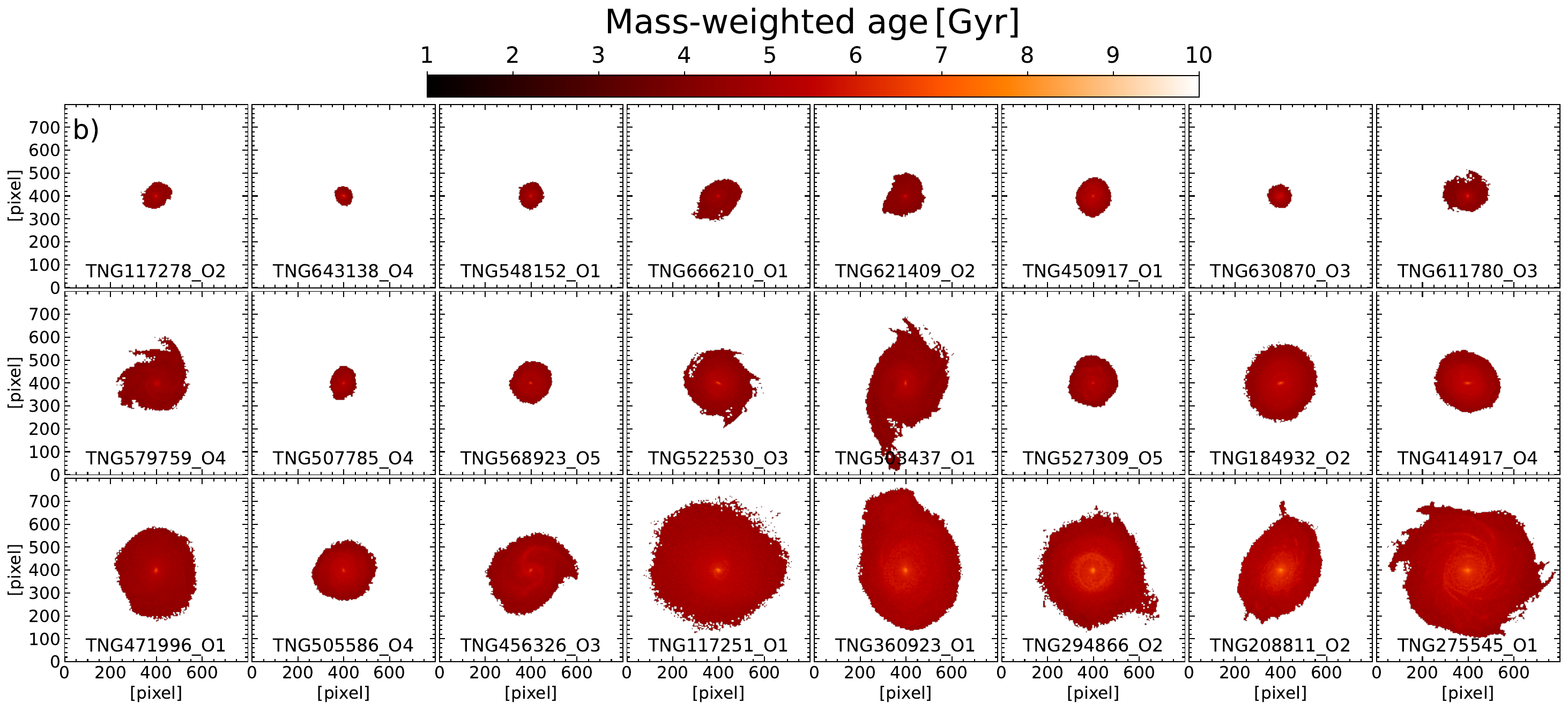}
\includegraphics[width=0.9\textwidth]{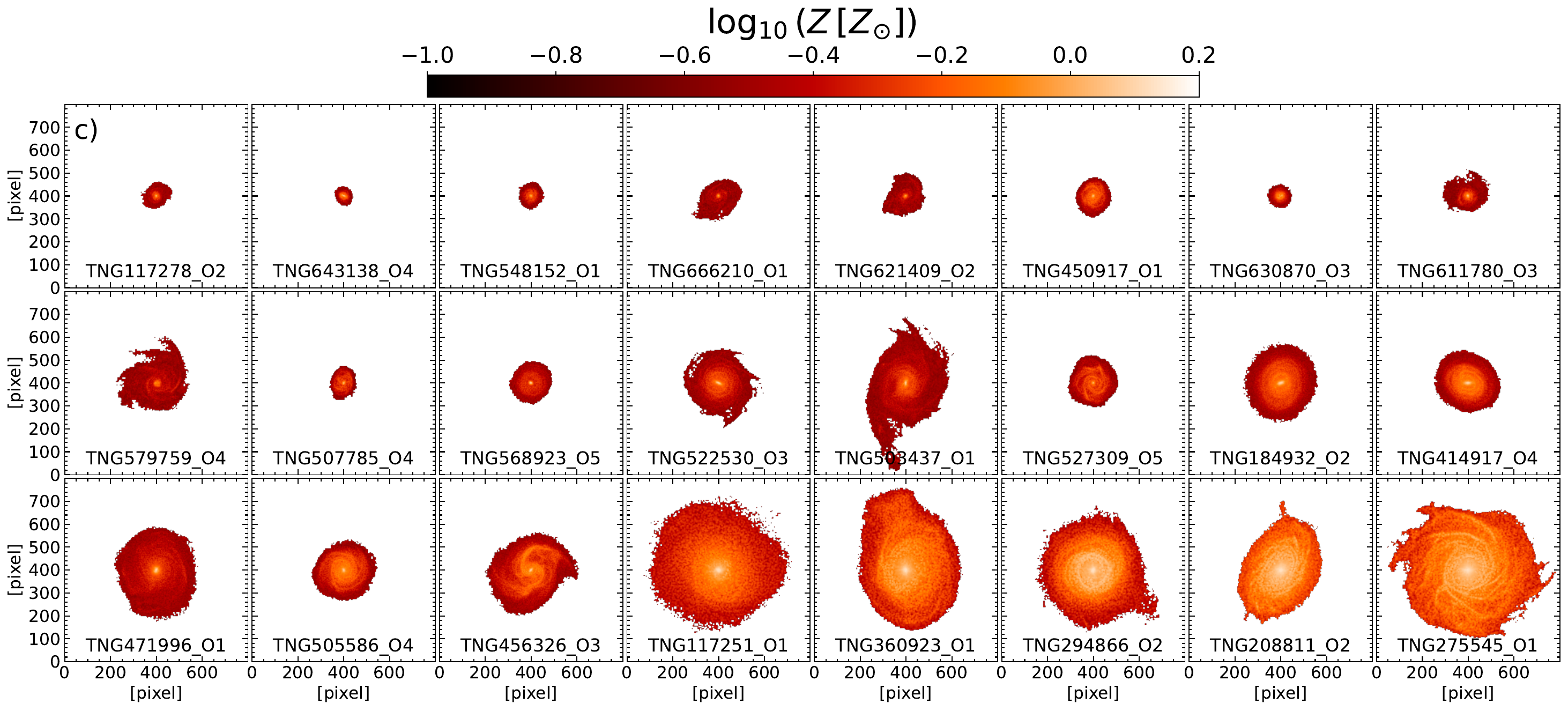}
\caption{Same as Fig.~\ref{fig:comb_maps_props_gle} but for the results obtained from the analysis on E data cubes.}
\label{fig:comb_maps_props_e}
\end{figure*}

Maps of the spatially resolved stellar population properties of 24 galaxies in our sample, derived from the analyses with the three data cubes (\gledc, \ledc, and E) are shown in Figs.~\ref{fig:comb_maps_props_gle}, ~\ref{fig:comb_maps_props_le}, and~\ref{fig:comb_maps_props_e}, respectively. The maps of the other galaxy TNG501725 are shown in Fig.~\ref{fig:maps_props_example}.

\end{appendix}
\end{document}